\documentclass[10pt]{article}

\usepackage{fullpage}
\usepackage{epic}
\usepackage{eepic}
\usepackage{latexsym}
\usepackage[dvips]{graphics, color}

\begin{document}

\title{Theory and Applications of X-ray Standing Waves in Real Crystals}
\author{I.A. Vartanyants\thanks{%
present address: Department of Physics, University of Illinois, 1110 W. Green
St., Urbana IL 61801; e-mail: vartaniants@mrlxp2.mrl.uiuc.edu} 
and M.V. Kovalchuk}
\maketitle

\begin{center}
\textit{A.V. Shubnikov Institute of Crystallography, Russian Academy of
Science, Leninsky pr. 59, 117333 Moscow, Russia}
\end{center}


\begin{abstract}
Theoretical aspects of x-ray standing wave method for investigation of the
real structure of crystals are considered in this review paper. Starting
from the general approach of the secondary radiation yield from deformed
crystals this theory is applied to different concreate cases. Various models
of deformed crystals like: bicrystal model, multilayer model, crystals with
extended deformation field are considered in detailes. Peculiarities of
x-ray standing wave behavior in different scattering geometries (Bragg,
Laue) are analysed in detailes. New possibilities to solve the phase problem
with x-ray standing wave method are discussed in the review. General
theoretical approaches are illustrated with a big number of experimental
results.
\end{abstract}
\eject

\begin{center}
\textbf{CONTENTS}
\end{center}

\textbf{1. Introduction}

\bigskip\ 

\textbf{2. X-ray dynamical diffraction in real crystals}

2.1 Takagi-Taupin equations

2.2 Susceptibilities

\bigskip\ 

\textbf{3. Theory of x-ray standing waves in a real crystal (general
approach)}

\bigskip\ 

\textbf{4. XSW in a perfect crystal}

4.1 XSW with a big and small depth of yield (extinction effect)

4.2 Multicomponent crystals

4.3 Crystals with an amorphous surface layer

\bigskip\ 

\textbf{5. Bicrystal model (Bragg geometry)}

5.1 Theory

5.2 Experiment

\bigskip\ 

\textbf{6. XSW in Laue geometry}

6.1 Theory

6.2 Experiment

\bigskip

\textbf{7. Model of a multilayer crystal}

7.1 Theory

7.2 Applications

\bigskip

\textbf{8. Crystals with an extended deformation field}

8.1 Crystals with the uniform strain gradient. Bent crystals

8.2 Vibrating crystals

\bigskip

\textbf{9. Phase problem}

\bigskip

\textbf{10. Conclusions}

\bigskip

\textbf{11. Appendix}

Secondary radiation yield from a multilayer crystal (analytical approach)

\eject

\section{Introduction}

A new field in the physics of x-ray diffraction has appeared and
successfully developed during last 30 years. It is based on studying and
using x-ray standing waves (XSW) that are formed in a perfect crystal under
conditions of dynamical diffraction. Apart from general physical interest
involving the enormously sharp change in the interaction of x-rays with
atoms in the crystal and on its surface, this field, as has now become
clear, is highly promising for analyzing the structure of crystals and its
adsorbates at the atomic level.

Actually a standing wave that has the same period as the crystal lattice is
extremely sensitive to the slightest deviation of the atomic planes (or
individual atoms) from their correct position in the perfect crystal (or on
its surface). Thus XSW method is particularly useful in its application for
structural analysis. For this technique, an x-ray interference field (XIF)
is produced by the superposition of, typically, two plane x-ray waves. In
this case we have the following expression for the amplitude of the electric
field in the crystal: 
\begin{equation}
\mathbf{E}(\mathbf{r})=\mathbf{E}_0e^{i\mathbf{k}_0\mathbf{r}}+\mathbf{E}%
_he^{i\mathbf{k}_h\mathbf{r}},  \label{c1.1}
\end{equation}
where $\mathbf{k}_0$ is an incident wave vector, $\mathbf{k}_h=\mathbf{k}_0+%
\mathbf{h,}$ and $\mathbf{h}$ is the reciprocal lattice vector multiplied by 
$2\pi .$ The field intensity is determined by the square of the modulus of
the amplitude $\mathbf{E}(\mathbf{r})$ and is equal to 
\begin{equation}
I(\mathbf{r})=|\mathbf{E}_0|^2\left[ 1+\frac{|\mathbf{E}_h|^2}{|\mathbf{E}%
_0|^2}+2\frac{|\mathbf{E}_h|}{|\mathbf{E}_0|}\cos \left( \mathbf{hr}+\alpha
(\theta )\right) \right] ,  \label{c1.2}
\end{equation}
where $\alpha (\theta )$ is the phase of the ratio $\mathbf{E}_h/\mathbf{E}%
_0 $. The spatial position of the planar wave field is determined by the
phase $\alpha (\theta )$ between the two (electric) field amplitudes.
Generated via Bragg reflection, employing a diffraction vector $\mathbf{h}$,
the x-ray standing wave exists within the overlap region of the incident and
reflected x-ray wave (Fig.1) and the phase $\alpha (\theta )$ and thus the
position of the wave field is a function of the angle $\theta $ measured
from exact Bragg angle, varying by half a diffraction plane spacing within
the total reflection range. Thus, atomic positions can be scanned by the XIF
and exactly determined if the yield of the element specific photoelectrons
or x-ray fluorescence photons is recorded as a function of the glancing
angle.

Structural analysis by the XSW technique represents actually a Fourier
analysis but, in contrast to diffraction techniques, the \emph{atomic}
distribution of an \emph{elemental sublattice} is sampled. The two important
parameters which are determined by an XSW measurement are called coherent
fraction ($F_c^h$) and coherent position ($P_c^h$) and represent the {$%
\mathbf{h}$--th} amplitude and phase, respectively, of the Fourier
decomposition of the distribution of atoms under consideration. The XSW
method is particularly powerful for the analysis of the structure of
adsorbates on crystalline substrates since the position of the adsorbate
atom within the surface unit cell can be determined with high accuracy for
low adsorbate coverages. In case several elements are present on the
surface, $F_c^h$ and $P_c^h$ can be obtained for each elemental sublattice
within one XSW measurement.

An effect involving the existence of a standing wave and the variation of
the total field at the atoms of the crystal lattice has been known for a
long time (see for example \cite{L60,BC64,P78}). However in the conditions
of a classical x-ray diffraction experiment, when the intensity of the
reflected and transmitted waves is measured separately it is manifested very
weakly. This is mainly due to the fact, that the cross sections of the
inelastic scattering channels are considerably smaller, than the cross
section of elastic scattering \cite{IT92}. The standing wave in the crystal
reveals itself in a traditional x-ray diffraction experiment only in the
form of an anomalous angular dependence of the absorption (the anomalous
transmission effect in the Laue case as discovered by Borrmann \cite{B41})
and also the weak asymmetry of the reflectivity curve in the Bragg case \cite
{L60,BC64,P78}.

Batterman \cite{B62} was the first who made an attempt to see the standing
wave and its behavior by measuring the $GeK_\alpha $ fluorescence emitted by
crystal atoms. Despite expectations, the measured curve has the angular
dependence similar to the inverted reflectivity curve of the x-rays. The
structure of the wave field manifested itself very weakly only at the edges
of the total reflection region. It was soon understood \cite{B64}, that this
behavior was due to the fact, that the depth of yield $L_{yi}$ of the
fluorescence radiation exceeds by far the penetration depth of x-rays in the
crystal. This penetration depth while the dynamical diffraction of x-rays is
of the order of extinction length $L_{ex}$. As a result all the radiation
absorbed in the crystal gives rise to a fluorescence signal. Its amount,
following the law of conservation of energy is equal to $1-P_R(\theta )$,
where $P_R(\theta )$ is the reflectivity curve. The secondary radiation (SR)
yield is proportional to the wave field intensity (\ref{c1.2}) at the atoms
only if the condition ($L_{yi}<<L_{ex}$) is fulfilled. Later there were
proposed methods for revealing the structure of the wave field by measuring
the fluorescence yield from the impurity atoms introduced in the lattice of
the crystal matrix at a very small depth \cite{GBB74} or measuring the
fluorescence signal at grazing exit angles \cite{B64}. Evidently the
condition $L_{yi}<L_{ex}$ is satisfied in this cases. Moreover, for a
monolayer of atoms absorbed on the surface of crystal this condition for the
escape depth of the SR is surely satisfied.

The above mentioned problem does not exist in measurements of the
photoelectron emission since electrons escape from a thin subsurface layer
with a thickness of fractions of a micrometer. Already in the first works on
the measurement of the photoelectron emission, carried out in the former
Soviet Union in the early 70's \cite{SKP70,SK72}, the dispersion like
angular dependence corresponding to the behavior of an x-ray standing wave (%
\ref{c1.2}) was observed. What was understood from the very beginning that
this angular dependence of the photoelectron yield curve contain essential
information about the structure of the surface layers. Later this field of
research was developing intensively in several scientific centers of the
former Soviet Union (see for details review paper \cite{KK86} and a book 
\cite{AAI89}).

Already in the middle sixties first attempts to measure different secondary
processes were made. For example measurements of thermal diffuse and Compton
scattering while the existence of the standing wave in the crystal were
reported \cite{AKK65, AKK66, A68} (see also later experimental paper \cite
{SZD88}). The angular dependence of the photoelectric current in the silicon
crystal with $p-n$ junction while the dynamical scattering of x-rays was
measured \cite{BS69,ZKK85}.

During last decade, due to the availability of the synchrotron radiation
facilities of last generation (ESRF, APS, Spring-8) XSW method has become a
useful and even in some cases a routine tool for investigating the surface
of the crystals and the structure of the adsorbates. Most of the results
obtained up to the beginning of 90-th, especially applications of XSW
technique to surface analysis were summarized in a review paper \cite{Z93},
an overview of the method was also given in a number of papers \cite
{MF92,P96,L96,W98}. However one of the important field of applications of
the XSW method for investigation of the structure of the real crystals
(containing different type of defects, implanted crystals, epilayers on the
surface of the perfect crystals, geterostructures etc.) has not been
reviewed up to now. Previous review on this subject \cite{KK86} was written
nearly fifteen years ago (see also the book \cite{AAI89}) and a big number
of new results are not summarized until now. At the same time still it is a
big interest to the foundations of the theory of XSW in real crystals (see
for example recent paper \cite{K98}). In our work we are planning to fill
this gap. Theoretical approach is illustrated by the experimental results
obtained in the Laboratory of Coherent Optics and Synchrotron Radiation of
the Institute of Crystallography RAS. Due to a limited size of this
manuscript we have no possibility to give a detailed consideration of all
results obtained in the field of XSW method in different research centers
all over the world. Some of them are just mentioned or even not mentioned,
but this, surely, does not mean that they are not relevant to the subject.
This can be a special subject of another review paper.

If XSW method in perfect crystals is based on the dynamical theory of x-ray
diffraction (see for e.g. books and reviews \cite{L60,BC64,P78,Z45,J67}) for
the description of the fields and the yield of the secondary radiation from
the real crystals it is most effective to use Takagi-Taupin theory \cite
{TTa,TTb,TTc} of the propagation of x-rays in the deformed crystals. For
convenience of the reader we start Chapter II with formulation of the main
results of this theory that will be used in the following parts of the work.
In the end of the same Chapter for the same reason we give the main
relationships for the description of the real and imaginary part of the
susceptibilities in crystals in x-ray wavelength region. For a recent review
of the dynamical theory of x-ray diffraction in a perfect and deformed
crystals see also \cite{A96a,A96b}. Chapter III gives a general mathematical
formalism for calculating the secondary radiation yield in a real crystal.
This chapter is based on the results of the paper \cite{AK78} and represents
the theoretical foundation for the remainder of this review paper. Next
Chapter IV is devoted to the theory of XSW in the case of perfect crystals.
Peculiarities of the secondary radiation yield with the big and small depth
of yield of the secondary radiation are discussed, in the next subsection
fluorescence and photoemission yield from the crystals containing different
type of atoms is analysed and in the end of the chapter crystals with
amorphous surface layer are discussed. Chapter V is devoted to very
important and often realized case of deformed crystal, that can be
approximated in the frame of bicrystal model. In the first subsection theory
of the secondary radiation yield from such a model crystal in the Bragg
geometry is presented and in the following subsection it is illustrated by a
numerous examples. In the next Chapter VI Laue geometry is considered,
peculiarities of x-ray standing wave behavior in this geometry are discussed
and illustrated by examples. In Chapter VII the bicrystal model is
generalized to the case of a multilayer model of the deformed layer and a
secondary radiation yield from such system is analyzed theoretically and
with its applications to the study of implanted crystals. Secondary
radiation yield from the crystals with extended deformation yield are
considered in Chapter VIII. There is given a detailed description of the
wave fields in the case of the crystals with the uniform strain gradient,
which includes the case of bent crystals and as a special case vibrating
crystals. Next Chapter IX describes one of the important applications of the
XSW analysis: the possibility to solve a phase problem while x-ray
scattering from deformed crystal. This approach opens the possibility to
determine uniquely the structure of the surface layer directly from the
scattering experiment. Last Section X presents a summary and an outlook for
the future applications of the XSW method in real crystals.

\eject

\section{X-ray dynamical diffraction in real crystals}

\subsection{Takagi-Taupin equations}

Directly from the Maxwell's equations for the electric field vector $\mathbf{%
E}(\mathbf{r},\omega )$ ($\omega $ is the frequency of the incident wave)
inside a crystal we can obtain the following wave propagation equation,

\begin{equation}
(\mathbf{\Delta +k^2})\mathbf{E}(\mathbf{r},\omega )-graddiv\mathbf{E}(%
\mathbf{r},\omega )=-k^2\frac{4\pi i}\omega \mathbf{j}(\mathbf{r},\omega ),
\label{c2.1}
\end{equation}
where $k=|\mathbf{k}|=\omega /c$ is the magnitude of the wave vector ($c$ is
the velocity of light), $\mathbf{j}(\mathbf{r},\omega )$ is the current
density induced by the electromagnetic field. This current in the case of
linear electromagnetic wave theory is, in fact, a linear function of $%
\mathbf{E}(\mathbf{r},\omega )$,

\begin{equation}
j_i(\mathbf{r},\omega )=\int d\mathbf{r}^{\prime }\sigma _{ik}(\mathbf{r},%
\mathbf{r}^{\prime },\omega )E_k(\mathbf{r}^{\prime },\omega ),  \label{c2.2}
\end{equation}
where $\sigma _{ik}(\mathbf{r},\mathbf{r}^{\prime },\omega )$ is the
nonlocal tensor of the conductivity of the crystal. In general case equation
(\ref{c2.2}) describes non-local coupling between $\mathbf{j}(\mathbf{r}%
,\omega )$ and $\mathbf{E}(\mathbf{r},\omega )$. It takes into account all
possible interactions (such as elastic Thompson scattering, photoelectric
absorption, Compton scattering and an inelastic scattering on thermal
phonons) between the electromagnetic wave and the crystal \cite{AK68}. The
main contribution to $\sigma _{ik}(\mathbf{r},\mathbf{r}^{\prime },\omega )$
is connected with elastic Thompson scattering and has a strictly local
character (the same is valid for the main inelastic process, that is
photoelectron absorption in the dipole approximation) so we can present the
tensor of the conductivity in the following way:

\begin{equation}
\sigma _{ik}(\mathbf{r},\mathbf{r}^{\prime },\omega )=\sigma (\mathbf{r}%
,\omega )\delta _{ik}\delta (\mathbf{r}-\mathbf{r}^{\prime }),  \label{c2.3}
\end{equation}
where $\delta _{ik}$ is the Kroneker symbol and $\delta (\mathbf{r}-\mathbf{r%
}^{\prime })$ is the Dirac $\delta -$function{.}

For further consideration, if it is not specially noted, we will assume
local coupling (\ref{c2.3}). According to (\ref{c2.3}), the right hand side
of equation (\ref{c2.1}) takes the form,

\begin{equation}
\frac{4\pi i}\omega \mathbf{j(r,}\omega \mathbf{)}=\chi \mathbf{(r,}\omega 
\mathbf{)E(r,}\omega \mathbf{),}  \label{c2.4}
\end{equation}
where $\chi \mathbf{(r,}\omega \mathbf{)=(}4\pi i/\omega )\sigma (\mathbf{r}%
,\omega )$ is the crystal susceptibility, related with the permittivity $%
\varepsilon \mathbf{(r,}\omega \mathbf{)}$ of the crystal by usual equation: 
$\varepsilon \mathbf{(r,}\omega \mathbf{)=}1+\chi \mathbf{(r,}\omega \mathbf{%
)}$ {\footnote{%
Further we shall ommit $\omega $ dependence in $\mathbf{E(r,}\omega \mathbf{)%
}$ and $\chi \mathbf{(r,}\omega \mathbf{)}$}}.

In a perfect (ideal) crystal susceptibility $\chi \mathbf{(r)}$ is a
periodic function with the period of the crystal lattice $\chi \mathbf{(r)=}%
\chi \mathbf{(r+a),a}$ is the translation vector. It can be therefore
expanded as a Fourier series,

\begin{equation}
\chi ^{(id)}\mathbf{(r)=}\sum\limits_h\chi _h^{(id)}\exp (i\mathbf{hr}),
\label{c2.5}
\end{equation}
where $\mathbf{h=}2\pi \mathbf{H,}$ $\mathbf{H}$ is the reciprocal lattice
vector.

We shall assume now, that some part of a crystal lattice (in most of
applications it is a thin surface part of the crystal (see Fig.2)) is weakly
deformed due to epitaxial growth, implantation or some other type of
deformation or defects. It is convenient to describe this weak deformation
field of a crystal lattice by two functions. The first one is the
deformation vector $\mathbf{u}(\mathbf{r})$, which determines the
displacements of atoms in a crystal from the position of perfect lattice and
the second one is the static Debye-Waller factor $e^{-W(r)}$ which takes
into account the random displacements of the atoms from the equilibrium
positions in the $\mathbf{h}$ direction.

In the case of weak deformations, that means that relative displacements are
small on interatomic distances,

\begin{equation}
\left| \frac{\partial u^i}{\partial x_k}\right| <<1,  \label{c2.6}
\end{equation}
the susceptibility of the crystal $\chi (\mathbf{r})$ is defined from that
of a perfect one according to relation \cite{TTb},

\begin{equation}
\chi (\mathbf{r})=\chi ^{(id)}(\bf{r-u} (\bf{r})).
\label{c2.7}
\end{equation}

The Fourier components of the susceptibility in the weakly deformed crystal
(now depending from the coordinate $\mathbf{r}$) can be defined according to
Eq. (\ref{c2.7}) as

\begin{equation}
\chi _h(\mathbf{r})=\chi _h^{(id)}\exp \left[ -i\mathbf{hu}(\mathbf{r}%
)\right] e^{-W(r)}.  \label{c2.8}
\end{equation}

We shall look for the solution of equation (\ref{c2.1}) in the form of the
expansion analogous to the Bloch waves,

\begin{equation}
\mathbf{E(\bf{r}})=\sum\limits_h\mathbf{E}_h(\mathbf{r})\exp (i%
\mathbf{k}_h\mathbf{r}),  \label{c2.9}
\end{equation}

\begin{equation}
\mathbf{k}_h=\mathbf{k}_0+\mathbf{h}.  \label{c2.10}
\end{equation}
Here $\mathbf{k}_0$ and $\mathbf{k}_h$ are the incident and diffracted wave
vectors and the sum has to be taken over all reciprocal lattice vectors $%
\mathbf{h}$. In the case of the weakly deformed crystal, when inequality (%
\ref{c2.6}) is satisfied the amplitudes $\mathbf{E}_h(\mathbf{r})$ in the
expansion (\ref{c2.9}) are slowly varying functions of coordinate (on the
contrary to the Bloch waves in a perfect crystal, when they does not depend
on $\mathbf{r}$). This amplitudes vary significantly on the distances much
bigger, then the X-ray wavelengths (of the order of extinction length $%
L_{ex} $ that will be defined later). Therefore, if we neglect the second
derivatives of $\mathbf{E}_h(\mathbf{r})$ {\footnote{%
In the case of the strong deformation fields, when condition (\ref{c2.6}) is
not satisfied, second derivatives of the amplitudes $\mathbf{E}_h(\mathbf{r}%
) $ also have to be taken into account \cite{A96}.} }we can obtain from (\ref
{c2.1}) the following set of equations,

\begin{equation}
\frac \partial {\partial \mathbf{s}_h}\mathbf{E}_h(\mathbf{r})=\frac{ik}%
2\sum\limits_{h^{\prime }}\left[ \chi _{hh^{\prime }}(\mathbf{r})-\alpha
_{h^{\prime }}\delta _{hh^{\prime }}\right] \mathbf{E}_{h^{\prime }}(\mathbf{%
r}),  \label{c2.11}
\end{equation}
where

\begin{equation}
\alpha _h=\frac{\mathbf{k}_h^2-\mathbf{k}_0^2}{\mathbf{k}_0^2};\frac
\partial {\partial \mathbf{s}_h}=(\mathbf{s}_h\bigtriangledown );\mathbf{s}%
_h=\frac{\mathbf{k}_h}{\left| \mathbf{k}_h\right| },  \label{c2.12}
\end{equation}
In Eq. (\ref{c2.11})

\begin{equation}
\chi _{hh^{\prime }}(\mathbf{r})=\chi _{h-h^{\prime }}^{(id)}\exp \left[
-i\left( \mathbf{h}-\mathbf{h}^{\prime }\right) \mathbf{u}(\mathbf{r}%
)\right] e^{-W(r)},  \label{c2.13}
\end{equation}
and both the displacement field $\mathbf{u}(\mathbf{r})$ and the
Debye-Waller factor $W(r)$ are slowly varying functions of coordinate $%
\mathbf{r}$.

Equations (\ref{c2.11}) are the general case of the so-called Takagi-Taupin
(TT) equations \cite{TTa,TTb,TTc} for the determination of the amplitudes of
the wave fields in the weakly deformed crystals {\footnote{%
In the case of the crystal with statistically distributed defects another
approach of so-called statistical dynamical theory was elaborated (see for
review \cite{K96} and also papers \cite{B89,GV99}).}}. In the limit of a
perfect crystal we have in Eq. (\ref{c2.13}) for functions $\mathbf{u(r)}%
\equiv 0$ and $e^{-W(r)}\equiv 1$ and in this case Eqs. (\ref{c2.11}) will
define the wave field in an ideal crystal lattice.

Taking into account that susceptibility of the crystals in x-ray range of
wavelengths is small ($\chi _h\sim 10^{-5}\div 10^{-6}$) it is possible to
remain in equations (\ref{c2.11}) only the waves $\mathbf{E}_h(\mathbf{r})$
satisfying Bragg condition,

\begin{equation}
\left| \alpha _h\right| \sim \left| \chi _h\right| .  \label{c2.14}
\end{equation}

For definite directions of the incident x-rays condition (\ref{c2.14}) can
be fulfilled simultaneously for a number of waves. It is so-called case of
multiple wave diffraction (see for e.g. book \cite{C84} and review papers 
\cite{C96,H96}).

From the other hand it is possible to find directions for which the
condition (\ref{c2.14}) can be fulfilled only for one reciprocal lattice
vector $\mathbf{h}$, it is so-called case of the two-wave diffraction.
Further we shall restrict ourself only for this case. Moreover we shall
consider, that the deformation field in a crystal $\mathbf{u}(z)$ and the
static Debye-Waller factor $e^{-W(z)}$ depend only from one coordinate $z$,
which is the distance from the entrance surface to the depth of the crystal
and we shall neglect its dependence along the surface.

The x-ray amplitude of the total wave field in such a crystal in the
two-wave approximation is the coherent superposition of the incident and
diffracted waves and according to (\ref{c2.9}) is given by

\begin{equation}
\mathbf{E(r})=\sum\limits_s\left[ \mathbf{e}_{0s}E_{0s}(z)e^{i\mathbf{k}_0%
\mathbf{r}}+\mathbf{e}_{hs}E_{hs}(z)e^{i\mathbf{k}_h\mathbf{r}}\right] ,
\label{c2.15}
\end{equation}
where $\mathbf{e}_0$ and $\mathbf{e}_h$ are the unit polarization vectors
and $s$ is the polarization index. In the x-ray diffraction theory they are
usually defined (see Fig.3) respectively to the so-called scattering plane
i.e. the plane containing the vectors $\mathbf{k}_0$ and $\mathbf{k}_h$.
Polarization vectors normal to the scattering plane are called $\sigma -$%
polarized (in the case of two-wave diffraction $\mathbf{e}_{0\sigma }||%
\mathbf{e}_{h\sigma }$) and polarization vectors lying in the scattering
plane are called $\pi -$polarized (in this case polarization vectors $%
\mathbf{e}_{0\pi }$ and $\mathbf{e}_{h\pi }$ are misaligned by the angle $%
2\theta _B$).

Now directly from the TT equations (\ref{c2.11}) for the scalar amplitudes $%
E_0(z)$, $E_h(z)$ and for the fixed polarization $s$ we have, 
\begin{eqnarray}
\frac{dE_{0s}(z)}{dz} &=&\frac{i\pi }{\lambda \gamma _0}\left[ \chi
_{00}E_{0s}(z)+\chi _{0h}Ce^{i\varphi (z)-W(z)}E_{hs}(z)\right] ,  \nonumber
\\
\frac{dE_{hs}(z)}{dz} &=&\frac{i\pi }{\lambda \gamma _h}\left[ \left( \chi
_{hh}-\alpha \right) E_{hs}(z)+\chi _{h0}Ce^{-i\varphi
(z)-W(z)}E_{0s}(z)\right] .  \label{c2.16}
\end{eqnarray}
Here $\varphi (z)=\mathbf{hu}(z)$; $\gamma _{0,h}=\cos (\mathbf{n\cdot k}%
_{0,h})$ are the direction cosines, $\mathbf{n}$ is the inward normal to the
entrance surface of the crystal and $\lambda $ is the wavelength of
radiation. For Bragg geometry of diffraction $\gamma _0>0,\gamma _h<0$ and
for the Laue diffraction $\gamma _0>0,\gamma _h>0$. The parameter $\alpha $
is characterizing the deviation of the wave vector \textbf{k}$_0$ from the
exact Bragg condition,

\begin{equation}
\alpha =\frac{k_h^2-k_0^2}{k_0^2}\approx -2\sin 2\theta _B\left( \theta
-\theta _B\right) ,  \label{c2.17}
\end{equation}
where $\theta _B$ is the Bragg angle; $C$ is the polarization factor defined
as,

\begin{equation}
C=\left\{ 
\begin{array}{c}
1,\sigma - \mbox{polarization} \\ 
\cos 2\theta _{B,}\pi - \mbox{polarization}
\end{array}
\right. .  \label{c2.18}
\end{equation}
In most of the situations considering only the strongest elastic scattering
and the photoelectric scattering process in dipole approximation we have for
the Fourier components of the susceptibility in Eq. (\ref{c2.16}): $\chi
_{00}=\chi _{hh}=\chi _0,\chi _{0h}=\chi _{-h}\equiv \chi \overline{_h}$ and 
$\chi _{h0}=\chi _h$.

Takagi-Taupin equations (\ref{c2.16}) have to be supplemented by the
boundary conditions, that for a crystal of thickness $L$ have the following
form for the different geometries of diffraction

\begin{equation}
\left. E_{0s}(z)\right| _{z=0}=E_s^{(in)},\left. E_{hs}(z)\right| _{z=L}=0
\label{c2.19}
\end{equation}
for Bragg geometry and

\begin{equation}
\left. E_{0s}(z)\right| _{z=0}=E_s^{(in)},\left. E_{hs}(z)\right| _{z=0}=0
\label{c2.20}
\end{equation}
for Laue geometry.

Having in mind further applications it is convenient to transform from the
set of equations (\ref{c2.16}) to a single nonlinear equation in the form of
the Rikatti equation for the amplitude function

\begin{equation}
R(z,\theta )=\frac 1{\sqrt{\beta }Y}\left( \frac{E_{hs}(z,\theta )}{%
E_{0s}(z,\theta )}\right) e^{i\varphi (z)},  \label{c2.21}
\end{equation}
where $\beta =\gamma _0/|\gamma _h|$ for Bragg and $\beta =\gamma _0/\gamma
_h$ for Laue geometries of diffraction and $Y=\sqrt{\chi _h/\chi _{\overline{%
h}}}=\mid Y\mid \exp (i\Phi _Y)$ (for centrosymmetric crystal with
monoatomic lattice $\mid Y\mid =1,\Phi _Y=0$). Substituting new function $%
R(z,\theta )$ (\ref{c2.21}) into (\ref{c2.16}) we obtain 
\begin{equation}
\mp iL_{ex}\frac{dR(z,\theta )}{dz}=2[-y(\theta )-iy_0+y_\varphi ]R(z,\theta
)+C_1[1\pm R^2(z,\theta )].  \label{c2.22}
\end{equation}
Here the upper sign correspond to Bragg diffraction and the lower one for
the Laue. We also have introduced the following notations: the angular
deviation from the exact Bragg position is measured by the dimensionless
parameter,

\begin{equation}
y(\theta )=\sqrt{\beta }\frac{\sin 2\theta _B\cdot \left( \theta -\theta
_B\right) }{X_r}\pm \frac{\chi _{0r}(1\pm \beta )}{2\sqrt{\beta }X_r},
\label{c2.23}
\end{equation}
parameters

\begin{equation}
y_0=\pm \frac{\chi _{0i}(1\pm \beta )}{2\sqrt{\beta }X_r}\mbox{ and }%
y_\varphi (z)=\pm \frac{L_{ex}}2\frac{d\varphi (z)}{dz}  \label{c2.24}
\end{equation}
define attenuation of x-rays due to the photoelectric absorption and the
shift of the Bragg position due to deformation in a crystal;

\begin{equation}
C_1=C\left( 1-ip\right) e^{-W(z)},p=-\frac{X_i}{X_r};  \label{c2.25}
\end{equation}
$L_{ex}$ is an extinction length defined as {\footnote{%
We want to note, that our choice of extinction length differ from commonly
used by the factor $\pi .$}},

\begin{equation}
L_{ex}=\frac{\lambda \gamma _0}{\pi \sqrt{\beta }X_r}.  \label{c2.26}
\end{equation}
Here we have also introduced the following parameters $X_r=Re\sqrt{\chi
_h\chi _{\overline{h}}}$ and $X_i=Im\sqrt{\chi _h\chi _{\overline{h}}}$. Now
boundary conditions for equation (\ref{c2.22}) are defined on one surface.
For the Bragg case of diffraction we have $R(z)|_{z=L}=0$ and $R(z)|_{z=0}=0$
for Laue case.

The reflectivity is usually defined for Bragg case as 
\begin{equation}
P_R(\theta )=(1/\beta )\left| E_h(0,\theta )/E_0(0,\theta )\right| ^2
\label{c2.27}
\end{equation}
now has the following form,

\begin{equation}
P_R(\theta )=\left| Y\cdot R(0,\theta )\right| ^2.  \label{c2.28}
\end{equation}

It is easy to obtain solutions of the equation (\ref{c2.22}) in the case of
a perfect thick crystal ($\mu _0L>>\gamma _0$, where $\mu _0$ is a normal
absorption coefficient defined as $\mu _0=k\chi _{0i}$). In this case $%
\varphi (z)=0,e^{-W(z)}=1$ and Eq. (\ref{c2.22}) reduces to an equation with
constant coefficients. So, for thick perfect crystal solution does not
depend on the thickness of a crystal, that is we have $dR/dz=0$. Now from (%
\ref{c2.22}) for Bragg case we obtain directly

\begin{equation}
R_0(\theta )=-\frac 1{C_1}\left[ \left( -y-iy_0\right) +\sqrt{%
(y+iy_0)^2-C_1^2}\right] ,  \label{c2.29}
\end{equation}
where for the square root it is chosen the branch with the positive
imaginary part.

For the amplitude of the refracted wave $E_{0s}(z,\theta )$ we have directly
from the TT equations (\ref{c2.16}) (and taking into account definition (\ref
{c2.21}))

\begin{equation}
\frac{dE_{0s}(z,\theta )}{dz}=\left[ \frac{i\pi \chi _0}{\lambda \gamma _0}-i%
\frac{C_1}{L_{ex}}R(z,\theta )\right] E_{0s}(z,\theta ).  \label{c2.30}
\end{equation}
Formal solution of this equation can be written in the following form,

\begin{equation}
E_{0s}(z,\theta )=E_s^{(in)}\exp \left[ \frac{i\pi \chi _0}{\lambda \gamma _0%
}z-i\frac 1{L_{ex}}\int\limits_0^zdz^{\prime }C_1R(z^{\prime },\theta
)\right]  \label{c2.31}
\end{equation}
and we have for the intensity of the incident wave,

\begin{equation}
I_0(z,\theta )=|E_{0s}(z,\theta )|^2=I_0^{(in)}\exp \left\{ -\frac{\mu _0}{%
\gamma _0}z+\frac 2{L_{ex}}Im\left[ \int\limits_0^zdz^{\prime
}C_1R(z^{\prime },\theta )\right] \right\} .  \label{c2.32}
\end{equation}
In the case of a perfect crystal, $R(z,\theta )\equiv R_0(\theta )$ and we
have from (\ref{c2.32}),

\begin{equation}
I_0(z,\theta )=I_0^{(in)}\exp \left( -\frac{\mu _{in}(\theta )}{\gamma _0}%
z\right) =I_0^{(in)}\exp \left\{ -\frac{\mu _0}{\gamma _0}z+\frac{2z}{L_{ex}}%
Im\left[ C_1R_0(\theta )\right] \right\} ,  \label{c2.33}
\end{equation}
where $\mu _{in}(\theta )$ is an interference absorption coefficient. This
expression takes not only into account normal attenuation of x-rays out of
the angular region of the dynamical diffraction ($y>>1$)

\begin{equation}
I_0(z,\theta )=I_0^{(in)}\exp \left\{ -\frac{\mu _0}{\gamma _0}z\right\}
\label{c2.34}
\end{equation}
but also takes into account a dynamical ''extinction'' effect coming from
the multiple scattering of x-rays on atomic planes in the narrow angular
region of the dynamical diffraction \cite{L60,BC64,P78}. In the region of
the total reflection for $y\simeq 0$, we obtain from (\ref{c2.33})

\begin{equation}
I_0(z,\theta )=I_0^{(in)}\exp \left\{ -\frac{2C}{L_{ex}}z\right\} .
\label{c2.35}
\end{equation}
Here we have taken into account also that y$_0<<1$ and $\mu _0z<<z/L_{ex}$.
From this expression we can see that for the angular position $y=0$ x-rays
are effectively attenuated on the typical distances $z\sim L_{ex}$ that for
the energies $E\sim 1\div 10keV$ are of the order of microns and are much
smaller then normal attenuation distances $z\sim \gamma _0/\mu _0$ that for
the same energies can be of the order of tenth and hundreds of microns (see
e.g. \cite{IT92}).

As we can see from the expression (\ref{c2.35}) extinction depth $L_{ex}$ is
one of the important parameters of the theory that give an effective
attenuation distance for x-rays while the dynamical diffraction. In our
further treatment all other distances will be compared with $L_{ex}$.

Here we want to make several remarks. The amplitudes $E_{0s}(z,\theta )$ and 
$E_{hs}(z,\theta )$ in TT equations (\ref{c2.16}) are complex numbers with
its amplitude and phase. Due to the fact that the dynamical scattering is a
coherent scattering process this two amplitudes are connected with each
other and, for example, in the case of a perfect crystal on its surface we
have from (\ref{c2.21}) for the ratio of these amplitudes on the surface of
the crystal

\begin{equation}
\left. \frac{E_{hs}(z,\theta )}{E_{0s}(z,\theta )}\right| _{z=0}=\frac{%
|E_h(\theta )|}{|E_0(\theta )|}e^{i\alpha (\theta )}=\sqrt{\beta }%
YR_0(\theta ),  \label{c2.36}
\end{equation}
where $R_0$ is defined in (\ref{c2.29}).

Typical behavior of the reflectivity $P_R(\theta )$ and of the phase $\alpha
(\theta )$ in the diffraction region is shown on Fig. 4. In this small
angular region typically of several arcsec the reflectivity $P_R(\theta )$
is of the order of unity and the phase $\alpha (\theta )$ of the wave field
changes from $-\pi $ to $0$ {\footnote{%
Note, that we have defined the E-field as (a) $E_{0,h}e^{i(\mathbf{k}_{0,h}%
\mathbf{r}-\omega t)}$ (see Eq. (2.15)), whereas frequently (b) $%
E_{0,h}e^{i(\omega t-\mathbf{k}_{0,h}\mathbf{r})}$ is used. However, this
only introduces different phase convention $\alpha _a=-\alpha _b$ if we
denote the phase $\alpha $ resulting from the case (a) and (b) with $\alpha
_a$ and $\alpha _b$.}}. Just this fast change of the phase makes x-ray
standing wave method so sensitive to any additional phase shifts.

\subsection{Susceptibilities}

The Fourier components of the susceptibility $\chi _0$ and $\chi _h$ (see
expansion (\ref{c2.5})) are in general complex valued \cite{P78}

\begin{equation}
\chi _h=\chi _{hr}+i\chi _{hi}.  \label{c2.37}
\end{equation}
The real part $\chi _{hr}$ correspond to elastic scattering of x-rays and
imaginary part $\chi _{hi}$ accounts for absorption effects. The values of $%
\chi _{hr}$ and $\chi _{hi}$ are calculated from quantum mechanics (see
Fig.5, where the values of $\chi _{0r}$ and $\chi _{0i}$ are calculated for 
$Si$ and $Ge$ for different energies) and for crystals without
center of symmetry may themselves be complex \cite{P78}. For hard x-ray
energy range $(E\sim 1\div 10keV)$ $\chi _{hr}$ is negative and for the most
of elements is of the order of $10^{-6}$. It is convenient to present it in
the following form \cite{P78},

\begin{equation}
\chi _{hr}=-\left( \frac{r_0\lambda ^2}{\pi \Omega }\right) F_{hr},%
F_{hr}=\sum\limits_j\left( f_j+\Delta f_j\right) e^{-W_j^T}e^{-i\mathbf{%
h\rho }_j}.  \label{c2.38}
\end{equation}
Here $r_0=e^2/mc^2=2.818\cdot 10^{-15}m$ is the classical electron radius, $%
\Omega $ is the unit cell volume and $F_{hr}$ is the structure factor for
the reciprocal lattice vector $\mathbf{h}$. Expression (\ref{c2.38}) is
written for an arbitrary unit cell of a crystal, summation is made over all
atoms of the unit cell, $\mathbf{\rho }_j$ is the coordinate of the $j-$th
atom in a unit cell; $e^{-W_j^T}$ is the thermal Debye-Waller factor that
takes into account the attenuation of the elastic scattering of x-rays due
to a thermal vibrations of the atoms. In equation (\ref{c2.38})

\begin{equation}
f_j(\mathbf{h})=\int n(\mathbf{r})e^{i\mathbf{hr}}d\mathbf{r}  \label{c2.39}
\end{equation}
is an atomic scattering factor for the $j-$th atom in a unit cell. It is
determined by the electron density $n(\mathbf{r})$ in an atom and $\Delta
f_j $ is an account for the dispersion corrections to an atomic scattering
factor. The values of this parameters are tabulated in International Tables
for X-ray Crystallography \cite{IT92}.

As it was already mentioned above, the imaginary part of the susceptibility $%
\chi _{hi}$ takes into account absorption effects. For hard x-rays $(E\sim
1\div 10keV)$ its value (see Fig.5) is two orders of magnitude smaller then
the real part $\chi _{hr}$ $(\chi _{hi}\sim 10^{-7}\div 10^{-8})$ and, for
our choice of the phase in the plane wave (\ref{c2.9}), it is positive. It
can be shown \cite{AK68,O64} that in general case the imaginary part of
susceptibility $\chi _{hi}$ contains contributions from all the inelastic
processes: the photoelectric absorption, Compton scattering and thermal
diffuse scattering

\begin{equation}
\chi _{hi}=\chi _{hi}(Ph)+\chi _{hi}(CS)+\chi _{hi}(TDS).  \label{c2.40}
\end{equation}

The imaginary part of the Fourier component of the susceptibility $\chi
_{hi} $ in a crystal in analogy to (\ref{c2.38}) can also be presented as a
sum of contributions of different atoms

\begin{equation}
\chi _{hi}=\left( \frac \lambda {2\pi \Omega }\right) \sum\limits_j\sigma
_je^{-W_j^T}e^{-i\mathbf{h\rho }_j},  \label{c2.41}
\end{equation}
where $\sigma _j$ are the cross sections of the different inelastic
processes for the $j-$th atom in the unit cell and their values can be
obtained from \cite{IT92,SI70}.

As it was pointed out previously in general case the susceptibility of a
crystal is a tensor and has a non-local character. Being interested in
diffraction and taking into account relationships between the values of the
cross sections of the different processes (see Fig.6)

\begin{equation}
\sigma _T>>\sigma _{Ph}>>\sigma _C\geq \sigma _{TDS},  \label{c2.42}
\end{equation}
where $\sigma _T,\sigma _{Ph},\sigma _C,\sigma _{TDS}$ are the cross
sections of the elastic Thompson scattering, photoelectric absorption,
Compton scattering and thermal diffuse scattering we can neglect in (\ref
{c2.16}) small non-local corrections to $\chi _h$ and treat susceptibilities
as scalar values. However analysing the yield of the secondary radiation
while the dynamical diffraction of x-rays this corrections may be essential
and can not be neglected. For example, while considering Compton and thermal
diffuse scattering it is necessary to account tensor character of $\chi
_{hi} $ and the angular dependence of the corresponding cross sections of
the inelastic scattering (see for e.g. \cite{AA81,GV99a,GV99b}). Different
situation is realized for practically valuable case of fluorescence
radiation and photoelectron emission (according to (\ref{c2.42}) it is the
main inelastic process). So far as these processes are caused by
photoelectron absorption, the total value of each in dipole approximation
does not depend from the direction of propagation of the radiation in an
isotropic crystal the imaginary part of the susceptibilities can be treated
as scalar values and without angular dependence. Further, if not mentioned
specially, we shall consider mainly this case.

Takagi-Taupin equations (\ref{c2.16}) were obtained in the dipole
approximation. Small quadrupole corrections in imaginary part of $\chi _{hi}$%
, if necessary, can be also taken into account. They will bring to
renormalization of the polarization factor $C$, that can become essential
for scattering near adsorption edges and backscattering (see for details 
\cite{W66,VZ97,VZ99}).

\eject

\section{Theory of x-ray standing waves in a real crystal (general approach)}

In this Chapter we shall obtain, using the approach of Afanasev and Kohn 
\cite{AK78}, the general expression for the yield of the secondary radiation
in the case of the dynamical diffraction of X-rays from the deformed crystal
lattice. The amplitudes of the waves $E_0(z,\theta )$ and $E_h(z,\theta )$
in such a crystal can be obtained from the TT equations (\ref{c2.16}) (we
shall consider deformations that depend only from $z$).

To find the intensity of the secondary radiation yield at the depth $z$ in a
crystal, one must determine the number of absorbed quanta in a layer with
the thickness $dz$ per unit area and unit time (Fig. 7). It is proportional
to the loss of the energy field in this layer. From the equation of the
field energy balance we have for the number of absorbed quanta

\begin{equation}
\hbar \omega \frac{dN(z)}{dz}=-div\mathbf{S}(z)=-\frac{dS_z}{dz},
\label{c3.1}
\end{equation}
where $\mathbf{S}(z)$ is the energy flow (Poynting vector), averaged over
the time period of the field oscillations and over the elementary cell of
the crystal. In (\ref{c3.1}) we have taken into account that $\mathbf{S}(z)$
depend only from $z$.

According to the definition of the energy flow vector 
\begin{equation}
\mathbf{S}(z)=\frac c{8\pi }\left[ \mathbf{s}_0|E_0(z)|^2+\mathbf{s}%
_h|E_h(z)|^2\right] ,  \label{c3.2}
\end{equation}
where $\mathbf{s}_0=\mathbf{k}_0/|k_0|$ and $\mathbf{s}_h=\mathbf{k}_h/|k_h|$
are the unit x-ray propagation vectors, we can obtain for the number of
absorbed quanta,

\begin{equation}
\frac{dN(z,\theta )}{dz}=-\frac c{8\pi \hbar \omega }\left[ \gamma _0\frac{%
d|E_0(z,\theta )|^2}{dz}+\gamma _h\frac{d|E_h(z,\theta )|^2}{dz}\right] .
\label{c3.3}
\end{equation}
Taking into account TT equations (\ref{c2.16}) we have 
\begin{eqnarray}
\frac{dN_m(z,\theta )}{dz} &=&\frac{ck}{8\pi \hbar \omega }%
\{E_0^{*}(z,\theta )\chi _{00,i}(m)E_0(z,\theta )+  \nonumber \\
&&\ E_h^{*}(z,\theta )\chi _{hh,i}(m)E_h(z,\theta )+  \label{c3.4} \\
&&\ 2Re[E_0^{*}(z,\theta )\chi _{0h,i}(m)E_h(z,\theta )\exp (i\varphi
(z)-W(z))]\}.  \nonumber
\end{eqnarray}
The number of absorbed quanta is determined only by the imaginary part of
the susceptibility $\chi _{hi}$ that is account for absorption effects.
According to (\ref{c2.40}) it is possible to separate the influence of the
different processes into the yield of the secondary radiation. The index $m$
introduced in (\ref{c3.4}), characterizes this contribution of a certain
secondary process, which is under investigation.

The total number of the secondary quanta emitted from the crystal is equal
to 
\begin{equation}
N_m(\theta )=\int\limits_0^\infty dzP_{yi}^m(z)\frac{dN_m(z,\theta )}{dz},
\label{c3.5}
\end{equation}
where $P_{yi}^m(z)$ is the probability function of the yield of the
secondary radiation of the type $m$ from the depth $z$.

Equations (\ref{c3.4}-\ref{c3.5}) are general and give the solution for the
problem of the angular dependence of the secondary radiation yield, when
x-ray standing wave exist in a crystal. They are valid for any type of
inelastic process such as photoeffect, fluorescence radiation, Compton
scattering and thermal diffuse scattering. They can be applied as well for
investigation of secondary electrons, i.e. Auger electrons and electrons
ejected due to absorption of fluorescence radiation. One must only define
the values of the Fourier components of the susceptibility $\chi
_{00,i},\chi _{hh,i}$, and $\chi _{0h,i}$ appropriately, as well as the
probability function $P_{yi}^m(z)$. The amplitudes $E_0(z,\theta )$ and $%
E_h(z,\theta )$, naturally, does not depend on the type of the inelastic
process that is experimentally registered, but are determined only by the
diffraction process on a real crystal. If, for example, the deformation
field $\mathbf{u}(z)$ and the level of amorphization $W(z)$ are known, then
the amplitudes $E_0(z,\theta )$ and $E_h(z,\theta )$ can be obtained
directly from the TT equations (\ref{c2.16}). On Fig. 8 results of
calculations of the reflectivity curves and photoeffect yield for the
silicon crystal with the known profile of deformation (also shown on Fig. 8)
are presented. The series of curves correspond to both the entire layer
(upper curves) and to parts of it.

In most of the applications of XSW method it is assumed, that the yield of
the inelastic process under investigation is completely determined by the
intensity of the wavefield at the atoms positions. However, according to
result summarized in Eqs. (\ref{c3.4}--\ref{c3.5}), it is not always
fulfilled. In fact there are two main effects that are taken into account.
This is first of all the deformation of a crystal lattice described by an
additional phase factor $\varphi (z)$ (due to the displacement of atomic
planes) and the static Debye-Waller factor $e^{-W(z)}$ (due to the random
displacements of atoms) in the third term of (\ref{c3.4}). The second effect
is coming from non-locality of some of the inelastic processes. This can be,
for example, effects of higher order multipole interactions for the
photoeffect processes \cite{VZ97,VZ99} or non-local character of such
processes as Compton scattering or thermal diffuse scattering \cite
{AA81,GV99a,GV99b}.

Being interested in future mainly by the fluorescence and photoelectron
yield in dipole approximation we can write Eq. (\ref{c3.4}) in the following
form (further we will omit index $m$ and assume, that $\chi _{00,i}=\chi
_{hh,i}\equiv \chi _{0i}$ and $\chi _{0h,i}\equiv \chi _{\overline{h}i}$), 
\begin{eqnarray}
\frac{dN(z,\theta )}{dz} &=&\frac{c\mu _0}{8\pi \hbar \omega }|E_0(z,\theta
)|^2\{1+\frac{|E_h(z,\theta )|^2}{|E_h(z,\theta )|^2}+  \nonumber
 \\
&&\ \ +2Re[\varepsilon _{\overline{h}}\frac{E_h(z,\theta )}{E_0(z,\theta )}%
e^{i\varphi (z)-W(z)}]\},  \label{c3.6}
\end{eqnarray}
where 
\begin{equation}
\varepsilon _{\overline{h}}=\frac{\chi _{\overline{h}i}}{\chi _{0i}}=\frac{%
\sum\limits_j\sigma _je^{-W_j^T}e^{-i\mathbf{h\rho }_j}}{\sum\limits_j\sigma
_j}.  \label{c3.6a}
\end{equation}
Now substituting into (\ref{c3.6}) the expression for the amplitude $%
R(z,\theta )$(\ref{c2.21}) we have for the normalized intensity yield of the
secondary process 
\begin{eqnarray}
\kappa (\theta ) &=&\frac{I(\theta )}{I(\infty )}=\frac 1{I(\infty
)}\int\limits_0^\infty dzP_{yi}(z)|E_0(z,\theta )|^2\{1+\beta
|Y|^2|R(z,\theta )|^2+  \nonumber   \\
&&+2\sqrt{\beta }e^{-W(z)}CRe[\varepsilon _{\overline{h}}YR(z,\theta )]\},
\label{c3.7}
\end{eqnarray}
where intensities are usually normalized by their values far from the region
of the Bragg diffraction $I(\infty )=\int\limits_0^\infty
dzP_{yi}^m(z)|E_0(z,\infty )|^2$.

Equation (\ref{c3.7}) together with equations (\ref{c2.22}, \ref{c2.28}) and
(\ref{c2.33}) completely determine the scheme of calculation of the angular
dependence of the yield of the secondary radiation in the most general case
under the condition that a plane wave is incident on the crystal. In a real
experimental situation the experiment is performed in a double-crystal
scheme with a first crystal-monochromator. In this case for the comparison
of theoretical calculations with the experimental results the convolution
between the curve of the secondary radiation yield and the reflectivity
curve of the monochromator crystal has to be calculated. If an asymmetric
reflection is used in both crystals, i.e. asymmetry factor $\beta $ in the
sample crystal and the asymmetry factor of the monochromator crystal $\beta
_1$ do not equal to unity, then we have for the convoluted curve of the SR
yield 
\begin{equation}
\overline{\kappa }(\theta )=\frac{I(\theta )}{I(\infty )}=\frac{%
\sum\limits_s\int\limits_{-\infty }^{+\infty }dy_1P_R^{(s)}\left( y_1\right)
\kappa ^{(s)}\left( y(\theta )+\sqrt{\beta \beta _1}y_1\right) }{%
\sum\limits_s\int\limits_{-\infty }^{+\infty }dy_1P_R^{(s)}\left( y_1\right) 
}.  \label{c3.8}
\end{equation}
Here $\theta $ is the angle between the reflecting planes of the
crystal-monochromator and the sample crystal and summation in Eq. (\ref{c3.8}%
) is performed over the different polarization states. Convolution with the
reflectivity curve (\ref{c2.28}) of the sample $\overline{P_R}(\Delta \theta
)$ is defined in the same way.

Finally, in this Section we have formulated the main equations for the SR
yield excited by XSW while the dynamical diffraction of x-rays in real
crystals. In the remainder of the work we will analyse different physical
applications to this general formalism.

\eject

\section{XSW in a perfect crystal}

In this Chapter we shall consider an effects of registration of the
different inelastic processes in a simple case of a perfect crystal. Though
it is the most simple case of the real crystal, the main peculiarities of
the XSW field and SR yield can be already revealed and understood in this
case. In the end of the Chapter the case of amorphous layer on the top of
the perfect one is shortly discussed as well.

\subsection{XSW with a big and small depth of yield (extinction effect)}

In the case of the perfect crystal the amplitude $R(z,\theta )\equiv
R_0(\theta )$ (see Eq. (\ref{c2.29})) and it does not depend from the
coordinate $z$. From (\ref{c3.7}) we obtain for the wave field intensity, 
\begin{equation}
I(\theta )=\{1+\beta |Y|^2|R_0(\theta )|^2+2\sqrt{\beta }CRe[\varepsilon _{%
\overline{h}}YR_0(\theta )]\}\int\limits_0^\infty dzP_{yi}(z)|E_0(z,\theta
)|^2.  \label{c4.1}
\end{equation}

In the limit when we can neglect the angular dependence in the last integral
and in addition if approximation $\varepsilon _{\overline{h}}=\chi _{%
\overline{h}i}/\chi _{0i}=1$ is valid we are coming to the well known
expression for the wave field intensity (\ref{c1.2}). From the above
expression we can see that the shape of the curve of the SR from the perfect
crystal is mainly determined by two factors. First of all it depend on the
depth of yield $L_{yi}$ of the SR that is determined by the probability
yield function $P_{yi}(z)$ and, secondly, in the case of detecting SR
process from the multicomponent crystal its shape essentially depend from
the complex factor $\varepsilon _{\overline{h}}$. In the beginning we shall
consider effects of the depth of yield.

For the future analysis it is convenient to take the probability yield
function $P_{yi}(z)$ in the form of an exponential function:

\begin{equation}
P_{yi}(z)=\exp (-\mu _{yi}^{eff}z).  \label{c4.2}
\end{equation}
This form of the probability function is exact for the yield of the
fluorescence radiation with $\mu _{yi}^{eff}=\mu _{yi}^{fl}/\gamma _{fl}$,
where $\mu _{yi}^{fl}$ is an attenuation coefficient of the specific
fluorescence line, that is measured in experiment and $\gamma _{fl}$ is the
cosine of the exit angle of this fluorescence radiation. For the case of
photoeffect integrated over all the directions of the photoelectron yield
this form of $P_{yi}(z)$ with $\mu _{yi}^{eff}=2.3/L_{yi}^{ph}$, where $%
L_{yi}^{ph}$ is an average escape depth of electrons is a good approximation 
\cite{KKL85} to probability yield function obtained from the Monte-Carlo
simulations \cite{L79,KLK86}. For the escape depth of the electrons with an
initial energy $E_i$ (in $keV$) the following approximation formula can be
used 
\begin{equation}
L_{yi}^{ph}(E_i)=780E_i^2/\rho \ln (E_i/E_0),  \label{c4.3}
\end{equation}
where $\rho $ is the density of the material in $g\cdot cm^{-3}$and $%
E_0\approx 0.39keV$.

Taking now into account that according to Eq. (\ref{c2.33}) for the perfect
crystal $|E_0(z,\theta )|^2=|E_0^{(in)}|^2\exp (-\mu _{in}(\theta )z/\gamma
_0)$, we obtain from (\ref{c4.1}) for the yield of the secondary radiation
from a perfect crystal

\begin{equation}
I(\theta )=\frac{I^{SW}(\theta )}{\mu _{yi}^{eff}+\mu _{in}(\theta )/\gamma
_0},  \label{c4.4}
\end{equation}
where

\begin{equation}
I^{SW}(\theta )=1+\beta |Y|^2|R_0(\theta )|^2+2\sqrt{\beta }CRe[\varepsilon
_{\overline{h}}YR_0(\theta )].  \label{c4.5}
\end{equation}

Now we can easily analyse different limits of the depth of yield parameter
on the angular dependence of the SR curve. For example, in the limit when
the escape depth of the secondary radiation $L_{yi}$ is much smaller then
the minimum penetration depth $L_{ex}$ of the standing wave field into a
crystal ($L_{yi}<<L_{ex}$) $\mu _{yi}^{eff}>>\mu _{in}(\theta )$ and from
Eq. (\ref{c4.4}), we obtain

\begin{equation}
I(\theta )=\frac{I^{SW}(\theta )}{\mu _{yi}^{eff}}\left[ 1-\frac 1{\gamma _0}%
\frac{\mu _{in}(\theta )}{\mu _{yi}^{eff}}\ldots \right] .  \label{c4.6}
\end{equation}

The angular dependence of the SR yield in this limit is mainly determined by
the intensity variation of the standing wave field through the atomic planes
(Eq.(\ref{c4.5})). Its shape for the monoatomic crystal will totally
coincide with the intensity variation of the wave field (\ref{c1.2}). At the
same time, following Eq. (\ref{c4.6}) the maximum change in the shape of the
intensity curve is due to an extinction effect that can be seen only in the
central region of the total reflection. It leads to a weak variation in the
slope of the linear part, i.e., to a weak decline in the intensity yield at
this angles. It is interesting to note here, that this actual form of the
standing wave curve contain information about the real escape depth of the
secondary radiation, that can be obtained by fitting experimental data to
theoretical calculations in the form of equation (\ref{c4.6}). This effects
were observed experimentally in the case of the fluorescence radiation \cite
{PG83} (change of the depth of yield $L_{yi}$ was obtained by the change of
the exit angle of the fluorescence yield) and in the case of photoeffect 
\cite{BMK84a} (see Fig. 9).

Now we shall consider opposite limit of the big depth of yield ($%
L_{yi}>>L_{ex}$). This is the typical situation in the case of measuring
fluorescence radiation from the atoms of a crystal lattice or for the
measuring of the inelastic scattering on thermal phonons. For the analysis
of this limit it is useful to use in Eq. (\ref{c4.1}) the following
expression (see for e.g.\cite{BC64}), 
\begin{equation}
\mu _0\left\{ 1+\beta P_R(\theta )+2\sqrt{\beta }Re\left[ \varepsilon _{%
\overline{h}}R_0(\theta )\right] \right\} =\mu _{in}(\theta )\left[
1-P_R(\theta )\right]  \label{c4.7}
\end{equation}
and now we obtain for the yield of the SR,

\begin{equation}
I(\theta )=\left[ 1-P_R(\theta )\right] \frac{\mu _{in}(\theta )/\mu _0}{\mu
_{yi}^{eff}+\mu _{in}(\theta )/\gamma _0}\simeq \left[ 1-P_R(\theta )\right]
\left[ 1-\gamma _0\frac{\mu _{yi}^{eff}}{\mu _{in}(\theta )}\right] .
\label{c4.8}
\end{equation}
In the case of the fluorescence radiation we have for the intensity 
\[
I^{fl}(\theta )=\left[ 1-P_R(\theta )\right] \left[ 1-(\gamma _0/\gamma
_{fl})(\mu _{yi}^{fl}/\mu _{in}(\theta ))\right] . 
\]
As we can see from Eq. (\ref{c4.8}) now in the limit of the big depths of
yield $\mu _{yi}<<\mu _{in}(\theta )$ the shape of the SR curve has the form
of the reverse curve of the reflectivity $I(\theta )\simeq \left[
1-P_R(\theta )\right] $. This result has a simple physical explanation.
Really, in the angular region of the dynamical scattering x-rays does not
penetrate deeper than the extinction depth $L_{ex}$, so, SR can be excited
only from this depths. Due to the law of the energy conservation the yield
of the SR has to be equal to Eq. (\ref{c4.8}). However, if the term $\gamma
_0\mu _{yi}^{eff}/\mu _{in}(\theta )$ in expansion (\ref{c4.8}) is becoming
comparable with unity, then on the curves of the secondary radiation one can
see small asymmetry due to the behavior of $\mu _{in}(\theta )$. This
behavior of the fluorescence radiation was for the first time measured and
understood in the pioneer works of B. Batterman \cite{B62,B64} and then
repeated in many other works (see for e.g. Fig. 10).

It is clear, that using the depth of yield $L_{yi}$ as a parameter we will
obtain a number of curves that lie between two limiting cases described by
Eq. (\ref{c4.6}) and (\ref{c4.8}). On Fig. 11 the calculated curves of the
angular dependence of the SR yield from a perfect $Si$ crystal are
presented. For calculations it was used the case of (400) diffraction of $%
CuK_\alpha $ radiation ($L_{ex}=3.60\mu m$) for different values of the
parameter $L_{yi}$.

\subsection{Multicomponent crystals}

In a perfect crystals with the complicated elementary cell containing
different type of atoms XSW method give a unique possibility to investigate
position and degree of disorder of different type of atoms. Experimentally
the most effective way to do it is to register the characteristic
fluorescence radiation from different atoms. The nodes and antinodes of
x-ray standing wave are located in the different way for different
sublattices of the crystal (on Fig. 12a (111) diffraction planes of $GaAs$
crystal are shown). In this situation the angular dependence of the
fluorescence yield for different type of atoms will have the different shape
and also will differ from the typical curves of the monoatomic crystals (see
for e.g. Fig.12b for the same case of $GaAs$ crystal).

Peculiarities of the angular dependence of the fluorescence yield in a
multicomponent perfect crystal are determined in fact solely by the factor $%
\varepsilon _{\overline{h}}$ in (\ref{c4.1}). If we are interested by the
yield of the fluorescence radiation from the atom of the sort $a$ from a
multicomponent crystal then we have for the factor $\varepsilon _{\overline{h%
}}$ in (\ref{c4.1}) (see also (\ref{c2.41})), 
\begin{equation}
\varepsilon _{\overline{h}}^a=\frac{\chi _{\overline{h}i}}{\chi _{0i}}%
=\left( \frac{\sigma _a(\mathbf{h})}{\sigma _a(0)}\right) \frac{%
\sum\limits_jc_j^ae^{-W_j^a(h)}S_j^a(\overline{\mathbf{h}})}{%
\sum\limits_jc_j^a},  \label{c4.9}
\end{equation}
where $c_j^a$ is the concentration of the atoms of the sort $a$ in the
sublattice $j$, $\sigma _a(0)$and $\sigma _a(\mathbf{h})$are the cross
sections of the corresponding processes, $e^{-W_j^a(h)}$ are the
Debye-Waller factors (here they are the sum of thermal and static
displacements) and $S_j^a(\overline{\mathbf{h}})=\sum\limits_{j^{\prime
}}e^{i\mathbf{h\rho }_{j^{\prime }}^a}$are the structure factors
corresponding to the positions of the atoms of the sort $a$ in the unit
cell. The total cross sections of the photoexcitation in dipole
approximation are isotropic and we have for $\sigma _a(\mathbf{h})/\sigma
_a(0)=1$, however if quadrupole contributions are valuable we have for the
ratio of cross sections in Eq. (\ref{c4.9})

\begin{equation}
\frac{\sigma _a(\mathbf{h})}{\sigma _a(0)}=1-\frac{\sigma _a^Q}{\sigma _a}%
\left( 1-\frac{C^Q}C\right) ,  \label{c4.10}
\end{equation}
where $\sigma _a=\sigma _a^D+\sigma _a^Q,\sigma _a^D$ and $\sigma _a^Q$ are
the total, dipole and quadrupole cross sections correspondingly, $C$ is the
polarization coefficient (\ref{c2.18}) and parameter $C^Q$ is equal to $\cos
2\theta _B$ for $\sigma -$ polarization and $\cos 4\theta _B$ for $\pi -$
polarization (see for details \cite{VZ97,VZ99}). This can be important while
measuring fluorescence radiation near absorption edges or in a
backscattering geometry, where quadrupole contribution can be essential.
Expression (\ref{c4.9}) is simplified for the case of two-component crystals
when different type of atoms occupy different sublattices (this is the case
of $GaAs$, $InSb$ and etc. crystals). In this case we have from Eq. (\ref
{c4.9}) 
\begin{equation}
\varepsilon _{\overline{h}}^a\simeq e^{-W^a(h)}S^a(\overline{\mathbf{h}}%
)=e^{-W^a(h)}\mid S^a(\overline{\mathbf{h}})\mid e^{i\varphi ^a(\mathbf{h})},
\label{c4.11}
\end{equation}
where dipole approximation is assumed. Structure factors $S_j^a(\overline{%
\mathbf{h}})$ are complex quantities with their amplitude $\mid S^a(%
\overline{\mathbf{h}})\mid $ and phase $\varphi ^a(\mathbf{h})$.
Substituting this values of $\varepsilon _{\overline{h}}(a)$ into expression
(\ref{c4.5}) we obtain for the intensity of standing wave on the positions
of the atoms of the sort $a$,

\begin{equation}
I^{SW}(\theta )=1+\beta |Y|^2|R_0(\theta )|^2+2\sqrt{\beta }C\mid
YR_0(\theta )\mid F_c^a(h)\cos \left[ \alpha (\theta )+\varphi ^a(\mathbf{h}%
)+\Phi _Y\right] ,  \label{c4.12}
\end{equation}
where 
\begin{equation}
F_c^a(h)=e^{-W^a(h)}\mid S^a(\overline{\mathbf{h}})\mid .  \label{c4.13}
\end{equation}
This is an important result. According to Eq.(\ref{c4.12}) the angular
dependence of the SR yield directly depend on the \textbf{phase} $\varphi ^a(%
\mathbf{h})$ of the structure factor $S_j^a(\overline{\mathbf{h}})$.
Moreover, due to the fact that this phase enter the interference term in (%
\ref{c4.12}) the angular dependence of the field intensity $I^{SW}(\theta )$
is very sensitive to the value of this phase. So, measuring this angular
dependence one can determine with high accuracy the \textbf{phase} and the 
\textbf{amplitude} $\mid S^a(\overline{\mathbf{h}})\mid $ of the structure
factor for different sublattices (or different sort of atoms) in
multicomponent crystals. In this way the $\mathbf{h}-$th Fourier component
of the structure factor can be totally determined \cite{HMZ85}. This is
illustrated for the case of the $GaAs$ crystal on Fig.12. Of course
according to our previous discussion this initial curve will be modified
when the final depth of yield of the fluorescence radiation (Eq. (\ref{c4.4}%
)) will be taken into account.

This idea of measuring the SR (fluorescence or photoelectron yield) in the
XSW field in multicomponent perfect crystals was successfully realized in a
number of experiments. For example, the polarity of $GaP$ crystals was
obtained in experiments \cite{T76} while monitoring fluorescence radiation
and in experiments \cite{TK79} $GaL$ photoelectrons were measured using a
cylindrical energy analyzer in a high-vacuum chamber. Due to a high
sensitivity of XSW method it has become possible to measure the change of
the phase of the structure factor as a function of energy near absorption
edges in noncentrosymmetric crystals (see for e.g. \cite{BMK84b,BM85} and
also theoretical paper \cite{A86}). In the paper \cite{LST84} it was
demonstrated the possibility of determination of the positions of the Ga and
Gd atoms in the unit cell (it contain 160 atoms) of the perfect garnet
crystals while monitoring characteristic fluorescence radiation in the XSW
field. In a recent papers the fluorescence radiation from different atoms in
the unit cell of High-T$_c$ crystal was measured \cite{KKO97}, positions of $%
Cd,Zn,Se$ and $Te$ atoms in $Cd_{1-x}Zn_xSe_yTe_{1-y}$ single crystals were
obtained \cite{KSZ97}.

Analysis of the photoelectron yield (especially if Auger electrons are
monitored) while scanning XSW in multicomponent crystals is more complicated
comparing to the fluorescence yield. Really, even if detectors with high
energy resolution are available then the contribution to the total
photoelectron yield in the registered energy range is in general the sum of
contributions of primary photoelectrons ejected from different atomic
subshells and from different sort of atoms. Finally we have \cite{AAI89}

\begin{equation}
Y_{el}(E,\theta )=\sum\limits_{N,a}g_{Na}Y_{Na}(E,\theta ),  \label{c4.14}
\end{equation}
where for the case of perfect crystals $Y_{Na}(E,\theta )$ is determined by
Eq. (\ref{c4.1}) with the factor $\varepsilon _{\overline{h}}=\varepsilon _{%
\overline{h}}^a(N)$ equal to that of (\ref{c4.9}) and with $%
P_{yi}(z)=P_{Na}(z,E)$, that determines now the probability of the
photoelectron escape with energy $E$ ejected at the depth $z$ from the $N-$%
th subshell of the $a-$th sort of atom. Parameter $g_{Na}$ determines the
fraction of electrons ejected from the $N-$th subshell of the $a-$th sort of
atom to the total number of such electrons,

\begin{equation}
g_{Na}=\frac{n_a\sigma _{Na}(0)}{\sum\limits_{N,a}n_a\sigma _{Na}(0)},
\label{c4.15}
\end{equation}
where $n_a$ is a number of atoms of the $a-$th sort in a unit cell. Taking
into account, that typically the photoelectron yield depth of yield is much
smaller, then the extinction depth ($L_{yi}<<L_{ex}$) we can use for the
photoelecton yield expression (\ref{c4.5}) with the parameter $\varepsilon _{%
\overline{h}}$ equal to, 
\begin{equation}
\varepsilon _{\overline{h}}=\sum\limits_{N,a}g_{Na}\varepsilon _{\overline{h}%
}^a(N)P_{Na}(E),  \label{c4.16}
\end{equation}
where $P_{Na}(E)=\int\limits_0^\infty P_{Na}(z,E)dz$.

\subsection{Crystals with an amorphous surface layer}

We shall assume now, that the top of the crystal contain thin amorphous
layer with the thickness $L_{am}$. We shall also assume, that the depth of
yield of the SR is much smaller, then the extinction depth, but bigger then
the thickness of amorphous layer $L_{am}<L_{yi}<<L_{ex}$ (this can be the
photoelectron yield for example). It is clear, that in amorphous layer there
is no interference between incident and diffracted beam and we can neglect
small attenuation of x-rays in this thin layer. Taking all this into account
and performing integration in (\ref{c3.7}) separately in amorphous and in
the perfect part of a crystal we obtain, 
\begin{equation}
\kappa (\theta )=p_{am}\kappa ^{am}(\theta )+(1-p_{am})\kappa ^{id}(\theta ),
\label{c4.17}
\end{equation}
where

\begin{equation}
\kappa ^{am}(\theta )=1+\beta P_R(\theta ),  \label{c4.18}
\end{equation}

\begin{equation}
\kappa ^{id}(\theta )=I^{SW}(\theta )=1+\beta P_R(\theta )+2\sqrt{\beta }%
CRe[\varepsilon _{\overline{h}}YR_0(\theta )]  \label{c4.19}
\end{equation}
and $p_{am}$ is amorphous fraction of the crystal defined as

\begin{equation}
p_{am}=\int\limits_0^{L_{am}}P_{yi}(z)dz.  \label{c4.20}
\end{equation}
In (\ref{c4.17}-\ref{c4.20}) we are assuming, that the depth of yield
function is normalized to unity $\int\limits_0^\infty P_{yi}(z)dz=1.$

From expressions (\ref{c4.17}-\ref{c4.19}) we see, that in general for
different thickness of amorphous layers we obtain different shape of the
curves of the photoelectron yield all lying between two curves $\kappa
^{id}(\theta )$ for perfect crystal and $\kappa ^{am}(\theta )$ for
amorphous layer $L_{am}>L_{yi}$. Such measurements were made in \cite
{KS77,SKZ81} and are presented on Fig.13.

In general escape depth of electrons $L_{yi}(E_i)$ from the crystal depend
from the initial energy $E_i$ (see for e.g. Eq. (\ref{c4.3})). Assuming now
exponential probability yield function in the form (\ref{c4.2}) 
\[
P_{yi}(E_i)=\frac 1{L_{yi}(E_i)}\exp (-\frac z{L_{yi}(E_i)}) 
\]
we obtain for the amorphous fraction of the crystal $p_{am}$%
\begin{equation}
p_{am}=1-\exp [-L_{am}/L_{yi}(E_i)].  \label{c4.21}
\end{equation}
From this expression we have for the ratio $L_{am}/L_{yi}(E_i)$%
\begin{equation}
L_{am}/L_{yi}(E_i)=-\ln (1-p_{am}).  \label{c4.22}
\end{equation}
Now, if one of the parameters: thickness of amorphous layer $L_{am},$ or
escape depth of electrons $L_{yi}(E_i)$ is known, then another one can be
obtained from Eq. (\ref{c4.22}).

This idea was realized in the paper \cite{BMK84a} (see Fig. 14), when the
photoelectron yield for different group of electrons with different loss of
energy was measured with the low-resolution gas-proportional counter. In
this experiments escape depth $L_{yi}=1.2$ $\mu m$ for the initial energy of
the electrons $E_i=13.3$ $keV$ in \textit{Si} crystal with amorphous \textit{%
SiO}$_2$ layer was obtained.

\eject

\section{Bicrystal model (Bragg geometry)}

\subsection{Theory}

The calculation scheme of the angular dependence of the SR yield $\kappa
(\theta )$ for the Bragg geometry of x-ray diffraction in the crystals with
the deformed surface layer was developed in \cite{KK81}. In general case
when the deformation profile of the crystal lattice has an arbitrary profile
the main problem is to solve nonlinear differential equation (\ref{c2.22})
for the amplitude $R(z,\theta )$ (\ref{c2.21}). For the arbitrary dependence
of the functions $\varphi (z)=\mathbf{hu}(z)$ and $e^{-W(z)}$ from the depth 
$z$ Eq. (\ref{c2.22}) and integrals in (\ref{c2.32}) and (\ref{c3.7}) can be
calculated only numerically (see for e.g. Fig. 8).

In this Chapter we will discuss the simplest model of the deformed crystal,
so-called bicrystal model, that allows to find analytical solution. In the
frame of this model crystal consists of two parts a thick perfect substrate
and the deformed layer of thickness $L_d$ with the linear dependence of the
deformation field $u(z)=(\Delta d/d)z$, where parameter $\Delta d/d$ is a
constant difference of the interplanar distance in the layer comparing to
that in the substrate (Fig. 15). In this layer parameter $y_\varphi
(z)=y_\varphi =\pi (L_{ex}/d)(\Delta d/d)$ (\ref{c2.24}) and static
Debye-Waller factor $e^{-W(z)}=e^{-W_0}$ has constant values. So, in this
model deformed layer has a different interplaner distance comparing to that
of the perfect substrate. It is additionally partially and uniformly
amorphysized and sharp transition (in fact step function) between the layer
and the substrate is assumed. Though this model is very simplified it turned
out to be very practical in the analysis of the SR yield in a number of
experiments.

Analytical solution for the angular dependence of the SR yield $\kappa
(\theta )$ and reflectivity $P_R(\theta )$ for Bragg diffraction and
bicrystal model was obtained for the first time in \cite{KKL85}. This result
is a particular case of a more general approach of a multilayer crystal
consisting of a number of layers with the different parameters that will be
discussed in details in Chapter VII (see also Appendix). According to this
approach SR yield for bicrystal model can be obtained from Eq. (\ref{a21})
with the number of the layers $N=2$%
\begin{equation}
\kappa _{tot}^a(\theta )=\kappa _{L_d}^a(\theta )+\left| T(L_d,\theta
)\right| ^2e^{-\mu _{yi}L_d}\kappa _0^a(\theta ).  \label{c5.1}
\end{equation}
Here, normalized SR yield $\kappa _{tot}^a(\theta )$ is a sum of two
contributions: from the deformed layer $\kappa _{L_d}^a(\theta )$ and a
perfect substrate $\kappa _0^a(\theta )$. This functions are defined in Eq.(%
\ref{a18}) and transmission coefficient $\left| T(L_d,\theta )\right| ^2$
can be determined from (\ref{a11}). Deformed layer is characterized by
constant values of deformation $\Delta d/d$, static Debye-Waller factor $%
e^{-W_0}$, coherent fraction $F_c^a$ and coherent position $P_c^a$. For the
perfect substrate $\Delta d/d=0$, $e^{-W_0}=1$ and, in general, it can has
the values of coherent fraction $F_c^a$ and coherent position $P_c^a$
different from that of the layer. The angular dependence of the reflectivity 
$P_R(\theta )$ for Bragg diffraction is obtained from (\ref{c2.28}) and (\ref
{a1}-\ref{a3}) with the amplitude $R_{d_n}=R_0(\theta )$, where $R_0(\theta
) $ is the value of the amplitude for the perfect substrate defined in (\ref
{c2.29}).

We want to note here that obtained result is exact and is valid for any
relationship between the values of $L_d$, $L_{ex}$ and $L_{yi}$. It is also
valid for the analysis of the angular dependence of the photoelectron yield
or fluorescence radiation in the XSW field. Calculations performed with an
exact expression (\ref{c5.1}) give possibility to take into account minor
effects of extinction and variation of phase on the depth of yield of the
SR. The convenience of the analytical approach in comparison with the direct
numerical calculations lies in the possibility of explicitly separating the
dependence on the various parameters.

Essential simplification of Eq. (\ref{c5.1}) can be obtained in the special
case, when the following condition $L_{yi}<<L_d<<L_{ex}$ is satisfied. It
can be easily fulfilled, for example, for the photoelectron yield. In this
case in the angular region of the dynamical diffraction from the substrate $%
|y(\theta )|<1$ the SR yield (\ref{c5.1}) can be presented in the following
way 
\begin{equation}
\kappa (\theta )\approx 1+\beta P_R^{id}(\theta )+2\sqrt{\beta }C\left|
R_0(\theta )\right| F_c^a(h)\cos \left[ \alpha _Y(\theta )+\varphi
_h^a+\varphi _0\right] ,  \label{c5.2}
\end{equation}
where $P_R^{id}(\theta )=|YR_0(\theta )|^2$ is the reflectivity from a
perfect substrate (\ref{c2.28}--\ref{c2.29}); $F_c^a(h)=|Y\varepsilon _{%
\overline{h}}^a|e^{-W_0}$ is the \textit{coherent fraction}; $\alpha
_Y(\theta )=\alpha (\theta )+\arg (Y)$ where $\alpha (\theta )$ is the
argument of the complex amplitude $R_0(\theta )$; $\varphi _h^a=2\pi
P_c^a=\arg (\varepsilon _{\overline{h}}^a)$, where $P_c^a$ is the \textit{%
coherent position}; and the phase $\varphi _0=\varphi (0)=2y_\varphi
L_d/L_{ex}$ gives the total phase shift due to deformation in the layer. In
fact this result is a special case of a more general result obtained for the
first time in \cite{AK78} and is valid for any type of deformation if the
condition $L_{yi}<<L_d<<L_{ex}$ is satisfied.

\subsection{Experiment}

Now we will show on a number of examples how this theoretical approach can
be applied for the analysis of different experiments with the use of XSW for
the investigation of the real surface structure.

For the first time it was applied for the analysis of homoepitaxial films on 
\textit{Si} surface \cite{KKL85} (see also \cite{KVK87}). In this
experiments photoelectron yield was measured from a set of specially
prepared \textit{Si }single crystals with the grown homoepitaxial \textit{Si}
films doped with \textit{Ge}. The concentration of \textit{Ge} varies from
sample to sample from $3.7\times 10^{19}$ to $1.5\times 10^{20}$ atoms $%
cm^{-3}$. The film thickness was $L_d=1.5\mu m$. The $CuK_\alpha $ radiation
and a (444) non-dispersive double crystal diffraction arrangement with an
asymmetric monochromator were used. In this case $L_{yi}=0.45\mu m,$ $%
L_{ex}=10.5\mu m$ and consequently the condition $L_{yi}<<L_d<<L_{ex}$ is
fulfilled. The experimental results are presented in Fig. 16. The doping
with the \textit{Ge} atoms changes the plane spacing $\Delta d$ uniformly in
the disturbed layer, which leads to a change of the surface displacement.
Hence the value of the phase on the surface $\varphi _0=|\mathbf{h}%
|u(0)=2\pi (\Delta d/d)(L_d/d)$ also changes. This factor along with the
change of Debye-Waller factor $e^{-W_0}$ in the surface layer is responsible
(according to Eq. (\ref{c5.2})) for the variation of the shape of the
photoelectron yield curve. Table 1 presents the values of $\varphi _0$ and $%
W_0$ obtained by fitting of the experimental curves on Fig. 16 to the
theoretical ones calculated using Eq. (\ref{c5.2}). In this experiment it
was demonstrated for the first time high sensitivity of the XSW method for
measuring small displacements in the surface layer (in fact, relative
difference of the lattice parameter in the layer comparing to that in the
substrate). We want to note here, that though lattice parameter changes were
comparably not so big from sample to sample the phase difference for high
order (444) reflection was of the order of $\pi $ that brought to a
significant change of the shape of the curves from sample to sample (see
Fig. 16).

In the paper \cite{KKK88} the films of \textit{In}$_{0.5}$\textit{Ga}$_{0.5}$%
\textit{P} of different thickness ($\sim 0.01\mu m$ and $\sim 0.6\mu m$
grown on the surface of \textit{GaAs} (111) single crystal were
investigated. The fluorescence radiation from the \textit{In} and \textit{P}
atoms excited by the XSW field in the substrate (in the case of thin film)
and in the film (for thick film) was measured (see Figs. 17, 18).

In the case of thin film condition $L_{yi}=L_d<<L_{ex}$ is fulfilled and we
can use Eq. (\ref{c5.2}) to obtain the phase shift $\varphi _0$ of the
surface layer. Taking also into account relationships (\ref{c4.9}-\ref{c4.11}%
) the angular dependence of the fluorescence yield from the thin film can be
analysed from the simple expression, 
\begin{equation}
\kappa _{fl}^a(\theta )=1+\frac{|E_h(\theta )|^2}{|E_0(\theta )|^2}+2C\frac{%
|E_h(\theta )|}{|E_0(\theta )|}F_c^a(h)\cos \left[ \alpha (\theta )+\varphi
_h^a+\varphi _0\right] ,  \label{c5.3}
\end{equation}
were parameter $F_c^a(h)$ (coherent fraction) is equal to $F_c^a(h)=|S_{%
\overline{h}}^a|\exp (-W^a(h))F_0$, $S_h^a=|S_h^a|\exp (i\varphi _h^a)$ is a
structure factor of the atoms of the sort $a$, $\alpha (\theta )$ is the
phase of the complex ratio $E_h/E_0$ (see Eq. (\ref{c2.36}) and $%
F_0=e^{-W_0} $ is the average amorphization of the surface layer.

From the fitting of the experimental curve to the theoretical one (Eq. (\ref
{c5.3})) it was obtained the value of $F_0=0.25$ and of the phase shift $%
\varphi _0=0.55$ that correspond to the average deformation in the film $%
\Delta d/d\approx 6\cdot 10^{-3}$. The same approach was used in \cite{AIT85}
for the analysis of the film \textit{In}$_x$\textit{Ga}$_{1-x}$\textit{As}$%
_y $\textit{P}$_{1-y}$ grown on the substrate \textit{InP} (100). Small
deformation $\Delta d/d\approx 2.3\cdot 10^{-4}$ in a thicker film $%
L_d\approx 0.2\mu m$ was measured with XSW method.

In the case of the thick film x-ray standing wave field is formed in the
film itself. Fluorescence yield measured in the angular position of the film
maximum has different behavior for \textit{In} and \textit{P} atoms due to
different position of the atoms in the unit cell. In this case the
simplified approach of Eq. (\ref{c5.3}) is not valid any more and general
approach based on Eq. (\ref{c5.1}) was used for fitting (see Fig. 18). As a
result thickness of the film $L_d=0.62\mu m$, deformation $\Delta d/d\approx
2\cdot 10^{-3}$ and amorphization factor $F_0=0.8$ were obtained. However as
it is seen from Fig. 18(b) coincidence between the experimental and
theoretical curves is not perfect, that can be due to simplified bicrystal
model, that does not take into account transition layer between the film and
the substrate. In this situation the theoretical model with transition layer
proposed in \cite{AGM89} can be useful for the analysis of the angular
dependence of the fluorescence yield from the geterostructures.

Thin epitaxial films of \textit{CaF}$_2$ grown on \textit{Si} (111) surface
where characterized by impurity luminescence probes, x-ray diffractometry
and x-ray standing wave technique \cite{AKK91,AHK92}. Molecular beam epitaxy
was used to grow 10nm thick films. The \textit{CaF}$_2$\textit{/Si}
interface was formed at $770^{\circ }C$. Due to different growth conditions
strain field in each \textit{CaF}$_2$ film is different. This is well seen
from the reflectivity (222) curves measured from the different samples (Fig.
19(a)). The angular dependence of the \textit{CaK}$_\alpha $ fluorescence
excited by the XSW field in the \textit{Si} (111) substrate was also
measured (Fig. 19(b)). In this case we again have the situation, when $%
L_{yi}=L_d<<L_{ex},$ and expression (\ref{c5.3}) can be applied for the
analysis. Finally the strain field in the film was obtained independently
from the photoluminescence study and x-ray rocking curve analysis and are
summarized in the Table 2. X-ray standing wave analysis gave an additional
information: the distance between the first (as counted from the substrate)
atomic plane of the film and the diffracting plane of the substrate nearest
to the interface (denoted as $P^{hkl}$) and the static Debye-Waller factor $%
e^{-W_0}$ of the film.

Comprehensive study of different type of garnet crystals with different kind
of films on the surface were analysed with XSW method monitoring
photoelectron yield and fluorescence radiation in \cite{ZZK88,NKM91,KKS92}
(first measurements of fluorescence radiation from garnet single crystals
were performed in \cite{LST84}). Peculiarity of the garnet crystals is a
comparably complicated unit cell containing 160 atoms (for the distribution
of different atoms in the unit cell of garnet crystal in the (111) direction
see Fig. 20(a)). Another interesting property is that due to the fact, that
garnet crystals contain heavy atoms with $Z>>1$ propagation of x-rays in
this crystals is highly reduced and fluorescence yield from this atoms is
absorbed on a short distances of about $1\div 10\mu m$.

The angular dependence of the total photoelectron yield from the gallium
gadolinium garnet (GGG) crystals excited by the XSW field was measured for
the first time in \cite{ZZK88} (Fig. 20(b)). Due to a small value of the
absorption depth $L_0=\sin \theta _B/\mu _0=5.44\mu m$, that is comparable
with the extinction length $L_{ex}=3.88\mu m$ in this crystal the reflection
coefficient for (888) diffraction of \textit{CuK}$_\alpha $ radiation has a
small $\sim 47\%$ maximum value. At the same time the photoelectron yield
curve has a phase sensitive dispersion like shape.

In the same paper the angular dependence of the photoelectron yield for the
GGG crystal with the epitaxial film (thickness of the film $L_d=2\mu m$) of
the iron yttrium garnet (FYG) was measured (Fig. 21). The same \textit{CuK}$%
_\alpha $ radiation and (888) reflection as in the previous case was used.
For the theoretical description of the photoelectron yield in this case of
thick film the general theory of bicrystal model described in the first part
of this chapter can be applied. However for understanding the physics of
formation of the photoelectron yield while the dynamical diffraction some
simplified considerations can be used. Lattice parameters of the FYG film
and of the GGG substrate does not differ essentially (the same is valid for
their Bragg angles), however the Fourier components of the susceptibility $%
\chi _h$ are quite different. For example, extinction depth for the FYG
crystal is equal to L$_{ex}=5.68$ $\mu m$ that is essentially bigger then
the thickness of the film ($L_d<<L_{ex}$), so reflection from the film is
kinematical. In the kinematic approximation we have for the reflectivity, 
\begin{equation}
P_R(\theta )=\frac{C_p}{y^2(\theta )}\sin ^2\left[ \frac{L_d}{L_{ex}}%
y(\theta )\right] ,  \label{c5.4}
\end{equation}
where angular deviation parameter $y(\theta )$ is defined in (\ref{c2.23})
and $C_p=(1+C^3)/(1+C)$ is the polarization factor. In this approximation we
have for the photoelectron yield, 
\begin{equation}
\kappa _{ph}(\theta )=1+K\sin ^2\left[ \frac{L_d}{L_{ex}}y(\theta )\right]
\left[ -\frac 2{y(\theta )}\mid \varepsilon _{\overline{h}}\mid +\frac
1{y^2(\theta )}\right] .  \label{c5.5}
\end{equation}

Parameters in equations (\ref{c5.4}-\ref{c5.5}) in the case of experiment 
\cite{ZZK88} has the following values $K=0.75,\mid \varepsilon _{\overline{h}%
}\mid =0.8$ and we obtain for $P_{R\max }=0.092,$ $\kappa _{\min }=0.75,$ $%
\kappa _{\max }=1.33.$ This simple approximation fit quite well to the
experiment. At the same time the photoelectron yield at the angular position
of the substrate peak has the shape similar to the reflectivity curve (this
is quite similar to the discussed before case of amorphous layer, however
the physics of the process is different). Period of the standing wave formed
in the substrate is different from that of the film (measured photoelectrons
are excited only in the film). Due to this difference there is no
correlation between the positions of the atomic planes on the escape depth
of the electrons and the positions of the nodes and antinodes of the
standing wave. In this experiment the escape depth of the electrons is about
of $L_{yi}=0.22$ $\mu m$. On this depth due to the difference in the period
of the film lattice and the substrate atom is shifted on $1.44\AA $ and at
the same time the period of the standing wave for (888) reflection is equal
to $a\sqrt{3}/24=0.89\AA .$ Finally, due to a big variation of phase on the
period of the standing wave field the third term in the expression for the
photoelectron yield $\kappa _{ph}(\theta )$ (\ref{c5.2}) cancel out and we
obtain for $\kappa _{ph}(\theta )=1+P_R(\theta ).$

In the same paper the fluorescence yield from the Gallium-Neodymium Garnet (%
\textit{GNG}) crystal and the films of \textit{FYG} (thickness of the film $%
L=1.6\mu m$) on the top of \textit{GGG} crystal with XSW method were
investigated. Due to a big depth of yield of the fluorescence radiation ($%
L_{yi}^{Ga}=20.4\mu m$ for \textit{GaK}$_\alpha $ radiation and $%
L_{yi}^{Nd}=15.1\mu m$ for \textit{NdL}$_\alpha $ radiation) comparing to
the extinction length ($L_{ex}=1.54\mu m$) the angular dependence of the
fluorescence yield curves in the XSW field is determined by extinction
effect discussed in the section 4.1. However as we can see from Fig. 22(a)
the fluorescence yield from \textit{Nd} and \textit{Ga} atoms has different
asymmetry on the tails of the curves. This is due to the fact, that for
reflection (444) the phase of the standing wave on the neighbour atomic
layers (on the distance of $1/24$ of the period) differs by $\pi .$ At the
same time average coherent positions for \textit{Ga} and \textit{Nd} atoms
coincide, but the effective coherent fractions $F_c$ are different: $%
F_c^{Ga}=0.321$ and $F_c^{Nd}=1.628$.

For the theoretical analysis of the fluorescence yield curves from the 
\textit{FYG} films on the \textit{GGG} crystal substrate the bicrystal model
described in this section was used. While fitting in this case it was
additionally taken into account the change of the susceptibility of the film 
$\chi _{film}$ comparing to that of the substrate $\chi _{substrate}$ as
well as the change of the interplanar distance $\Delta d/d.$ Results of the
fitting are presented on Fig. 22(b). In this case \textit{YK}$_\alpha $
fluorescence radiation from the atoms of the film and \textit{GaK}$_\alpha $
radiation from the atoms of the substrate in the case of (444) diffraction
of AgK$_\alpha $ radiation were measured. Different shape of the curves
reflects different conditions of the fluorescence yield formation in XSW
field. Extinction depth is equal to $L_{ex}=2.3\mu m$ and is bigger then the
thickness of the film. The depth of yield of the \textit{YK}$_\alpha $
radiation is restricted by the thickness of the film and coincide with the
volume of XSW formation in the film (the same as on Fig. 21, but with bigger
value of $L_d/L_{ex}$). At the same time (when Bragg conditions for the film
are fulfilled) the \textit{Ga} fluorescence yield is decreased because
x-rays being reflected in the film hardly penetrate to the substrate, where
Ga atoms are located. Right maximum on the reflectivity curve on Fig. 22(b)
correspond to the reflection of x-rays from the substrate. In this case the
angular dependence of the \textit{GaK}$_\alpha $ fluorescence yield has a
big dip due to an extinction effect described before (due to a big escape
depth of the fluorescence radiation comparing to an extinction depth). At
the same time the yield of the \textit{YK}$_\alpha $ radiation in this
angular region has a maximum of the same origin as described on Fig. 21.

In the papers \cite{NKM91,KKS92} XSW method was used to study the positions
of $Bi^{3+}$ ions in the lattice of a heteroepitaxial film of
yttrium-bismuth iron ($Y_{3-x}Bi_xFe_5O_{12}$) garnet. The main problem was
to determine quantitatively the distribution of bismuth over the different
dodecahedral positions which are occupied by yttrium in a pure yttrium iron
garnet. This information is essential for understanding the growth-induced
magnetic anisotropy of this garnet crystals. The stated problem was solved
by detailed analysis of the angular dependence of the $Bi^{3+}$ fluorescence.

Ideal garnets have cubic symmetry with cations entering octahedral (a),
tetrahedral (d) and dodecahedral (c) sublattices. In the $%
Y_{3-x}Bi_xFe_5O_{12}$ system $Bi^{3+}$ and $Y^{3+}$ ions enter only the $c$
sites. For the [001] growth direction there are two inequivalent groups of
dodecahedral sites, the first consists of 16 sites, which are denoted as $%
c_{xy}$, the second consists of 8 sites denoted as $c_z$ \cite{N84}. The
distribution of $c_{xy}$ and $c_z$ sites within the elementary cell for this
growth direction together with the distribution of the nodes and antinodes
of (004) X-ray standing wave is shown in Fig. 23. Note that the sites
belonging to the $c_{xy}$ and $c_z$ groups lie in different layers and
consequently they can be, in principle, distinguished by the XSW method.

For the analysis of the fluorescence yield from this garnet samples a
bicrystal model consisting of the thin film and the substrate was used. If
the depth of yield of the fluorescence radiation $L_{yi}$ is much smaller
than the extinction length $L_{ex}$ the angular dependence of the intensity
of fluorescence radiation in multicomponent crystal is given by Eq. (\ref
{c5.3}). For the reflection (004) the inequivalent groups of sites $c_z$ and 
$c_{xy}$ have different structure factors $S_h(c_z)=-1$ (8 sites), $%
S_h(c_{xy})=+1$ (16 sites). The structure factor of the $Bi^{3+}$ ions,
which enter the dodecahedral sublattice in $Y_{3-x}Bi_xFe_5O_{12}$ garnet
films, is influenced by the distribution of these ions between $c_z$ and $%
c_{xy}$ positions. The fraction of $Bi^{3+}$ ions in $c_z$ positions will be
denoted by parameter $p$%
\[
p=\frac{N_z^{Bi}}{N_z^{Bi}+N_{xy}^{Bi}}, 
\]
where $N_z^{Bi}$, $N_{xy}^{Bi}$ are the numbers of $Bi^{3+}$ ions in $c_z$
and $c_{xy}$ sites, respectively. Then the structure factor of $Bi^{3+}$
ions for the (004) reflection is $S_h^{Bi}=1-2p$ and due to Eq. (\ref{c5.3})
the fluorescence yield of $Bi^{3+}$ ions will depend on the value of the
parameter $p$. The uniform distribution of $Bi^{3+}$ ions between $c_z$ and $%
c_{xy}$ sites corresponds to $p=1/3$. The calculated angular curves of the
reflectivity and $Bi^{3+}$-ion fluorescence yield for values of $p$ in the
interval $0$ to $1$ are shown in Fig. 24. The angular yield curves of the $%
Bi^{3+}$ fluorescence demonstrate, as expected, a strong dependence on the
distribution of these ions. For values of the parameter $p=0$ (all $Bi^{3+}$
ions in $c_{xy}$ sites) and $p=1$ (all $Bi^{3+}$ ions in $c_z$ sites) the
amplitude of the structure factor of the $Bi^{3+}$ ions attains its maximum (%
$\left| S_h^{Bi}\right| =1$) and the phase $\varphi _h^{Bi}$ of the
structure factor equals $0$ and $\pi $, respectively. The angular yield
curves for these limiting cases show reversed positions for maxima and
minima of the yield. When $\left| S_h^{Bi}\right| $ decreases the ratio of
the fluorescence to the background becomes smaller. Note that on the curve
corresponding to $p=0.33$ (uniform distribution of $Bi^{3+}$ ions over $c_z$
and $c_{xy}$ sites) a clear maximum occurs at the high-angle side (with
respect to the maximum of Bragg diffraction on the film). At the same time
the reflectivity curves (Fig. 24(a)) has very small dependence on the value
of the parameter $p$.

The diffraction curve for the sample studied is shown in Fig. 25. The
angular distance between two peaks (corresponding to the substrate and the
film) allows a direct determination of the lattice mismatch between the film
and substrate $\Delta d/d=2.5\times 10^{-3}$. In Fig. 26 the experimental
angular dependencies of the fluorescence of $Bi^{3+}$ (Fig. 26(a)), $Y^{3+}$
(Fig. 26(b)), $Fe^{3+}$ (Fig. 26(c)) and $Gd^{3+}$ (Fig. 26(c)) are shown
together with the reflectivity curve (Fig. 26(e)) in the angular range
corresponding to the diffraction on the film. Results can be understood on
the basis of the theory described in the first part of this Chapter. Three
main factors influence the shape of the fluorescence yield curve, namely the
amplitude and the phase of the structure factor of the sublattice of the
ions under study and the extinction effect for a given thickness of the
film. The experimentally observed shape of the curve is determined by the
fact that the $Bi^{3+}$ sublattice structure factor is close to zero and the
extinction effect for a given thickness of film leads to the decrease of
intensity. The best fit (shown on Fig. 26 by solid curve) was obtained for
the following values of the parameter $p=0.44\pm 0.02$, thickness of the
film $L_d=3\mu m$, and Debye-Waller factors of the film and the substrate $%
\exp (-W_f)=\exp (-W_s)=0.93$. Small mosaicity of the sample is also taken
into account by additional convolution of the calculated curves with the
Gaussian function with $\sigma =3.8 \mbox{ arcsec}$. The obtained value of
the parameter $p$ represents the main result of the papers \cite{NKM91,KKS92}%
. Due to interference nature of the XSW method it gives the possibility to
obtain the value of this parameter with high accuracy (see Fig. 26a), where $%
Bi^{3+}$ fluorescence yield curve with the value of $p=0.33$ corresponding
to the uniform distribution of $Bi^{3+}$ ions) is also shown for comparison.

The same bicrystal model was used in the analysis of a recent experiment 
\cite{KZC98} where the lattice constant difference of the isotopically pure $%
^{76}Ge$ grown as a film on the top of a natural $Ge$ single crystal was
measured as a function of the temperature. The variation of XSW in the film
was monitored with the photoelectron detector.

\eject

\section{XSW in Laue geometry}

Up to now we were considering formation of Standing Waves in crystal and
secondary radiation yield for the Bragg geometry of diffraction.
Registration of SR yield in Laue geometry has it's own peculiarities
different from Bragg geometry of scattering and will be discussed in this
Chapter. XSW in the Laue case with the registration of SR yield from the
exit surface have been studied in several papers concerning lattice-atom
fluorescence yield \cite{A67}, external \cite{PY86} and internal \cite{ZKK85}
photoeffect. A Laue-case interferometer was used for the location of
chemisorbed atoms \cite{MFB84}. In this Chapter we will present mainly
theoretical and experimental results obtained in \cite{KKK87,KKK90a}, where
the formation of the fluorescence-yield angular curves in the Laue case and
the possibilities of this geometry for the location of impurity atoms were
studied. The advantage of the Laue geometry of diffraction (Fig.19) is the
possibility to study the position of impurity atoms distributed through the
bulk of a crystal. In this case the sensitivity of XSW in the Bragg geometry
decreases due to extinction effect (see discussion in Chapter 4 of this
review). Moreover, in the Laue case it is easy to use different reflections
(including asymmetrical ones) to study the impurity-atom positions in
different crystallographic directions.

\subsection{Theory}

The general theory of SR yield described in Chapter 3 is valid for the Laue
geometry as well. The yield of the SR is determined by the expression
similar to (\ref{c3.7}) 
\begin{eqnarray}
\kappa ^a(\theta ) &=&\frac 1{I(\infty )}\int\limits_0^LdzP_{yi}^a(z)\left\{
\left| E_0(z,\theta )\right| ^2+\left| E_h(z,\theta )\right| ^2+\right. 
\nonumber \\
&&\ \ \left. +2CF_c^a(z)Re\left[ E_0^{*}(z,\theta )E_h(z,\theta )\exp
[i\varphi ^a(z)]\right] \right\} ,  \label{c6.1}
\end{eqnarray}
where 
\begin{equation}
F_c^a(z)=e^{-W^a(z)}|\varepsilon _{\overline{h}}^a|  \label{c6.2}
\end{equation}
is coherent fraction and $L$ is the crystal thickness. The type of the SR
and the type of atoms emitting it are characterized by index $a$. The phase $%
\varphi ^a(z)=2\pi P_c^a+\mathbf{hu}(z)=\varphi _{\overline{h}}^a+\mathbf{hu}%
(z)$ and the factor $\exp [-W^a(z)]$ correspond only to the atoms of type $a$%
, $\varphi _{\overline{h}}^a=2\pi P_c^a$ is the phase of the complex value $%
\varepsilon _{\overline{h}}^a$ (\ref{c4.9}) and $P_c^a$ is coherent
position. The function $P_{yi}^a(z)$ describes the yield of SR emitted by
atoms of type $a$ located at depth $z$. In the case of Laue geometry and the
fluorescence yield from the exit surface $P_{yi}^a(z)=c_a\exp [\mu
_{yi}^{eff}(z-L)]$, where $\mu _{yi}^{eff}=1/(\gamma _{fl}L_{yi}^{fl})$, $%
L_{yi}^{fl}$ is the depth of the fluorescence yield, $c_a$ is the
concentration of the impurity atoms (see Eq. (\ref{c4.2})).

The field amplitudes $E_{0s}(z,\theta )$ and $E_{hs}(z,\theta )$ for the
crystal with the deformation field $\mathbf{u}(z)$ and Laue geometry of
diffraction (in this case parameter $\gamma _h>0$) can be obtained from the
general TT equations (\ref{c2.16}). In this Chapter it will be discussed in
details the situation, when crystal can be considered as a set of two layers
(bicrystal model) with different lattice parameters, perfection and atomic
compositions. A crystal lattice of each layer is characterized by a constant
value of the parameters $\Delta d/d$, $\exp (-W)$ and $\chi _{0,h},$ which
describe the change of the plane spacing, decrease of the coherent
scattering amplitude and the composition, respectively.

Let us consider like in the previous Chapter a layer of thickness $L_d$ at
the depth $z_0<z<z_0+L_d$. The phase $\varphi (z)$ in (\ref{c2.16}) depends
linearly on $z$ because $\Delta d/d$ is a constant within a layer 
\begin{equation}
\varphi (z)=\varphi (z_0)+2y_\varphi \Delta z/\overline{L}_{ex},
\label{c6.3}
\end{equation}
where $\Delta z=z-z_0$, $y_\varphi $ and $\overline{L}_{ex}$ are constant, $%
\overline{L}_{ex}$ is an extinction length averaged over the crystal bulk.
The solution of TT equations (\ref{c2.16}) should satisfy the given values
of $E_{0s}(z,\theta )$ and $E_{hs}(z,\theta )$ on the entrance boundary of
the layer.

As it was shown in Chapter 2, if the amplitude function $R(z,\theta )$ is
defined in the form (\ref{c2.21}), then it satisfies non-linear equation (%
\ref{c2.22}) (for Laue case lower signs has to be taken) with a boundary
condition $R=R_0$ on the upper boundary of a layer at $z=z_0$. The solution
of Eq. (\ref{c2.22}) in the Laue geometry with constant parameters can be
found analytically (see Eq. (\ref{a1}) from Appendix, where in the case of
one layer the value of $R_n$ on the upper boundary has to be changed by $R_0$%
). Intensity of the transmitted beam $I_0(z,\theta )=\left| E_0(z,\theta
)\right| ^2$ on the depth $z$ defined in (\ref{c2.32}) also can be found
analytically (see Eqs. (\ref{a9}-\ref{a12}) from Appendix).

The contribution of the layer with the thickness $L_d$ to the SR yield in
the frame of these model is analysed in details in Appendix and is
summarized in Eqs. (\ref{a18}-\ref{a20}), where thickness of the layer $d_n$
has to be changed by $L_d$. This equations have to be complemented by the
boundary conditions on the entrance surface of crystal, where $E_0=1$ and $%
R=0$. X-ray transmission ($P_T(\theta )$) and reflectivity ($P_R(\theta )$)
are given by 
\begin{equation}
P_T(\theta )=\left| E_0(L,\theta )\right| ^2,P_R(\theta )=P_T(\theta )\left|
YR(L,\theta )\right| ^2.  \label{c6.10}
\end{equation}

Finally the problem of the SR yield from the exit surface of the crystal in
Laue geometry is completely defined. If measurements are made in a
double-crystal scheme with the first crystal-monochromator convolution with
the first crystal reflectivity curve (according to (\ref{c3.8})) has to be
taken into account.

\subsection{Experiment}

Theory of the SR yield in Laue geometry of diffraction described in the
previous subsection was applied in \cite{KKK87,KKK90a} for the analysis of
the experimental results.

In the first series of measurements $Si$ single crystals of
different thickness with (100) surface orientation uniformly doped during
growth with germanium ($N_{Ge}=7.5\times 10^{25}$ $m^{-3}$) were used. $%
GeK_\alpha $ fluorescence from the exit surface of the silicon crystals in
the Laue geometry and (111) reflection was measured by an energy-dispersive $%
Si(Li)$ detector (see Fig. 27). The angular dependence of $%
GeK_\alpha $ fluorescence yield from a thick crystal with thickness $2.2$ $%
mm $ (Fig. 28a) shows a large maximum slightly shifted to the low-angle side
with respect to the maximum of the diffracted intensity. On the contrary, in
the case of a thin crystal with thickness $0.49$ $mm$, the fluorescence
curve (Fig. 28b) shows a dip near the center of the reflection range and a
weak maximum at the high-angle side.

According to our previous theoretical approach in this case the whole
crystal can be considered as one layer with thickness $L_d=L$ and
fluorescence yield can be determined by Eq. (\ref{a18}). In experiments the
following conditions were also fulfilled $L_0>>L_{yi}>>L_{ex}$, where $%
L_0=\gamma _0/\mu _0$. Since $L_{yi}>>L_{ex}$, the third (interference) term
in (\ref{a18}) is very small ($\left| \psi _3\right| <<\left| \psi _1\right|
\left| \psi _2\right| $) and can be neglected. On the other hand, since $%
L_0>>L_{yi}$ the normalized fluorescence yield does not depend on $L_{ex}$
and $L_{yi}$. Taking this considerations into account from (\ref{a18}) an
approximate expression can be obtained 
\begin{equation}
\kappa (L,\theta )\simeq \sum\limits_{k=1,2}I_k(\theta )\exp \left[ \Delta
\mu _k(\theta )L\right] ,  \label{c6.5}
\end{equation}
where 
\begin{eqnarray}
I_k(\theta ) &=&\frac{1+\beta x_k^2+Bx_k}{1+x_k^2},  \nonumber \\
x_{1,2} &=&\left[ -y\mp \sqrt{y^2+C^2}\right] \left/ C\right. ,  \nonumber \\
\Delta \mu _k(\theta ) &=&-x_k\frac{\left[ 2C\sqrt{\beta }\left| \varepsilon
_{\overline{h}}\right| /\chi _{0i}-x_k(1-\beta )\right] }{L_0(1+x_k^2)},
\label{c6.6} \\
B &=&2C\sqrt{\cdot \beta }\varepsilon _h^a\exp (-W^a)\cos (2\pi P_c^a). 
\nonumber
\end{eqnarray}
To derive (\ref{c6.5}) the following values of the parameters were used $%
Y=1, $ $y_\varphi =0$ and expansion for the complex parameters $x_k$ in the
power series of a small value $\chi _i/\chi _r$ were used.

The most significant feature of Laue-case X-ray diffraction is that two
types of standing waves are formed in a crystal (Fig. 27b). One of these
waves is a weakly absorbed field with the nodes on the atomic planes,
corresponding to the well known Borrmann effect. The other one is a strongly
absorbed field with the antinodes on the atomic planes. So, as follows from (%
\ref{c6.5}), the fluorescence yield angular curve differs distinguishably
for a thick crystal ($L>>L_0$) and a thin one ($L<L_0$). The experimental
curve shown in Fig. 28a corresponds to $L=4.3L_0$. In this case the SR yield
is excited only by the weakly absorbed X-ray standing wave field [$k=1$ in (%
\ref{c6.5})]. The other feature of the Laue case is that with rocking a
crystal through the reflection position, the standing wave fields do not
move with respect to the atomic planes (as in the Bragg case), but only
intensities of these fields increase or decrease. The weakly absorbed field
has a maximum intensity at $\theta =\theta _B$. Even for impurity atoms
lying strictly in the crystal nodes ($P_c^a=0$) one can observe an increase
of the fluorescence yield at $\theta \simeq \theta _B$ in comparison with a
background yield. At any displacement from the crystal node, the impurity
atom occurs in the region of the increased field intensity and fluorescence
yield increases even more.

The obtained experimental curve (Fig.20a) is in a good agreement with the
theoretical one for the model of substitutional impurity atom position. The
maximum normalized fluorescence yield is equal to $3.3$. For the impurity
atom between the reflecting planes ($P_c^a=0.5$) this value will be $13.9$.
Note, that the sensitivity to the position of the impurity atom increases
with increasing crystal thickness. Calculated dependencies of the normalized
fluorescence yield maximum on the impurity position with respect to (111)
and (022) reflecting planes are shown in Fig. 29 for different crystal
thickness. Crystals with intermediate thicknesses are more appropriate for
impurity location studies, due to a rapid decrease of the background
fluorescence yield with increasing of crystal thickness. It should be noted
that a randomly distributed fraction of impurities also leads to an increase
of the fluorescence yield. Indeed, the static Debye-Waller factor for
impurity atoms $\exp (-W^a)$ and the term $\cos (2\pi P_c^a)$ describing
coherent displacements are included in (\ref{c6.5}) as the multipliers.

For a thin crystal, because of the excitation of both standing wave fields
the situation is more complicated. Both weakly ($k=1$) and strongly ($k=2$)
absorbed fields can make a significant contribution to the fluorescence
yield curve, but at different angular positions: field $1$ at $\theta
<\theta _B$ and field $2$ at $\theta >\theta _B$. The main factor now is a
degree of interaction of impurity atoms with the standing wave fields but
not the anomalous X-ray transmission. If the impurity atom is located on
diffraction plane, it interacts with field $2$ more than with field $1$. So,
the maximum of the yield will be observed at $\theta >\theta _B$ (curve $1$
in Fig. 28b). The minimum on curve 1 occurs because field 1 does not
interact with impurities and field 2 is not excited in this angular region.
If impurity atoms are randomly distributed [in (\ref{c6.6}), $\exp
(-W^a)=B=0 $], this effect is compensated entirely by increasing the
interaction with field 1 (curve 2 in Fig. 28b). In Fig. 30, one can see the
fluorescence yield curves calculated for different positions ($P_c^a$) of
impurity atoms. It is obvious that with displacement of the impurity from
the diffraction plane interaction with the weakly absorbed field increases
and with the strongly absorbed field decreases. So, the fluorescence yield
increases at $\theta <\theta _B$ and decreases at $\theta >\theta _B$. The
experimental curves both for thick and thin crystals unambiguously show that
germanium is substitutional in silicon. Such behavior corresponds to the
isovalent nature of this impurity.

In the second series of measurements $Si$ crystals with thickness $%
0.35$ $mm$, (111) surface orientation with an epitaxial layer with thickness 
$1.6$ $\mu m$ on the exit surface were studied. During growth the epilayer
was doped with boron and simultaneously with germanium ($N_{Ge}=10^{28}$ $%
m^{-3}$). The experimental curve of $GeK_\alpha $ fluorescence yield from
the epilayer in the angular range of diffraction in the substrate is shown
in Fig. 31a. In this case fluorescence was excited by XSW formed in the bulk
of the crystal. An X-ray rocking curve in a wide angular range is shown in
Fig. 31b. From this rocking curve the deformation in the epilayer comparing
to that in the substrate can be estimated as $\Delta d/d=-3.8\times 10^{-3}$%
. In the conditions of X-ray diffraction on the epilayer, the so-called
'kinematical' standing waves are formed in this layer. The fluorescence
yield in this angular range is shown in Fig. 31b. Weak minimum and maximum
were clearly observed on the increasing background. The angular dependence
of the background is due to the difference between the plane wave and the
wave incident on the layer due to the diffraction in the bulk.

In this series of experiments the layer with thickness $L_d=1.6$ $\mu m$ is
much less than the extinction length $L_{ex}=12.6$ $\mu m$ and $L>L_0$.
Impurity atoms which are only in the layer 'see' both standing wave fields
and their interference in the angular range of the X-ray diffraction in the
substrate. The interference term oscillates with increasing $z$ and the
period of these oscillations is different at different values of the
incident angle: at $\theta =\theta _B$ the period is equal to $\pi L_{ex}$
and it decreases with increasing $|\theta -\theta _B|$. Since $L>>L_{ex}$,
the fluorescence yield angular curve for the case of the incident plane wave
strongly oscillates with the period less than the angular divergence of the
real incident beam. So, in the treatment of the experimental curve the
interference term can be averaged again giving a zero contribution, although
the reason for this averaging is quite different for that of the previous
case.

Taking all this into account, it is clear why the calculated curves in Fig.
31a for substitutional and randomly distributed impurities are very similar
to the curves in Fig. 28b, although the calculations were carried out in a
double-layer model including convolution. In this case opposite to the
previous one the experimental curve is in a good agreement with the model of
random distribution. This means that impurity atoms are in all possible
positions with respect to the standing waves formed in the substrate.
Nevertheless, impurities in the layer could be ordered, but with period
different from the period of the substrate reflection planes.

The problem of the impurity position can be solved by measuring the
fluorescence yield curve in the angular range of X-ray diffraction on the
epilayer. Since the reflection of X-rays by the epilayer is small ($%
L_d<<L_{ex}$), the standing wave formed in the layer has the intensity
weakly oscillating around the average value. It is the case of so-called
kinematical standing waves. The total reflectivity amplitude $R(z,\theta )$ (%
\ref{c2.21}) in the kinematical approximation can be presented in the
following form 
\begin{eqnarray}
R(z,\theta ) &=&Ce^{-W}e^{iqz}\frac{\sin (qz)}{iqL_{ex}}+  \label{c6.7} \\
&&\ R_0(\Delta \theta +\Delta \theta _B)e^{2iqz},  \nonumber
\end{eqnarray}
where the coordinate $z$ starts from the upper boundary, $q=y/L_{ex}=(\pi
\sin 2\theta _B/\lambda \gamma _h)\Delta \theta $, $\Delta \theta =\theta
-\theta _B$ is the angular deviation from the Bragg angle for the layer, $%
\Delta \theta _B=\theta _B-\theta _B^0$, where $\theta _B$ and $\theta _B^0$
are the Bragg angles for the layer and the substrate, $R_0$ is the
reflectivity amplitude of the substrate.

The fluorescence yield from the layer is defined by the general equation (%
\ref{c6.1}) but at $L_{yi}>>L_d$ one can put $P_{yi}^a(z)=1$ in (\ref{c6.1}%
). Moreover, one can consider that the amplitude of the 'incident' wave does
not depend on $z$ and this is the $0$ component of one of the standing waves
formed in the substrate. Since X-ray diffraction by the layer is observed at 
$\theta >\theta _B^0$ the 'incident' wave is the $0$ component of the
strongly absorbed field ($k=2$). In the case of the paper \cite{KKK90a} the
shift of the Bragg angle in the layer with respect to the substrate $\Delta
\theta _B$ is large enough. It allows to neglect the $h-$component of the
substrate standing wave and put $R_0(\theta )=0$ in the following
consideration.

For simplicity two different cases will be discussed: the first one is for
randomly distributed impurities ($F_c^a=0$) and the second one is for
impurities in crystal nodes ($P_c^a=0$).

In the first case, substituting (\ref{c6.7}) into (\ref{c6.1}) and
performing integration the following result is obtained 
\begin{equation}
\kappa (L_d,\theta )\simeq \left| E_0(\theta )\right| ^2L_d\left[ 1+\frac
\beta 2\left( \frac{Ce^{-W}}{qL_{ex}}\right) ^2\left( 1-\frac{\sin (2qL_d)}{%
2qL_d}\right) \right] .  \label{c6.8}
\end{equation}
As follows from (\ref{c6.9}), on the background yield which increases with
increase of angle $\theta $ in the range $\theta >\theta _B^0,$ a very weak
maximum should be observed with symmetrical shape and intensity proportional
to $(L_d/L_{ex})^2$.

In the second case, 
\begin{equation}
\kappa (L_d,\Delta \theta )\simeq \left| E_0(\theta )\right| ^2L_d\left[ 1+%
\frac{2\sqrt{\beta }F_c^aCe^{-W}}{qL_{ex}}\left( 1-\frac{\sin (2qL_d)}{2qL_d}%
\right) \right] .  \label{c6.9}
\end{equation}
From (\ref{c6.10}) quite different angular dependence of the fluorescence
yield follows. It is determined mainly by the interference term and after
substraction the background it has a dispersion form with the maximum at $%
\Delta \theta >0$ and the minimum at $\Delta \theta <0$ both being
proportional to $(L_d/L_{ex})$. Just this behavior is observed
experimentally (see Fig. 31b) pointing to the correct positions of germanium
in the layer lattice.

The theoretical curves shown in Fig. 31 were calculated using the general
approach described in the beginning of this Section. Both experimental
fluorescence yield curve and X-ray rocking curve are weaker and broader.
This effect can be explained by formation of the mosaic structure due to a
high concentration of germanium in silicon.

Finally, in this chapter we have demonstrated, that standing waves giving
useful structural information can be formed in thin and disordered crystals.
In the most of the previous papers it was considered that XSW fields can be
formed only in large and nearly perfect crystals.

\eject

\section{Model of a multilayer crystal}

\subsection{Theory}

The previous case of a bicrystal model in Bragg and Laue geometry of
scattering is the simplest (however quite useful for the analysis in many
situations as it was demonstrated in the previous Sections) model of a
crystal with deformed surface layer. At the same time there are many cases,
when the profile of deformation and amorphization has a more general
functional form then just a step function. In this case different approaches
can be used to solve the problem of SR yield from such a crystal. If the
profile of deformation and amorphization is known, then one can solve TT
equations either in the form of a set of equations (\ref{c2.16}) or in the
form of Eq. (\ref{c2.22}) for the amplitude $R(z,\theta )$ numerically and
then substitute the obtained solution into Eq. (\ref{c3.7}) for the angular
dependence of the SR yield $\kappa (\theta ).$ Such approach was proposed
for the first time in the paper \cite{KK81} for the analysis of the
photoelectron yield from the crystal with the deformed surface layer (see
Fig. 8). Later the same approach was used in \cite{AGM89} for the analysis
of the wave fields in a crystal with the deformed surface layer taken in the
form of a step Fermi function.

However, in many cases another approach \cite{Kohn} can be useful. One can
divide deformed (and amorphisized) layer with an arbitrary shape on a
several sublayers (see Fig. 32) with a constant structure parameters in each
layer (these are in our case deformation $\Delta d/d$ and amorphization $%
e^{-W}$) and to solve TT equations in each sublayer analytically. Then
substituting this solutions in the general equation (\ref{c3.7}) for the SR
yield the desired result for a multilayer crystal can be obtained.

The details of this approach are given in the Appendix. Analytical solution
of Eq. (\ref{c2.22}) for the amplitude $R(z,\theta )$ in the layer $%
z_n<z<z_n+d_n$ in the frame of the described model can be presented in the
form of Eq. (\ref{a1}). In this equation both geometries of diffraction
(Bragg and Laue) are considered simultaneously.

From the obtained solution it is easy to get standard recurrent equations.
Defining by $R_{in}$ the reflection amplitude for both cases on the
''known'' boundary ($R_{in}=R_{d_n}$ for Bragg case and $R_{in}=R_n$ for
Laue), by $R_{out}$ the reflection amplitude on the unknown boundary and by $%
r_{out}$ solution for $R_{in}=0$ (reflection amplitude by thin layer) we
have from (\ref{a1}-\ref{a4}) 
\begin{equation}
r_{out}=\pm \frac{1-e^{\sigma d_n}}{x_2-x_1e^{\sigma d_n}}  \label{c7.1}
\end{equation}
\begin{equation}
R_{out}=\frac{r_{out}+TR_{in}}{1\mp r_{out}R_{in}}, T=\frac{%
x_2e^{\sigma d_n}-x_1}{x_2-x_1e^{\sigma d_n}},  \label{c7.2}
\end{equation}
where $T$ is the transmission amplitude.

Directly from Eq. (\ref{c2.30}) solution for the direct beam $E_0(z,\theta )$
in the layer $z_n<z<z_n+d_n$ can be obtained in the form 
\begin{equation}
E_0(z,\theta )=T(z,\theta )E_0(z_0,\theta ),  \label{c7.3}
\end{equation}
where $T(z,\theta )$ is the transmission amplitude through the layer defined
in (\ref{a10}). Intensity transfer through the layer $I(\theta
)=|E_0(d,\theta )|^2$ is determined by Eq. (\ref{a11}).

According to (\ref{c3.7}) the normalized SR yield from the n-th layer in
general case of multiatomic unit cell for the sort of atoms $a$ and with the
exponential form of the yield function (\ref{c4.2}) from the entrance
surface can be presented in the following form 
\begin{eqnarray}
\kappa ^a(\theta ) &=&\chi
_{i0}^a\int\limits_{z_n}^{z_n+d_n}dzP_{yi}(z)|T(z,\theta )|^2\times 
\nonumber \\
&&\ \ \ \ \left\{ 1+\frac{|E_h(z,\theta )|^2}{|E_0(z,\theta )|^2}+2CRe\left[ 
\frac{E_h(z,\theta )}{E_0(z,\theta )}\varepsilon _{\overline{h}%
}^ae^{i\varphi ^a(z)-W^a}\right] \right\} .  \label{c7.4}
\end{eqnarray}
Substituting to this expression previously defined functions we obtain 
\begin{eqnarray}
\kappa ^a(\theta ) &=&\chi _{i0}^a\int\limits_0^ddzP_{yi}(z)|T(z,\theta
)|^2\times  \nonumber \\
&&\ \ \ \ \left\{ 1+\beta |Y|^2|R(z,\theta )|^2+2\sqrt{\beta }CF_c^aRe\left[
YR(z,\theta )e^{i\varphi _c^a}\right] \right\} .  \label{c7.5}
\end{eqnarray}
In this equation two important parameters of the theory i.e. \textit{%
coherent fraction} ($F_c^a$) and \textit{coherent position} ($P_c^a$) for
the atoms of the sort $a$ in the n-th layer are introduced. They are defined
as following 
\begin{equation}
F_c^a=|\varepsilon _{\overline{h}}^a|e^{-W^a}, \varphi _c^a=2\pi
P_c^a=\Delta \varphi ^a+\arg (\varepsilon _{\overline{h}}^a)  \label{c7.6}
\end{equation}
and have a constant value in each layer, but can change from layer to layer.

Taking into account analytical solutions for the amplitudes $R(z,\theta )$ (%
\ref{a1}), $T(z,\theta )$ (\ref{a10}) and exponential form for the
probability yield function $P_{yi}(z)$ (\ref{c4.2}) integral in expression
for $\kappa ^a(\theta )$ can be taken analytically (see Eq. (\ref{a18}) in
Appendix). Finally the total yield of the SR from the whole multilayer
crystal is the sum of the yields from every layer 
\begin{equation}
\kappa _{tot}^a(\theta )=\sum\limits_{n=1}^Nz_n(\theta )\kappa ^a(\theta ).
\label{c7.7}
\end{equation}
where $z_1=1$ and $z_n(\theta )=z_{n-1}(\theta )|T(d,\theta )|^2e^{-\mu
_{yi}d}$. So, the whole algorithm for calculation of the angular dependence
of the SR yield from the multilayer crystal in both (Bragg and Laue)
geometries is formulated. In the following several examples of application
of this approach will be given. We have to note here, that described in the
previous Sections bicrystal model is a particular case of this more general
approach with a number of layers $N=2$, where the second (thick) layer can
be considered as a substrate and the top (thin layer) as an epilayer.

\subsection{Applications}

Described above approach of multilayer crystal was used in the papers \cite
{VKS91,VAN96} where silicon crystals were implanted with $Fe$ and $Ni$ ions
and analysed with XSW technique and Rutherford backscattering (RBS) method.
The XSW technique was applied previously \cite{GBB74,HMZ85,AGM76} for
structure investigation of $Si$ crystals after implantation with
the different impurity atoms ($As,$ $Bi$). However, in this papers more
simplified model of the damaged crystal was assumed.

Below, following the papers \cite{VKS91,VAN96}, we will discuss in details
results of XSW analysis of silicon single crystals implanted with $Fe$ ions
(dose $5\times 10^{15}$ ions/cm$^2$ at an energy of $80$ keV). After ion
implantation samples were annealed at 750$^{\circ }$ in the $N_2$
atmosphere. The fluorescence yield from $Fe$ atoms was measured. Results of
this measurements on the samples before and after annealing are presented on
Fig. 33.

The RBS measurements were performed with 2 MeV $^4He$ ions. On Fig. 34 the
results of the damage analysis from the RBS spectra are presented. On this
figure the evolution of the $Fe$ atoms depth distribution is also shown. The
measure of the damage density $k$ is given by the ratio of the displaced
atom density to the crystal atom density , hence $k=100\%$ denotes a
complete amorphization. An important feature of the $Fe-$atom depth
distribution for an as implanted sample is its quasi-Gaussian shape with the
maximum (i.e. $Fe$ mean projected range) located at the depth of 70 nm and
range struggling of 80 nm. As it follows from the analysis the region of
amorphization extends over 190 nm. After annealing the $Fe$ distribution
become narrower, however, its maximum position did not change. Also the peak
area has remained the same. This effect is apparently due to the low
solubility of iron in silicon. In such a case the regrowth front pushes away
impurity atoms in concentrations exceeding the solubility limit. Due to this
process the impurity segregation occurs in a not yet regrown region.

For the theoretical fitting of the angular dependencies of the fluorescence
yield for implanted region of the silicon sample a ''layered'' model
described in the first part of this Chapter was used. In the frame of this
model each deformed layer was characterized by its constant value of
thickness $d,$ deformation $\Delta d/d$ and amorphization factor $\exp (-W)$%
. Distribution of foreign atoms in these layers was characterized by two
main parameters: $P_c^{Fe}-$ the coherent position, i.e. the position of the
mean plane of the impurity atoms with respect to the diffraction planes of
the crystal and $F_c^{Fe}$- coherent fraction, which describes the static
and thermal displacements of the atoms from the mean position $P_c^{Fe}$
(see for definition Eq.(\ref{c7.6}))$.$ Naturally, the contribution of each
layer to the fluorescence yield is determined also by the concentration
distribution $n_c^{Fe}$ of the impurity atoms in the surface layer of the
silicon sample. All other parameters were the same in all layers.

For the theoretical fitting of the fluorescence yield from the implanted
crystal before annealing 14-th layer model and after annealing 11-th layer
model were used. The profile of amorphization $e^{-W}$ of the silicon
surface and the distribution of implanted $Fe$ atoms $n_c^{Fe}$ for each
layer were taken from the results of the RBS experiment (Fig. 34). The best
fitting in the case of not annealed sample was obtained with the value of
the coherent fraction $F_c^{Fe}=0.00\pm 0.02$. This result, as it was
expected, mean that the implanted atoms occupy random positions in the
lattice cell. Parameters of the layers and results of the fitting for the
sample after annealing are summarized in Table 3. In addition to results
obtained from RBS it was proposed that some parts of implanted atoms after
annealing locate in substitutional positions of $Si$ atoms
(parameter $P_c^{Fe}=0$). During the fitting procedure the coherent fraction
of $Fe$ in the corresponding layers $F_c^{Fe}$ and the value of deformation $%
\Delta d/d$ of the surface layer were varied. To minimize the number of
fitting parameters the uniform distribution of $F_c^{Fe}$ and $\Delta d/d$
in the surface layer for the depths $\simeq 40\div 120$ nm was taken. The
best fitting result (see Fig. 33 and Table 3) gives for the average value of
the surface relaxation $\Delta d/d=(-9.6\pm 0.09)\times 10^{-4}$ and the
value of the coherent fraction $F_c^{Fe}=0.27\pm 0.02.$ This value of the
coherent fraction means that approximately $27\%$ (with respect to the
diffraction planes) of implanted $Fe$ atoms are in substitutional $%
Si$ positions. It is interesting to point here that from the independent
XSW analysis it was obtained the same value of substitutional fraction of $%
Fe $ atoms as after the RBS analysis.

The obtained results show the effectiveness and mutual complementarity of
the two methods for investigation of implanted crystals. The RBS experiments
directly give the depth profiles of damage density for $Si$
surface layer and the profile of implanted ions concentration. The XSW
analysis is the precise phase sensitive method. It gives directly the value
of the phase $\varphi (z)$ that in our case is equal to $\varphi (z)=\mathbf{%
hu}(z)+\varphi _c^{Fe},$ where $\mathbf{u}(z)$ is the deformation of the
surface layer (in the case of the uniform deformation we have $u_z=(\Delta
d/d)L$ and $\varphi _c^{Fe}=2\pi P_c^{Fe}.$ As one can see from this
expression the phase is the sum of deformation and the position of the
impurity atoms in the unit cell and additional information is necessary to
determine this two values independently. As it was shown in our analysis the
combination of XSW and RBS methods gives us the opportunity to solve, in
principle, this problem uniquely and determine positions of impurity atoms
in the unit cell as well as the total deformation (relaxation) of the
surface layer.

In the recent paper \cite{LSR99} for the analysis of the fluorescence yield
from the superlattice crystal was also used a similar multilayer model.

\eject

\section{Crystals with an extended deformation field}

Up to now we were mainly considering the cases when the deformation field is
located in the thin subsurface layer. However this condition is not always
fulfilled. There exist a big class of deformations in crystals, that extends
through the whole volume of the crystal. These are, for example, crystals
with the Uniform Strain Gradient (USG), bent crystals, vibrating crystals
and etc. In the following of this section we will show how x-ray standing
wave analysis can be performed in such crystals.

\subsection{Crystals with the uniform strain gradient. Bent crystals}

Theory of x-ray standing wave behavior in crystals with USG and as a special
case bent crystals was for the first time developed in papers \cite
{VKB93a,VKB93b,VKU94}. Later similar theoretical approach for the analysis
of the crystals with USG was performed in \cite{CMG96}. Investigation and
characterization of crystals with USG (bent crystals) by different
diffraction methods is of great practical interest - first of all, due to
their wide application as x-ray monochromators at synchrotron radiation
sources \cite{HSR,CMR87} or focusing spectrometers for x-ray plasma
diagnostics \cite{UFG93}, and secondly, due to the fact that a lot of
typical samples (e.g. implanted crystals, heterostructures) are often
curved. Below we will present results obtained in \cite{VKB93a,VKB93b,VKU94}.

The problem of the x-ray standing wave analysis in bent crystals is quite
complicated due to the fact that the deformation field in such crystals is
in general a two-dimensional function and extend through the whole volume of
the crystal. Below we will consider a special case of such field: crystal
with USG with the deformation field $\mathbf{hu}=Bz^2$, where $z$ is the
depth from the crystal surface. If the incident wave is spherical and the
source is on the Rowland sphere then cylindrically bent crystals are
particular case of crystals with USG \cite{TTb,CMG96}.

The general problem of two variables in this specific case is reduced to
determination of the amplitudes $\mathbf{E}_0(z)$ and $\mathbf{E}_h(z)$,
which depend on one variable $z$. The x-ray amplitude $R(z,\theta )$
defined, as before, in the form (\ref{c2.21}) satisfies equation (\ref{c2.22}%
). In the case of the crystal with the USG\ an effective angular deviation
parameter $y_{eff}(\theta ,z)$ is a linear function from the depth $z$: 
\begin{equation}
y_{eff}(\theta ,z)=y(\theta )+y_\varphi (z)=y(\theta )+C(z/L_{ex}),
\label{c8.1}
\end{equation}
where $y(\theta ),y_\varphi (z)$ are determined by the incident angle on the
surface of the crystal (\ref{c2.23}) and the deformation field (\ref{c2.24})
and $C$ is the dimensionless parameter defining the strength of this field.
For the cylindrically bent crystal the relationship between parameter $C$
and the curvature radius $R$ is given by \cite{TTc} 
\begin{equation}
C=-\frac{\lambda \gamma _0}{\pi X_r^2}\frac{1+\beta }{\beta R}\left[ 1-\beta
\gamma _h^2(1+K)\right] ,  \label{c8.2}
\end{equation}
where $K$ is an elastic constant.

For the analysis of the angular dependence of the SR yield in the crystal
with USG\ general equation (\ref{c3.7}) can be used. The type of the
secondary process (fluorescence, photoemission) will be determined, as
before, by the imaginary part of susceptibility $\chi _{0i}$ and probability
yield function $P_{yi}^a(z)$.

Analytical solutions for the wave fields $\mathbf{E}_0(z)$ and $\mathbf{E}%
_h(z)$ in the crystal with USG were obtained in \cite{CGP78}, however they
can be effectively analysed only in their asymptotic form. The problem of
x-ray Bragg diffraction and standing wave determination in a bent crystal
can be effectively solved with numerical methods. In the papers \cite
{VKB93a,VKB93b} the modified layer model described in the previous Section
was used (see Fig. 35). The whole deformed crystal was divided into layers,
each of thickness $\Delta z=L_{ex}/(mC)$, where $m$ is an integer. In each
layer the value of the angular deviation parameter $y_i$ was considered
constant and its value differs from layer to layer by the magnitude $\Delta
y=1/m$ (parameter $m$ was taken from the condition of small change of $%
\Delta y$ from layer to layer). Analytical solution of Eq. (\ref{c2.22}) for 
$R(z,y_i)$ in each such layer and, hence, in the whole crystal are given by (%
\ref{a1}). The angular dependence of the SR yield in the crystal with USG
can be obtained straightforwardly by substituting the solutions for $%
R(z,y_i) $ into Eq.(\ref{c3.7}), integrating it in each thin layer with the
exponential probability yield function and summing over all layers (see Eqs.
(\ref{c7.5}-\ref{c7.7})).

The angular dependencies of the reflectivity and SR yield were calculated
according to the described model for the different values of the parameter $%
C $. It was taken equal to $0$, $0.1$, $1$ and $10$, which, according to (%
\ref{c8.2}), correspond to the curvature radius $R$ equal to $\infty $, $550$
cm, $55$ cm and $5.5$ cm respectively. On Fig. 36 results of these
calculations in the specific case of (400) symmetrical Bragg diffraction, $%
CuK_\alpha $, $\sigma -$polarized radiation in a thick $Si$ crystal are
presented.

The behavior of the x-ray reflectivity (the broadening of the curves in the
central part, oscillations on the tails of the curves) correspond completely
to the behavior described for the first time in \cite{TTc}. The angular
dependencies of SR yield were calculated for the processes with the
different depths of yield. Two cases were considered: fluorescence yield of $%
 Si K_\alpha $ radiation (with $L_{yi}=70$ $\mu m$) and
photoelectron yield (with $L_{yi}=0.1$ $\mu m$) from the atoms of the
crystal matrix.

The behavior of SR at small values of parameter $C\leq 0.1$ (Fig. 36b) in
the range of the angular parameter $|y|\leq 1$ differs only slightly from
those for the perfect crystal (Fig. 36a). The photoemission yield ($%
L_{yi}<<L_{ex}$, $L_{ex}=3.6$ $\mu m$) for this angles has a dispersion like
behavior and is quite sensitive to the phase shift. On the contrary, for the
fluorescence yield ($L_{yi}<<L_{ex}$) due to extinction effect in the same
angular region the characteristic dip is observed, which suppresses the
phase sensitivity. However, already at small values of curvature
characteristic oscillations can be seen on the tails of SR curves. They
correspond to oscillations on x-ray reflectivity curve. With increasing
curvature one can see first of all a broadening of the phase-sensitive
region and secondly the increasing of the amplitude of the oscillations on
the tails of the photoemission yield curves (Fig. 36c). For large values of
curvature the SR yield curves become, on the average, flatter and flatter
and the period of oscillations is increased (Fig. 36d). These curves
correspond to a strongly deformed crystal. So, in the case of the curved
crystal we have an interesting possibility to study on the same sample the
standing wave behavior in both limits of a weakly and strongly deformed
crystal.

The x-ray standing wave field behavior in the bulk of the deformed crystal
is determined first of all by the type of deformation in the crystal (see
for e.g. \cite{KK81,AGM89}). Reflectivity $P_R(z,y)$ and x-ray standing
wavefield in the bulk of the crystal with USG were analysed in details \cite
{VKB93b}. The intensity of the wavefield on the depth $z$ has the form 
\begin{equation}
I(z,y)=\left| E_0(z,y)\right| ^2\left| 1+\sqrt{\beta }YR(z,y)e^{i\Delta
\varphi _c^a}\right| ^2  \label{c8.3}
\end{equation}
and the phase shift $\Delta \varphi _c^a$ is determined by the positions of
the impurity atoms relative to diffraction planes. The values of $%
E_0(z,\theta )$ and $R(z,\theta )$ were calculated according to the model
described before.

On Fig. 37a two-dimensional distribution of the reflectivity $P_R(z,y)$ in
the ($z,y$) plane is presented. Calculations were made for the parameter $%
C=1 $ and for the depths $z=0\div 2L_{ex}$. As it follows from the obtained
result for a crystal with USG the reflectivity at an arbitrary depth $z$
(while remaining invariable in its shape) is shifted in the angular range
according to (\ref{c8.1}). Such behavior of the reflectivity indicates that
at the depth $z$ there exist an angular region of the total reflection of
x-rays where the phase of the field amplitude changes from $0$ to $\pi $
(so-called phase-sensitive region). For the fixed incident parameter $%
y_0(\theta )$ that region is located at the depth $z_{eff}\sim \left|
y_0(\theta )L_{ex}/C\right| $ and has an effective thickness $\Delta
z_{eff}\sim \left| 2L_{ex}/C\right| $. Thus the size of this region is
determined by the value of the parameter $C$.

The two-dimensional distribution of the intensity $I(z,y)$ (\ref{c8.3}) in
the ($z,y$) plane for the same value of the parameter $C$ are shown on Fig.
38. On Fig. 38a intensity of the wavefield is calculated \textbf{on} the
diffraction planes ($\Delta \varphi _c^a=0$) and on Fig. 38b \textbf{between}
the diffraction planes ($\Delta \varphi _c^a=\pi $). Comparing these
intensity distributions one can notice that they differ both by the value
and by the position of oscillations the tails of the curves. This effect of
phase sensitivity of the standing wave intensity (\ref{c8.3}) in the bulk of
deformed crystal (contrary to the case of perfect crystal) can be understood
if the behavior of the transmitted wave intensity $T^2(z,y)=\left|
E_0(z,y)\right| ^2$ in the crystal with USG is analysed.

The two-dimensional distribution of the transmitted wave intensity $T^2(z,y)$
in the ($z,y$) plane for the parameter $C=1$ is presented on Fig. 37b.
According to the analysis performed in Section 4 the phase sensitivity of
the standing wave intensity in the bulk of the perfect crystal at depths $%
z\simeq L_{ex}$ is suppressed by the extinction effect. The situation is
quite different in the case of the crystals with USG (see Fig. 38b). This
more complicated picture is a result of interference of the incident and
diffracted waves with a deformation field of a crystal. This interference
gives rise to an increase of the transmitted wave on the edge of the
phase-sensitive region even at the depths $z\simeq L_{ex}$ (Fig. 37b). Just
due to this effect the phase sensitivity of the standing wave intensity in
the bulk of the crystal with USG is increased (Fig. 38). Therefore, although
for the perfect crystals applications of the standing wave method are
limited to thin subsurface layers in the case of the crystals with USG this
method can be extended to investigation of thin buried layers (monolayers)
of impurity atoms even located at the depths $z\geq L_{ex}$.

Different calculation approach was used in the paper \cite{VKU94} (see also (%
\cite{CMG96})) for the analysis of x-ray standing wave field on the surface
of the crystals with USG (deformation field was taken in the same form $%
\mathbf{hu}(\mathbf{r})=B\cdot z^2$, where $B$ is a constant). According to
this approach the amplitudes of the x-ray field $\mathbf{E}_0(\mathbf{r})$
and $\mathbf{E}_h(\mathbf{r})$ were obtained from the direct integration of
the TT equations in the form (\ref{c2.11}). For the details of this
calculation algorithm in the Bragg geometry of diffraction see, for example,
review paper \cite{G91}.

In the paper \cite{VKU94} the angular dependencies of x-ray reflectivity and
intensity of x-ray standing waves (\ref{c8.3}) on the surface of the crystal
with USG were calculated for the case of (400) symmetrical Bragg
diffraction, $CuK_\alpha $, $\sigma -$polarized radiation in a thick crystal
of silicon. Results of these calculations are presented on Fig. 39. The XSW
curves on the surface of the crystal were calculated as before for two
different values of the phase $\Delta \varphi _c^a=0$ and $\Delta \varphi
_c^a=\pi $ that correspond to coordinate $\mathbf{r}$ on the surface of the
bent crystal, that either coincide with the position of the diffraction
planes or to the position between diffraction planes. In the calculations
the USG constant $B$ was taken equal to $B=2\cos ^2\theta
_B(1/L_{ex}^L)\beta _c$, where $\beta _c=\pi /(2L_{ex}^L)$ is the critical
value used in the theory of x-ray propagation in bent crystals \cite{G91}
and $L_{ex}^L$ is the extinction length in the symmetric Laue case. This
value of constant $B$ correspond to parameter $C=1$ from Eq. (\ref{c8.2}).
The XSW curves in the deformed crystal comparing to that in the perfect one
(Fig. 39) shows broadening in the central part (in the so-called phase
sensitive region) and also a sharp oscillations on the tails that correspond
to oscillations on the reflectivity curve. The angular position of these
oscillations depend first of all on the strength of the deformation field
(determined now by the value of parameter $B$) and also on the phase shift
value $\Delta \varphi _c^a$. The origin of this oscillation is the waveguide
character of X-ray propagation in crystals with USG \cite{TTc}. It is
important to note here, that the standing wave picture obtained from the
direct numerical calculations of the TT equations for the crystals with the
USG are in a good agreement with the simplified multilayer model used in 
\cite{VKB93a,VKB93b} and described in the beginning of this Chapter.
However, from the point of the waste of the computer time the first method
is much more effective.

\subsection{Vibrating crystals}

Another case of x-ray standing waves in the crystals with extended
deformation field: namely vibrating crystals was considered in the paper 
\cite{NK98}. If in the previous subsection the deformation field in crystal
can be controlled by the strain gradient or radius of curvature in the case
of vibrating crystal this can be achieved by changing the amplitude and the
wave vector of ultrasonic excitations in the crystal. What is more important
for x-ray standing wave technique - extinction length in vibrating crystals
can be also tuned in controllable way. The periodic deviation of the
reflection planes from their ideal position in a perfect crystal
misorientates the inner crystalline structure of the crystal, effectively
''switching off'' the whole parts of the crystal from the dynamical
diffraction scattering. Increasing of the ultrasonic amplitude increases the
deformation field and due to its periodic structure creates new dynamical
reflexes (satellites) with extremely big effective extinction length. In a
limit of a big ultrasonic amplitude, one can achieve the kinematical limit
of diffraction when the initial reflecting power of the crystal is
distributed over the large angular interval.

Below, following \cite{NK98}, the angular dependence of the SR yield will be
considered for the special case of vibrating crystals, crystals disturbed by
longitudinal ultrasonic vibrations propagating normally to the crystal
surface. The advantage of this scheme is the uniform distribution of the
deformation field along the crystal surface ant its layered structure in the
bulk of the crystal.

In this case a vibrating crystal can be presented by a set of identical
layers with thickness equal to the period of the ultrasonic wave. Every
layer can be approximated by a combination of thinner layers with constant
value of interplaner distance (see Fig. 40). This periodic structure give
rise to an additional satellites in the reciprocal space with the position $%
\mathbf{h}+n\mathbf{K}_s$, where $\mathbf{K}_s$ is the wave vector of the
ultrasonic wave. With increasing the amplitude of the ultrasonic wave the
undisturbed part of the crystal (and the reflecting power of the main
reflex) decreases whereas the fraction corresponding to the satellites
increases. Below the special case of the so-called ''high-frequency''
ultrasonic wave, when the ultrasonic wavelength is smaller then the
extinction length of the given reflection will be considered. This gives the
possibility of investigating each satellite independently, because they do
not overlap in the angular range with each other and with the main reflex.

The displacement field of the ultrasonic standing wave will be taken in the
following form 
\begin{equation}
\mathbf{u}(z,t)=\mathbf{w}\cos \omega t\cos K_sz,  \label{c8.4}
\end{equation}
where $\mathbf{w}$ and $\omega $ are the amplitude and frequency of the
ultrasonic vibrations (the wave vector $\mathbf{K}_s$ is directed along the
normal to the crystal surface).

The dynamical symmetrical x-ray diffraction on a thick crystal with the
diffraction vector $\mathbf{h}$ directed along the normal to the crystal
surface (axis $z$) will be considered below. The x-ray wavefield amplitudes $%
E_0(z,\theta )$ and $E_h(z,\theta )$ in the displacement field (\ref{c8.4})
can be effectively found from the set of TT equations (\ref{c2.16}). In this
case Fourier component of the susceptibility of the deformed crystal $\chi
_h^d$ can be presented as an infinite sum 
\begin{equation}
\chi _h^d=\chi _he^{i\mathbf{hu(}z\mathbf{)}}=\chi _h\sum_{n=-\infty
}^\infty J_n(\mathbf{hw})i^n\exp (inK_sz),  \label{c8.5}
\end{equation}
where $J_n(\mathbf{hw})$ is the Bessel function of real argument. Each term
in the infinite sum (\ref{c8.5}) correspond to the satellite with the
diffraction vector $\mathbf{h}+n\mathbf{K}_s$. At small values of the
ultrasonic amplitude $\mathbf{w}$ only two first order satellites ($n=\pm 1$%
) are essential in the expansion (\ref{c8.5}). The amplitudes $E_0(z,\theta
) $ and $E_h(z,\theta )$ can be also expanded in the same way 
\begin{equation}
E_g(z,\theta )=\sum_{n=-\infty }^\infty E_{g,n}(z,\theta )\exp
(inK_sz),g=0,h.  \label{c8.6}
\end{equation}
Substituting expression (\ref{c8.6}) into TT equations (\ref{c2.16}) it is
possible to obtain a set of similar equations for the amplitudes $%
E_{0,n}(z,\theta )$ and $E_{h,n}(z,\theta )$ with the following substitution 
$\chi _h\rightarrow \chi _{h,n}$ and $\alpha \rightarrow \alpha _n$, where 
\begin{equation}
\alpha _n=\alpha +nK_s\frac{\lambda \gamma _h}\pi \mbox{ and }\chi
_{h,n}=i^nJ_n(\mathbf{hw})\chi _h  \label{c8.7}
\end{equation}
are the new values of angular deviation parameter and susceptibility of the
n-th satellite. After straightforward calculations analytical solutions for
the amplitudes of transmitted (only component $E_{0,0}(z,\theta )$ is
important) and diffracted $E_{h,n}(z,\theta )$ wave in the vicinity of the
n-th satellite can be obtained (see for details \cite{NK98}).

The x-ray wavefield in the vicinity of the n-th satellite has the form of a
standing wave with the period $2\pi /|\mathbf{h}+n\mathbf{K}_s|$%
\begin{equation}
E\simeq E_{0,0}e^{i\mathbf{k}_0\mathbf{r}}+E_{h,n}e^{i(\mathbf{k}_h+n\mathbf{%
K}_s)\mathbf{r}},  \label{c8.8}
\end{equation}
It is important to note here that the period of the standing wave does not
depend effectively on the order of the satellite, because the diffraction
vector is much bigger than the wave vector of ultrasonic wave. The only
parameter varying from satellite to satellite is the effective extinction
length. In the angular region of the n-th satellite the effective extinction
length is equal to 
\begin{equation}
L_{ex}^n=\frac{L_{ex}}{|J_n(\mathbf{hw})|},  \label{c8.9}
\end{equation}
where $L_{ex}$ is the extinction length in the undisturbed crystal.
According to (\ref{c8.9}) an effective extinction length is a function of
the ultrasonic amplitude $\mathbf{w}$ and can be tuned by the change of the
voltage applied to ultrasonic transducer.

The secondary radiation yield from the vibrating crystal can be obtained in
the frame of the general theory discussed in Chapter 3. Substituting the
obtained values of the wave fields $E_{0,0}(z,\theta )$ and $%
E_{h,n}(z,\theta )$ into Eqs. (\ref{c3.5}, \ref{c3.6}) the angular
dependence of the total secondary radiation yield $N_n(\theta )$ in
vibrating crystal on the n-th satellite can be presented in the form of the
infinite sum 
\begin{equation}
N_n(\theta )=\sum_{m=-\infty }^\infty N_{n,m}(\theta ).  \label{c8.10}
\end{equation}
Each term in this expansion can be calculated analytically if exponential
form (\ref{c4.2}) for the probability yield function $P_{yi}(z)$ is assumed.
Finally it was obtained \cite{NK98} that every term $N_{n,m}(\theta )$ in
expansion \ref{c8.10} is proportional to the relative contribution of the
scattering potential $J_m(\mathbf{hw})$ and through this Bessel function
depend on the value of the amplitude of the ultrasonic wave.

This theoretical approach was applied for the calculation of the
photoelectron yield (with the depth of yield $L_{yi}=0.2\mu m$) from $%
Si$ (111) crystal $CuK_\alpha $ radiation. The ultrasonic
wavelength $\lambda _s$ was taken equal to $0.7\mu m$. In this case general
equations can be simplified and in the limit of small ultrasonic amplitude
the photoelectron yield near two first satellites ($n=\pm 1$) is equal to
(higher order terms in expansion (\ref{c8.10}) can be neglected) 
\begin{equation}
N_1(\theta )\simeq N_{1,1}(\theta )+N_{1,0}(\theta )+N_{1,-1}(\theta ).
\label{c8.11}
\end{equation}
At the same time contributions to the main reflex from two first-order
satellites are completely compensated and the total SR yield can be
approximated by the expression 
\begin{equation}
N_0(\theta )\simeq N_{0,0}(\theta ).  \label{c8.12}
\end{equation}

Results of numerical calculations of the angular dependence of the
photoelectron yield for the different values of ultrasonic amplitude $%
\mathbf{hw}=0.1,0.5,0.9$ are presented on Fig. 41. As it follows from the
numerical calculations the photoelectron yield near the main reflex varies
very slowly, however, the yield excited in the angular range of the
satellites is very sensitive to the value of the ultrasound amplitude.

It is important to note here that approach discussed in this subsection
gives the intensity of the diffracted x-ray waves and the intensity of the
SR yield for a fix time. In ordinary x-ray scattering experiment on
vibrating crystals only the quantities averaged over the period of
vibrations are measured. However results discussed here can be, in
principle, observed in stroboscopic experiments, that are available on
synchrotron radiation sources. The coincidence between the time-periodic
flashes of the synchrotron radiation beam and the period of the ultrasonic
vibrations gives the possibility to scatter the x-ray wave on the ''frozen''
distorted lattice.

In the recent paper \cite{GKN99} developed above theory was extended for the
case of the yield of the thermal diffuse scattering in vibrating crystals.

\eject

\section{Phase problem}

It is well known, that in any experiment based on the scattering of
radiation by condensed matter, the measured quantity is an intensity, from
which one can determine only the modulus of the complex valued scattering
amplitude. In order to obtain comprehensive information about the structure
of the scattering object, however, it is also necessary to know the phase of
this amplitude. This is a general problem of scattering that arises in many
applications, for example, in the study of the structure of the surface
layers of single crystals by means of x-ray scattering. In this case the
structure of the surface layer is directly connected via Fourier transform
with the angular dependence of the scattering amplitude. At the same time a
standard x-ray diffraction experiment does not give an opportunity to
determine the phase of this amplitude.

To solve this problem of phase determination indirect methods are widely
used in x-ray crystallography (see for review of different methods a book 
\cite{V79}). During last years it was a rapid development of the direct
methods for the phase determination of the structure factors, for example,
by means of multibeam diffraction \cite{C84,C96,H96,KS86}. Another direct
method of the structure analysis and phase determination is based on XSW
method. As it was demonstrated in Chapter 4 coherent fraction $F_c^a(\mathbf{%
h})$ and coherent position $P_c^a(\mathbf{h})$ represent, in fact, the $%
\mathbf{h}-$th amplitude and phase of the Fourier decomposition of the
distribution of atoms under consideration (see also \cite{HMZ85,BM85}). Due
to availability of the synchrotron radiation sources of the third generation
new methods based on the utilization of the high coherency properties of the
synchrotron radiation for the phase determination has become possible \cite
{VPL97,RLV99}. In this case powerful iterative methods for the phase
retrieval are used \cite{F82}.

Here we will discuss another approach for the direct phase determination of
the scattering amplitude based on the simultaneous measurements of the
reflectivity and SR yield in the large angular range. This information can
be used for the direct and what is very important unique determination of
the profile of the structural changes of the crystal surface layer. The idea
of this method was for the first time proposed in \cite{AK78} and later
realized practically in \cite{VKK89}.

If the thickness $L_d$ of the deformed layer in the sample satisfies the
condition $L_d<<L_{ex}$, where $L_{ex}$ is the extinction length, then the
scattering of x-rays by this layer is kinematic and the scattering amplitude 
$R(\theta )$ is given by 
\begin{equation}
R(\theta )=\left| R(\theta )\right| e^{i\alpha (\theta
)}=R_0\int\limits_0^\infty dzF(z)e^{iq(\theta )z},F(z)=f(z)e^{i\varphi (z)}.
\label{c9.1}
\end{equation}
Here $R_0=-i\pi \chi _h/(\lambda \sin \theta _B)$, $q(\theta )=(4\pi
/\lambda )\cos \theta _B\Delta \theta $, $f(z)=\exp (-W(z))$ is the static
Debye-Waller factor and $\varphi (z)=\mathbf{hu}(z)$. According to
expression (\ref{c9.1}), outside the region of the total reflection the
scattering amplitude $R(\theta )$ is the Fourier transform of the complex
function $F(z)$. This function and thus the structure of the surface layer
can be reconstructed directly by the use of the inverse Fourier transform.
However, it is necessary to know both the amplitude and phase of $R(\theta )$%
.

The amplitude of $R(\theta )$ can be obtained from the angular dependence of
the reflectivity curve measured in a big angular range far from the exact
Bragg condition according to equation 
\begin{equation}
P_R(\theta )=C_p\left| R\left( \theta \right) \right| ^2,  \label{c9.2}
\end{equation}
where $C_p$ is a polarization factor equal to $C_p=(1+C^3)/(1+C)$ in a
double-crystal experimental scheme and parameter $C$ is defined in (\ref
{c2.18}).

The phase $\alpha (\theta )$ of the scattering amplitude $R(\theta )$ can be
determined by measuring SR yield with a small depth of yield $L_{yi}$
satisfying the condition: $L_{yi}<<L_d$ in a big angular range far from
exact Bragg condition. This can be, for example, photoelectron yield coming
from the submicron depths of the sample or the fluorescence yield of the
foreign atoms absorbed or implanted on the top of the sample. This angular
dependence can be obtained directly from the general equation (\ref{c3.7})
and is equal to 
\begin{equation}
\kappa (\Delta \theta )=1+C_p\left| R(\Delta \theta )\right| ^2+2C_p\left|
R\left( \Delta \theta \right) \right| \left| \varepsilon _{\overline{h}%
}\right| f(0)\cos \left[ \Phi _R(\Delta \theta )+\varphi (0)+\arg
(\varepsilon _{\overline{h}})\right] ,  \label{c9.3}
\end{equation}
where $f(0)$ and $\varphi (0)$ are the values of the static Debye-Waller
factor $f(z)$ and phase $\varphi (z)$ at $z=0$. As it follows from (\ref
{c9.2}) and (\ref{c9.3}) by simultaneous measurement of the reflectivity $%
P_R(\theta )$ and SR yield $\kappa (\theta )$ both the amplitude $\left|
R\left( \Delta \theta \right) \right| $ and the phase $\alpha (\theta )$ of
the scattering amplitude $R(\theta )$ is obtained uniquely in the whole
angular range. The values of $f(0)$ and $\varphi (0)$ can be found from the
fitting of the central part of curves $P_R(\theta )$ and $\kappa (\theta )$,
where conditions of the strong dynamical scattering are satisfied. Finally,
using inverse Fourier transform, the structure of the damaged layer can be
obtained directly.

In the paper \cite{VKK89} this approach was applied for the determination of
the depth profile of distortions of the surface layers of silicon crystals
after implantation. Samples were implanted by $B^{+}$ ions (with energy of
100 keV, dose of $10^{15}$ $cm^{-2}$) and then annealed ($T=800^{\circ }C$, $%
t=10$ $\min $) in an oxidized atmosphere. In experiment the (400) reflection
and $CuK_\alpha $ radiation were used. The photoelectron emission and the
x-ray reflectivity from the samples were measured both at large deviations
from the Bragg angle (Fig. 42a) and in the region of total reflection (Fig.
42b). Electrons from $K-$shell with energies from $5.9$ keV to $6.4$ keV
were excited and detected by a gas proportional counter with a resolution $%
\approx 15\div 20\%$. According to Eq. (\ref{c4.3}) an estimated depth of
yield of this electrons was equal to $L_{yi}=0.1$ $\mu m$.

It is important to note here, that reflectivity curves obtained in a
standard double-crystal scheme of measurements are the superposition of
coherent and incoherent scattering channels and therefore are inapplicable
for our theoretical analysis, where only coherent part of the scattered
radiation has to be used for the phase retrieval. This problem can be solved
by measuring the ''tails'' of the reflectivity curves in a triple-crystal
scheme. This arrangement makes it possible to separate the coherent part of
scattering from its incoherent part \cite{KKK90b}.

Photoemission curve in the region of the strong dynamical diffraction (Fig.
42b) is similar to the inverted curve from an ideal crystal. It coincides
with theoretical one with the values of the phase $\varphi (0)=2.3$ and
static Debye-Waller factor $f(0)=0.9$. Described above algorithm for the
phase reconstruction was applied to obtain the structural information about
the surface layer of the sample from the ''tails'' of experimentally
measured curves (Fig. 42a). Profiles of the deformation $\Delta d(z)/d$ and
Debye-Waller factor $f(z)$ obtained according to this approach are presented
on Fig. 43. From this Figure it is well seen, that the deformation in the
damaged layer is positive and reaches its maximum at a depth $L\approx 0.3$ $%
\mu m$. This result correlates well both with the distribution of $B^{+}$
ions in silicon lattice under the conditions of implantation and the theory
for defect formation in these crystals \cite{RR78}. We have to note here,
that all calculations were carried out in the limit $L_{yi}\rightarrow 0$.
The final depth of yield of the photoelectrons can be also taken into
account, however, this does not change result essentially.

Finally, in the paper \cite{VKK89} it was presented a method for a direct
solution of the phase problem while the x-ray scattering in the deformed
crystal. From simultaneous measurements of the photoelectron emission and
reflectivity curves in a big angular range near exact Bragg angle the
complex scattering amplitude can be determined completely and thereby the
structure of the deformed surface layer can be reconstructed uniquely.

\eject

\section{Conclusions}

Traditionally, since the discovery of x-rays, x-ray diffraction methods were
used to obtain structural information from different type of crystals,
including protein crystals \cite{V79}. In this review we have shown how
during last 20 years combination of the traditional x-ray diffraction and
x-ray standing wave technique has become one of the most sensitive methods
of studying structure of real crystals. Here we have treated in detail the
problems involved in theoretical studying and practical application of x-ray
standing wave method for the structural analysis of the surface layers.
Starting from the perfect crystal, crystals with amorphous surface layer
were analysed. Then theory and different applications of the simplest model
of deformed surface layer, so-called bicrystal model were presented. This
model was generalized for the case of a multilayer model. Analytical
approach for the secondary radiation yield from such crystal was analysed in
details and different applications of this approach were discussed. It was
shown, that both geometries of diffraction Bragg and Laue can be studied in
the frame of one theoretical approach. However x-ray standing waves and
secondary radiation yield in Laue geometry has its own peculiarities
different from Bragg geometry and they were discussed separately. This
theory was generalized for the case of the crystals with an extended
deformation field, such as crystals with a uniform strain gradient (bent
crystals) and vibrating crystals. We have also discussed the possibilities
of the direct measurements of the phase of the scattering amplitude from
deformed crystal and presented the way of the unique reconstruction of the
structure of this deformed layer.

During last decade new materials with high crystal quality has appeared such
as High-T$_c$ superconductors, quasicrystals, isotopically pure materials.
This opens the possibility to use XSW analysis in such crystals. For
example, High-T$_c$ superconducting crystals $Nd_{1.85}Ce_{0.15}CuO_{4-%
\delta }$ and ultrathin films of YBa$_2$Cu$_3$O$_7$ on SrTiO$_3$(001) were
studied with XSW method in \cite{KKO97,ZSF95}. Fluorescence radiation from $%
AlPdMn$ icosahedral quasicrystal excited by standing waves formed from x-ray
reflections along a twofold symmetry axis was measured in a recent paper 
\cite{JZC99}, lattice constants of the isotopically pure layers of $^{76}Ge$
grown on natural germanium were determined while the photoelectron yield XSW
studies in \cite{KZC98}. The effect of different methods of growth on the
interface quality of extremely thin buried $Ge$ $\delta $ layers in $%
Si$ were studied by means of different diffraction techniques
including XSW \cite{FBM96}. Photoelectron yield and fluorescence radiation
from $(AlAs)(GaAs)$ short-period superlattices was measured in \cite
{LSR99,AIM93,SLM95}.

One interesting approach in XSW analysis was developed in \cite{GBM}. Kossel
diffraction in perfect crystals can be viewed as the reverse process of
x-ray standing waves with the source of radiation located inside the
crystal. New experimental method was proposed in this paper for
determination of atomic positions in the bulk or at a surface of a crystal
by analysing the profile of angularly resolved Kossel lines.

In the last years different theoretical approaches were elaborated for the
analysis of XSW in real crystals. For example, x-ray standing waves in
crystals with statistically distributed defects were analysed theoretically
in \cite{B94,B95}. Analytical results for standing wave behavior in the
deformed crystal with a special type of deformation were obtained in \cite
{K98}. Standing waves in Laue geometry for a special type of scattering with
only imaginary part of the atomic scattering factor were analysed
theoretically in \cite{NFK99}.

As it was shown in our review in many situations it is important to reduce
the signal coming from the background and to enhance the surface
contribution. From this point it is quite favorable to make experiments in
grazing incidence geometry proposed for the first time by W.C. Marra, P.
Eisenberger and A.Y. Cho \cite{MEC79} and later developed as a useful tool
for surface sensitive XSW measurements in a number of papers \cite
{AM83,C85,JCS89,JB90} (see also a book \cite{AAI89}).

In this review paper we were discussing in details formation of XSW field in
the case of two-beam dynamical diffraction in real crystals. This approach
can be generalized for the case of multibeam dynamical diffraction. This was
done for the first time theoretically in \cite{KVK87,K88}. Angular
dependencies of the photoelectron yield in the multibeam region for the case
of four-beam ($220/400/\overline{2}20$) and three-beam ($444/335$) dynamical
diffraction in $Si$ perfect crystals were calculated. If in the
case of two-beam diffraction in one measurement information about atomic
distribution in the sublattice can be obtained only along $\mathbf{h}$%
-vector, in the case of the multibeam diffraction when second reciprocal
vector $\mathbf{g}$ is involved in diffraction process one can obtain
structural information simultaneously in both directions. For the first time
fluorescence measurement in the case of multibeam diffraction for a perfect
crystal was performed in \cite{GM87}, later photoemission yield from a
perfect $Si$ crystal for the case of ($111/220$) three-beam
diffraction was measured in \cite{KKK92,KKK93} (see also for review of this
results \cite{KKK94}). Here we would like to mention also another method of
direct measurement of the phase of the structure factor or total deformation
field in thin surface layers without measuring any SR yield. This method is
based on the measurement of the ''weak'' three-beam reflection far from the
multibeam region simultaneously with the strong one. In this case intensity
of this reflection can be presented in the form 
\begin{equation}
I_g\sim \frac 1{\alpha _g^2}\left| E_0+e^{i\mathbf{hu}(0)}\frac{\chi _{gh}}{%
\chi _{g0}}E_h\right| ^2,  \label{c10.1}
\end{equation}
where $\alpha _g$ is the angular deviation from exact multibeam Bragg
region. This expression is similar to (\ref{c5.2}) obtained for the SR yield
with the small depth of yield. This approach was for the first time applied
for the measurement of the phase of the structure factor in perfect crystals
in \cite{KS86} and then generalized for the case of deformed crystals
theoretically \cite{K88,KS92} and experimentally \cite{KKK96}.

Approach described in this review paper to the calculation and analysis of
the secondary radiation yield while the dynamical diffraction of x-rays was
quite general. However we have restricted ourself mainly with the cases of
fluorescence and photoelectron yield. Actually the potentialities of
applications of x-ray standing waves are much broader. They include for
example the study of such inelastic processes as thermal diffuse and Compton
scattering and internal photoeffect under conditions of dynamical
diffraction of the incident radiation. Every process has its own
peculiarities and contains information of a fundamentally new type.

For example, a unique information about high--order total photoelectron
cross sections, radial wave functions and electron wave phase shifts can be
obtained from XSW angular resolved and integrated photoelectron yield (see
for theoretical and first experimental results \cite
{VZ97,VZ99,VZ99a,FIJ98,JCW00}). We want to mention here also recent results 
\cite{WNP00} where the photoelectron yield in the field of XSW was measured
from the valence band of the Ge and GaAs crystals giving the possibility to
measure the bond polarity parameter.

Thermal diffuse and Compton scattering in the conditions of dynamical
diffraction were studied experimentally starting from the 60-th \cite
{AKK65,AKK66,A68} (see also theoretical work \cite{AA81}). In \cite{SZD88}
it was shown the possibility to determine the values of the phonon
eigenvectors directly from XSW experiment. In the theoretical papers \cite
{GV99a,GV99b} unified Green function approach to the problem of the yield of
the thermal diffuse scattering was elaborated and it was shown that surface
contribution in a special diffraction conditions can become important.
Compton scattering in the XSW field was studied intensively in 80-th \cite
{S81,GKK81,SBM81,S82,SM86,BL87,BKK88a,BKK88b} (see also for review of this
results \cite{GK99}). In this papers it was shown, that investigation of the
coherent Compton scattering opens the possibility for determination of
non-diagonal elements of the density matrix \cite{S82} and contribution of
the valence electrons to the atomic factor of the x-ray scattering \cite
{BL87,BKK88a,BKK88b}.

Measurements of the photoelectric current in a semiconductor crystals while
the dynamical scattering of x-rays \cite{BS69,ZKK85} opens the possibility
to study both the structural and the electrophysical characteristics of the
crystal simultaneously. In a series of papers \cite{AKF78,AML82} it was
demonstrated, that the depth of yield in this case is determined by the
diffusion length $L_d$ of minority carriers. X-ray luminescence studies of
phosphor crystals excited by x-ray standing waves were carried out in \cite
{ZKB92}.

In this review paper we were mainly discussing the possibilities of
generating the XSW in real crystals not in extreme conditions. However,
there are exist some ways for generating x-ray standing waves in so-called
extreme conditions with Bragg diffraction close to $\theta _B=90^{\circ }$
and total external reflection (so-called $0-$th order Bragg reflection). Due
to extreme broadening of the Darwin curve in the first case (the half-width
of the curve $\Delta \theta _{1/2}$ in this case is proportional to$\sqrt{%
\chi _h}$) this conditions of generating XSW can be effectively applied for
less perfect for e.g. mosaic crystals (see for modifications of the
dynamical theory \cite{CC82,GM82} and for XSW measurements \cite
{OKK86,WSM87,WSM88,NHO89} and also review paper \cite{W98}).

The case of the total external reflection can be treated as a special case
of dynamical diffraction with $0-$th order Bragg reflection. The amplitude
ratio for the reflected amplitude $E_R$ to the incident amplitude $E_0$ can
be obtained from the Fresnel formula \cite{BW} or directly from (\ref{c2.21}%
, \ref{c2.29}) with $\theta _B=0$, $\chi _h=\chi _0$ and approximation $\sin
\theta \approx \theta $%
\begin{equation}
\frac{E_R}{E_0}=\frac{\theta -\sqrt{\theta ^2+\chi _0}}{\theta +\sqrt{\theta
^2+\chi _0}}.  \label{c10.2}
\end{equation}
The peculiarity of XSW at the total external reflection is the strong
variation of the wavefield spacing from the reflection angle from about 50
\AA\ for $\theta \approx \theta _c$ (where $\theta _c=\sqrt{|\chi _0|}$ is
the critical angle) to infinity for $\theta \rightarrow 0$. For experimental
and theoretical results of XSW measurements in the total reflection
conditions see for e. g. \cite{BBS89,B91,ZKN93a,ZKN93b,K95}.

Another way to generate an x-ray interference field with long-period
wavefield spacing is the use of Bragg diffraction from Layered Synthetic
Microstructures (LSM) or organic Langmur-Blodgett (LB) multilayers.
Naturally, for generating standing waves these structures has to be
periodic. Alternating amorphous layers of high$-Z$ low$-Z$ materials (e.g. $%
W/C,$ $Rh/C$ or $Ni/C$ and \emph{etc}.) provides high maximum reflectivity.
The lattice plane spacing in such materials can typically range from 15 to
200 \AA . After the first demonstration by Barbee and Warburton \cite{BW84}
of the possibility to use these structures in XSW analysis it was applied
effectively in a number of papers \cite{NSI85,IMI85,BBB88,ZKN92,ZKN93c,ZKN94}%
.

In conclusion we can see that XSW method due to its unique properties of
combination of the x-ray dynamical diffraction with spectroscopic technique
has become a powerful tool in modern crystallography and even more general
in condensed matter physics.

\eject

\section{Appendix}

\bigskip 

\begin{center}
\textbf{Secondary radiation yield from a multilayer crystal }

\textbf{(analytical approach)}
\end{center}

\bigskip 

In Appendix we will give the details of the analytical approach (following 
\cite{Kohn}) for the angular dependence of the SR yield for a multilayer
crystal considered in Section VII. Here both geometries (Bragg and Laue)
will be discussed simultaneously. Evidently, bicrystal model considered in
Section V is a particular case of this more general approach.

Following this approach deformed layer with an arbitrary shape is considered
as a sum of several sublayers (see Fig. 24) with constant parameters
(deformation $\Delta d/d$ and amorphization $e^{-W}$). In this case TT
equations in each sublayer can be solved analytically. Substituting this
solutions in the general equation (\ref{c3.7}) angular dependence for the SR
yield in analytical form for a multilayer crystal can be obtained.

Solution of Eq. (\ref{c2.22}) with constant coefficients for the amplitude $%
R(z,\theta )$ in the n-th layer $z_n<z<z_n+d_n$ has the following form 
\begin{equation}
R(z,\theta )=\frac{F_u(z,\theta )}{F_d(z,\theta )}=\frac{x_1-x_2x_3e^{\mp
\sigma \Delta z_n}}{1-x_3e^{\mp \sigma \Delta z_n}},  \label{a1}
\end{equation}
where the upper sign belong to Bragg geometry and the lower one to Laue. In
this equation the following notations are used 
\begin{equation}
F_u(z,\theta )=x_1-x_2x_3e^{\mp \sigma \Delta z_n}, F_d(z,\theta
)=1-x_3e^{\mp \sigma \Delta z_n}  \label{a2}
\end{equation}
and 
\begin{eqnarray}
x_1 &=&\mp \frac 1{C_1}\left[ b+\sqrt{b^2\mp C_1^2}\right] ,x_2=\mp \frac
1{C_1}\left[ b-\sqrt{b^2\mp C_1^2}\right]  \nonumber  \label{c6.2a} \\
x_3 &=&\frac{x_1-R_{d_n}}{x_2-R_{d_n}}e^{\sigma d_n}\mbox{ (Bragg case),}
\label{a3} \\
x_3 &=&\frac{x_1-R_n}{x_2-R_n}\mbox{ (Laue case)}  \nonumber  \label{c6.2c}
\end{eqnarray}
with 
\begin{equation}
b=-y(\theta )-iy_0+y_\varphi , \sigma =\frac{2i}{L_{ex}}\sqrt{b^2\mp
C_1^2}, \  \Delta z_n=z-z_n  \label{a4}
\end{equation}
and parameters $y(\theta )$, $y_0$, $y_\varphi $ and $C_1$ are defined in (%
\ref{c2.23}-\ref{c2.25}). For the square root in Eqs. (\ref{a3}, \ref{a4})
the branch with the positive imaginary part is chosen. The amplitudes $R_n$
and $R_{d_n}$ in (\ref{a3}) are the values of $R(z,\theta )$ on the top ($%
\Delta z_n=0$) and on the bottom ($\Delta z_n=d_n$) of the n-th layer
boundaries.

Angular parameters $y(\theta ),y_0,y_\varphi $ defined in (\ref{c2.23}, \ref
{c2.24}) can be written in a more general and invariant form through the
extinction length $L_{ex}$ (\ref{c2.26}). Not taking into account the shift
of the origin on the axis $\Delta \theta $ (due to refraction) and defining
by $\delta \theta $ the angular shift due to deformation in the layer we
have 
\begin{equation}
y(\theta )=C_{yt}\Delta \theta ,\  y_0=\pm \frac{\mu _0L_{ex}}{4\gamma
_0}(1\pm \beta ),\  y_\varphi =C_{yt}\delta \theta .  \label{a7}
\end{equation}
In (\ref{a7}) parameter $C_{yt}$ gives connection between the angles
measured in angular units (\textit{arcsec)} and in dimensionless parameter $%
y $ 
\[
C_{yt}=\pi g\frac{L_{ex}}\lambda \frac \beta {\gamma _0}\sin 2\theta
_B,g=0.4848\cdot 10^{-5}. 
\]
If extinction length $L_{ex}$ is defined with the parameters of the
substrate, then parameters $C_{yt}$ and $L_{ex}$ in all layers are the same,
however, such parameters as $\delta \theta ,e^{-W},\mu _0$ and $p$ (see Eq. (%
\ref{c2.25})) can, in principle, differ from layer to layer.

From Eq. (\ref{c2.30}) formal solution for the direct beam $E_0(z,\theta )$
in the layer $z_n<z<z_n+d_n$ can be obtain in the following form 
\begin{equation}
E_0(z,\theta )=E_0(z_n,\theta )\exp \left[ \frac{i\pi \chi _0}{\lambda
\gamma _0}\Delta z_n-i\frac{C_1}{L_{ex}}\int\limits_{z_n}^zdz^{\prime
}R(z^{\prime },\theta )\right] .  \label{a8}
\end{equation}
Integral in the exponent with the amplitude (\ref{a1}) can be taken
analytically giving for the direct field amplitude 
\begin{equation}
E_0(z,\theta )=T(z,\theta )E_0(z_n,\theta ).  \label{a9}
\end{equation}
Here $T(z,\theta )$ is the transmission amplitude through the layer 
\begin{equation}
T(z,\theta )=\exp \left[ i\Phi (\theta )\Delta z_n/2\right] \left( \frac{%
1-x_3e^{\mp \sigma \Delta z_n}}{1-x_3}\right) ,\  \Phi (\theta )=\frac{%
2\pi \chi _0}{\lambda \gamma _0}-\frac{2C_1x_1}{L_{ex}}.  \label{a10}
\end{equation}
Intensity transfer through the layer is given by the function 
\begin{equation}
|T_n|^2=|T(d_n,\theta )|^2=e^{-M(\theta )d_n}\frac{|F_d(d_n,\theta )|^2}{%
|F_d(0,\theta )|^2},  \label{a11}
\end{equation}
where 
\begin{equation}
M(\theta )=Im[\Phi (\theta )]=\frac{\mu _0}{2\gamma _0}\left( 1\mp \beta
\right) \mp \sigma _r.  \label{a12}
\end{equation}
According to (\ref{c3.7}) the normalized SR yield from the n-th layer in
general case of multiatomic unit cell for the sort of atoms $a$ and with
exponential form of the yield function (\ref{c4.2}) from the entrance
surface can be presented in the following form 
\begin{eqnarray}
\kappa _n^a(\theta ) &=&\chi
_{i0}^a\int\limits_{z_n}^{z_n+d_n}dzP_{yi}(z)|T(z,\theta )|^2\times 
\nonumber \\
&&\ \ \ \left\{ 1+\frac{|E_h(z,\theta )|^2}{|E_0(z,\theta )|^2}+2CRe\left[ 
\frac{E_h(z,\theta )}{E_0(z,\theta )}\varepsilon _{\overline{h}%
}^ae^{i\varphi ^a(z)-W^a}\right] \right\} .  \label{a13}
\end{eqnarray}
Taking into account our definitions this can be written as 
\begin{eqnarray}
\kappa _n^a(\theta ) &=&\chi
_{i0}^a\int\limits_0^{d_n}dzP_{yi}(z)|T(z,\theta )|^2\left\{ 1+\beta
|Y|^2|R(z,\theta )|^2\right.  \nonumber \\
&&\ \ \ \left. +2\sqrt{\beta }CRe\left[ YR(z,\theta )\varepsilon _{\overline{%
h}}^ae^{i(\varphi ^a(z)-\varphi (z))-W^a}\right] \right\} .  \label{a14}
\end{eqnarray}
In the following it will be assumed, that $\varphi ^a(z)-\varphi (z)=\Delta
\varphi ^a=const$. This condition mean in fact, that due to macroscopical
displacements the unit cell is shifted as a whole. Substituting now
previously obtained solutions for the reflected $R(z,\theta )$ (\ref{a1})
and transmitted $T(z,\theta )$ (\ref{a10}, \ref{a11}) amplitudes into Eq. (%
\ref{a14}) SR yield from the n-th layer can be obtained 
\begin{eqnarray}
\kappa _n^a(\theta ) &=&\frac{\chi _{i0}^a}{|F_d(0,\theta )|^2}%
\int\limits_0^{d_n}dze^{-(M(\theta )+\mu _{yi})z}\times \left\{
|F_d(z,\theta )|^2\right.  \nonumber \\
&&\ \ \ \left. +K_r|F_u(z,\theta )|^2+Re\left[ K_iF_d^{*}(z,\theta
)F_u(z,\theta )\right] \right\} ,  \label{a15}
\end{eqnarray}
where functions $F_u(z,\theta )$ and $F_d(z,\theta )$ are defined in \ref{a2}
and 
\begin{equation}
K_r=\beta |Y|^2,\  K_i=2\sqrt{\beta }CYF_c^ae^{i\varphi _c^a}.
\label{a16}
\end{equation}
In this equation two important parameters of the theory i.e. \textit{%
coherent fraction} ($F_c^a$) and \textit{coherent position} ($P_c^a$) for
the atoms of the sort $a$ in the n-th layer are introduced. They are defined
as following 
\begin{equation}
F_c^a=|\varepsilon _{\overline{h}}^a|e^{-W^a},\  \varphi _c^a=2\pi
P_c^a=\Delta \varphi ^a+\arg (\varepsilon _{\overline{h}}^a),  \label{a17}
\end{equation}
Integral in the expression for $\kappa _n^a(\theta )$ can be taken
analytically and finally for the angular dependence of the SR yield from the
n-th layer we have 
\begin{equation}
\kappa _n^a(\theta )=\frac{\chi _{i0}^a\cdot d_n}{|1-x_3|^2}\left[ A_1\psi
_1+A_2\psi _2-Re(A_3\psi _3)\right] ,  \label{a18}
\end{equation}
where 
\begin{eqnarray}
A_1 &=&1+K_r|x_1|^2+Re(K_ix_1),  \nonumber \\
A_2 &=&|x_3|^2\left[ 1+K_r|x_2|^2+Re(K_ix_2)\right] ,  \label{a19} \\
A_3 &=&x_3\left[ 2(1+K_rx_1^{*}x_2)+K_ix_2+K_i^{*}x_1^{*}\right]  \nonumber
\end{eqnarray}
and 
\begin{eqnarray}
\psi _1 &=&(1-e^{-f_1})/f_1,\  \psi _2=(1-e^{-f_2})/f_2,\  \psi
_3=(1-e^{-f_3})/f_3,  \nonumber \\
f_1 &=&(M(\theta )+\mu _{yi})d_n,\  f_2=f_1\pm 2\sigma _rd,\  %
f_3=f_1\pm \sigma d.  \label{a20}
\end{eqnarray}
The total yield of the SR from the whole multilayer crystal is the sum of
the yields from every layer 
\begin{equation}
\kappa _{tot}^a(\theta )=\sum\limits_{n=1}^Nz_n(\theta )\kappa _n^a(\theta ),
\label{a21}
\end{equation}
where $z_1=1$ and $z_n(\theta )=z_{n-1}(\theta )|T(d_n,\theta )|^2e^{-\mu
_{yi}d_n}.$ In the Bragg case the reflectivity is equal to 
\begin{equation}
P_R(\theta )=|YR(0,\theta )|^2  \label{a22}
\end{equation}
and calculation is performed layer by layer from down to the top starting
from the $N$-th layer up to the first.

In the Laue case transmission coefficient can be calculated according to the
relationship 
\begin{equation}
P_T(\theta )=z_{N+1}(\theta )\exp (\sum_{n=1}^N\mu _{yi}^nd_n)  \label{a23}
\end{equation}
and reflectivity in the same geometry is equal to 
\begin{equation}
P_R(\theta )=|YR_{N+1}(\theta )|^2P_T(\theta ).  \label{a24}
\end{equation}
Finally, the whole algorithm for calculation of the angular dependence of
the SR yield from the multilayer crystal in both (Bragg and Laue) geometries
is defined.

\newpage\ 

\textbf{Acknowledgments}

The authors are thankful to S.I. Zheludeva for her interest and support
while the preparation of this manuscript. One of the authors (IAV) is
thankful to V.G.Kohn for illuminating discussions on the problems of
dynamical theory and XSW method . The authors are thankful to E. Tereshenko
and S. Verkoshansky for the technical support.

\bigskip 

\textbf{List of acronyms}

APS -- Advanced Photon Source

ESRF -- European Synchrotron Radiation Facility

FYG -- Iron-Yttrium garnet

GGG -- Gallium-Gadolinium garnet

GNG -- Gallium-Neodymium garnet

LB -- Langmur-Blodgett

LSM -- Layered synthetic microstructures

SR -- Secondary radiation

TT equations -- Takagi--Taupin equations

USG -- Uniform strain gradient

XSW -- X-ray standing wave

\eject

\eject

\textbf{Table 1.} Values of $\varphi _0$, $W_0$ and $\Delta d/d$ obtained as
a result of fitting of experimental data presented on Fig. 16. Calculations
were made for different concentrations of $Ge$ ($n_c^{Ge}$) in the epitaxial
film in the frame of bicrystal model.

\bigskip 

\begin{center}
\begin{tabular}{|l|l|l|l|}
\hline
$n_c^{Ge}$ (10$^{19}$ atoms cm$^{-3}$) & $W_0$ & $\varphi _0/\pi $ & ($%
\Delta d/d)\times 10^5$ \\ \hline
0 & 0.51(20) & --0.84(20) & --2.2(5) \\ \hline
3.7 & 0.25(20) & --0.55(20) & --1.4(5) \\ \hline
7 & 0.8(2) & --0.09(20) & --0.2(5) \\ \hline
10 & 0.8(2) & 0.37(20) & 0.9(5) \\ \hline
15 & 0.79(20) & 1.36(20) & 3.6(5) \\ \hline
\end{tabular}
\end{center}

\eject

\textbf{Table 2.} Experimental parameters measured by photoluminescence
(PL), x-ray rocking curve analysis (XRC) and x-ray standing wave method
(XSW) for $10$ $nm$ $CaF_2/Si (111)$ layers grown at different
temperatures.

\bigskip 

\begin{center}
\begin{tabular}{|l|l|l|l|l|}
\hline
T$_1$ ($^{\circ }$C) & 100 & 550 & 660 & 770 \\ \hline
$\Delta a/a$ $(\%)$ & 0.8 & 1.7 & 1.9 & 2.2 \\ \hline
$\Delta E$ $(cm^{-1})$ & --38 & --30 & -- & 100 \\ \hline
$\delta \lambda $ $(nm)$ & 0.7 & 1.4 & -- & 1.6 \\ \hline
$\varepsilon _{xx}^{PL}(300)$ $(\%)$ & --0.60 & --0.55 & -- & 0.54 \\ \hline
$\varepsilon _{xx}^{XRC}$ $(\%)$ & --0.58 & --0.42 & 0.007 & 0.55 \\ \hline
$\Delta \theta _{111}$ (arcsec) & --640 & --550 & --320 & --25
\\ \hline
$e^{-W}$ & 0.80 & 0.80 & 0.65 & 0.65 \\ \hline
$P_{XSW}^{111}$ $(\AA )$ & 2.91 & 2.86 & 2.83 & 2.91 \\ \hline
\end{tabular}
\end{center}

\eject

\textbf{Table 3.} Parameters of the layers obtained as a result of
theoretical fitting of the fluorescence yield from $Si$ samples
implanted by $Fe$ ions after annealing (Fig. 33b).

\bigskip 

\begin{center}
\begin{tabular}{|l|l|l|l|l|l|l|}
\hline
No. & d ($\mu m$) & $\Delta d/d$ ($\times 10^4$) & $e^{-W}$ & $F_c^{Fe}$ & $%
P_c^{Fe}$ & n$_c^{Fe}$ (\%) \\ \hline
1 & 0.015 & 0 & 0 & 0 & 0 & 4 \\ \hline
2 & 0.01 & 0 & 0 & 0 & 0 & 14 \\ \hline
3 & 0.01 & 0 & 0 & 0 & 0 & 24 \\ \hline
4 & 0.01 & --9.6(0.09) & 0.15 & 0.27(0.02) & 0 & 38 \\ \hline
5 & 0.01 & --9.6(0.09) & 0.34 & 0.27(0.02) & 0 & 86 \\ \hline
6 & 0.01 & --9.6(0.09) & 0.41 & 0.27(0.02) & 0 & 100 \\ \hline
7 & 0.01 & --9.6(0.09) & 0.54 & 0.27(0.02) & 0 & 86 \\ \hline
8 & 0.01 & --9.6(0.09) & 0.64 & 0.27(0.02) & 0 & 40 \\ \hline
9 & 0.01 & --9.6(0.09) & 0.77 & 0.27(0.02) & 0 & 30 \\ \hline
10 & 0.01 & --9.6(0.09) & 0.88 & 0.27(0.02) & 0 & 14 \\ \hline
11 & 0.02 & --9.6(0.09) & 0.97 & 0.27(0.02) & 0 & 7 \\ \hline
\end{tabular}
\end{center}

\eject

%
%
\begin{figure}[tbp]
\begin{center}
\includegraphics{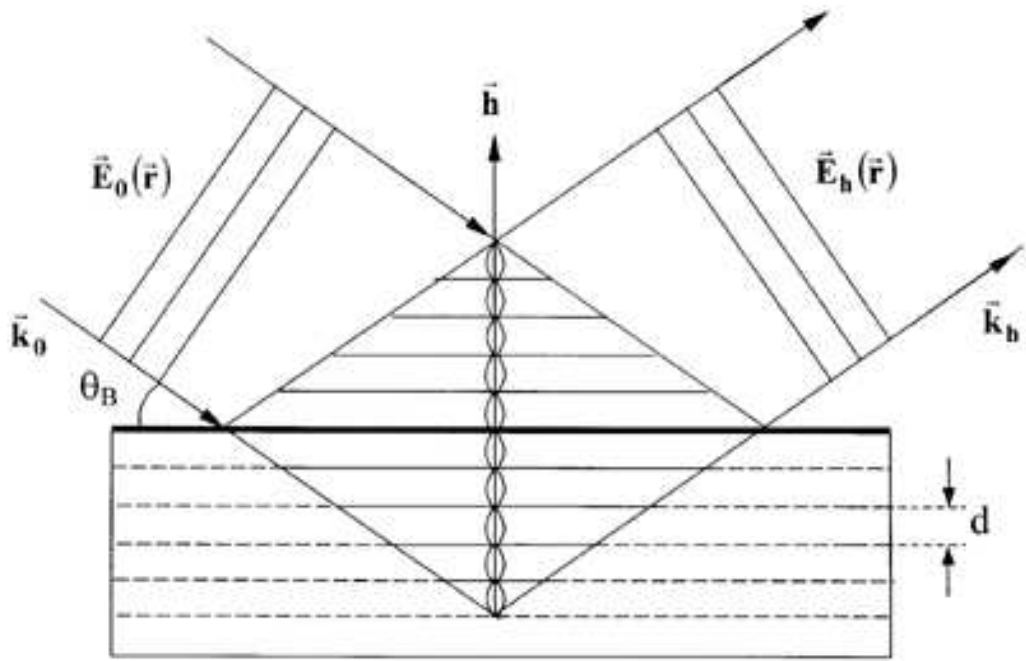}
\end{center}
\caption{
\textsl{\label{}Interference x-ray standing wave field as a result of
coherent superposition of the incident $\mathbf{E}_0(\mathbf{r})$ and
diffracted $\mathbf{E}_h(\mathbf{r})$ plane waves with the periodicity $%
d=2\pi /|\mathbf{h}|$.}
}
\end{figure}

\clearpage

%
%
\begin{figure}[tbp]
\begin{center}
\includegraphics{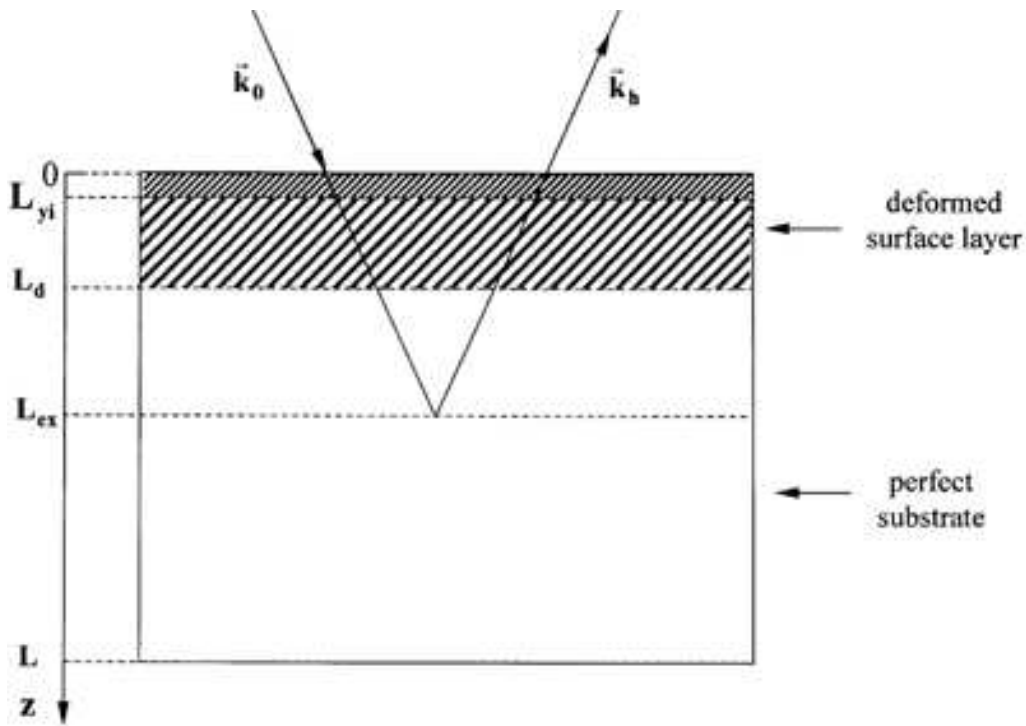}
\end{center}
\caption{
\textsl{\label{}Crystal of thickness $L$, with the deformed surface layer
of thickness $L_d$. While the Bragg diffraction x-rays penetrate into the
crystal on the depths of the order of $L_{ex}$. The depth of the SR yield is
determined by the parameter $L_{yi}$. It is shown the situation, when the
condition $L_{yi}<<L_d<<L_{ex}$ is satisfied.}
}
\end{figure}

\clearpage

%
%
\begin{figure}[tbp]
\begin{center}
\resizebox{13cm}{!}{
\includegraphics{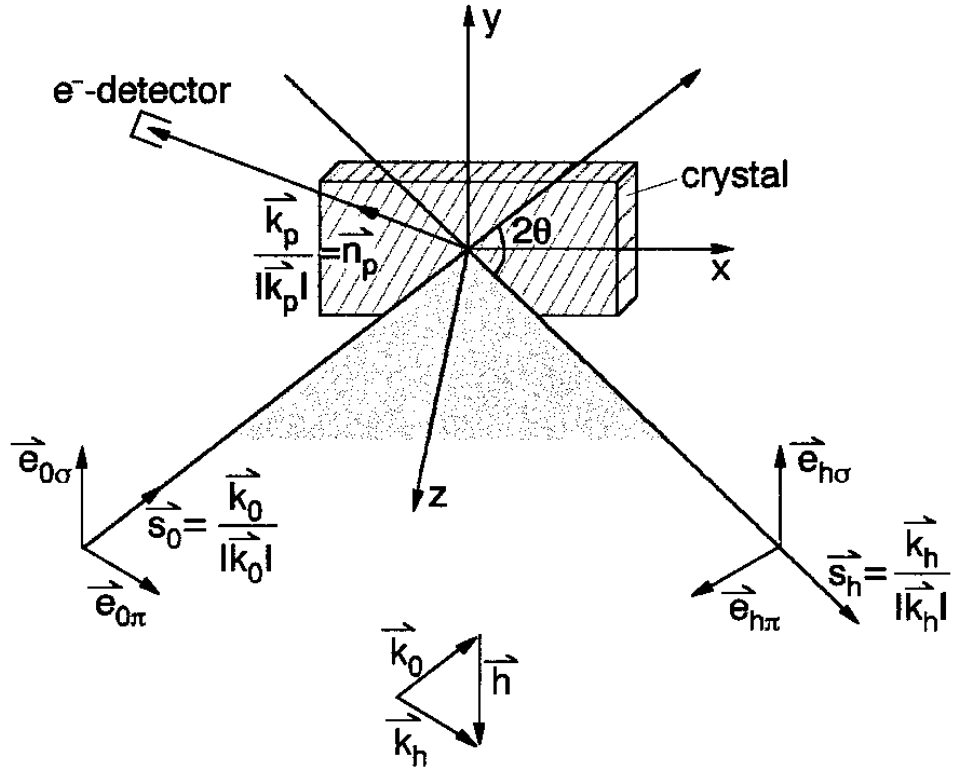}}
\end{center}
\caption{
\textsl{\label{}Shematical view of dynamical diffraction and photoelectron
yield measurements excited by two coherent electromagnetic waves $\mathbf{E}%
_0(\mathbf{r})$, $\mathbf{E}_h(\mathbf{r})$ characterised by the wave
vectors $\mathbf{k}_0$\textbf{, }$\mathbf{k}_h$ and polarization vectors $%
\mathbf{e}_{0\sigma }$, $\mathbf{e}_{0\pi }$ and $\mathbf{e}_{h\sigma }$, $%
\mathbf{e}_{h\pi }$. The angle $2\theta $ is the angle between the wave
vectors $\mathbf{k}_0$\textbf{\ }and\textbf{\ }$\mathbf{k}_h$. The direction
of the escaping photoelectron is determined by the vector $\mathbf{k}_{%
\mathbf{p}}$.The relationship of the vectors $\mathbf{k}_0$\textbf{, }$%
\mathbf{k}_h$ and $\mathbf{h}$ is indicated.}
}
\end{figure}

\clearpage

%
%
\begin{figure}[tbp]
\begin{center}
\includegraphics{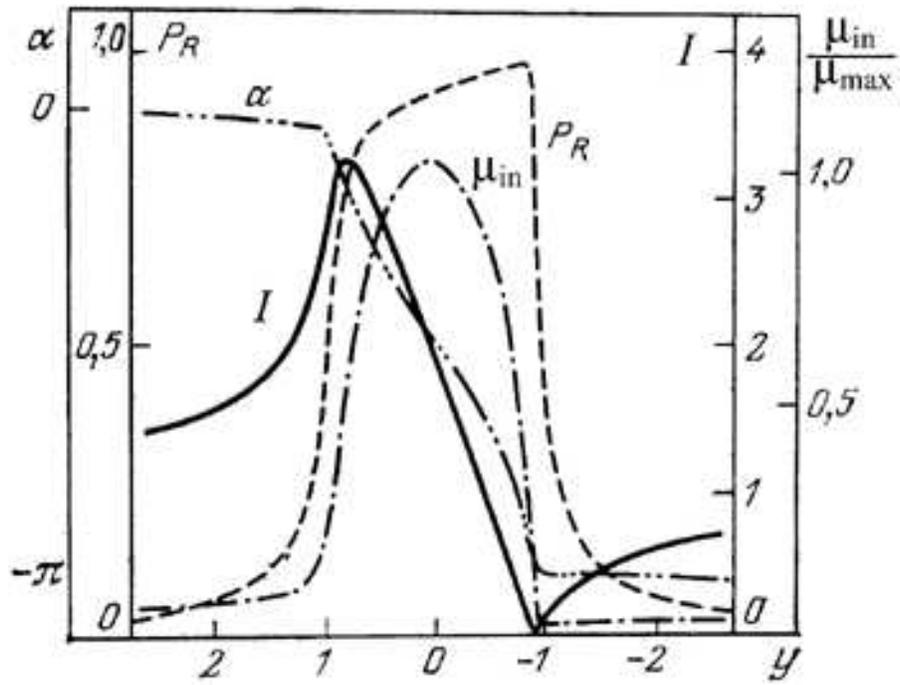}
\end{center}
\caption{
\textsl{\label{}Angular dependences of the intensity of the wave field at
the atomic planes ($I$, solid line), the reflectivity ($P_R$, dotted line),
the interference absorption coefficient ($\mu _{in}$ dot-dash line), and the
phase $\alpha $ (dash-double-dot line).}
}
\end{figure}

\clearpage

%
%
\begin{figure}[tbp]
\begin{center}
\resizebox{13cm}{!}{
\includegraphics{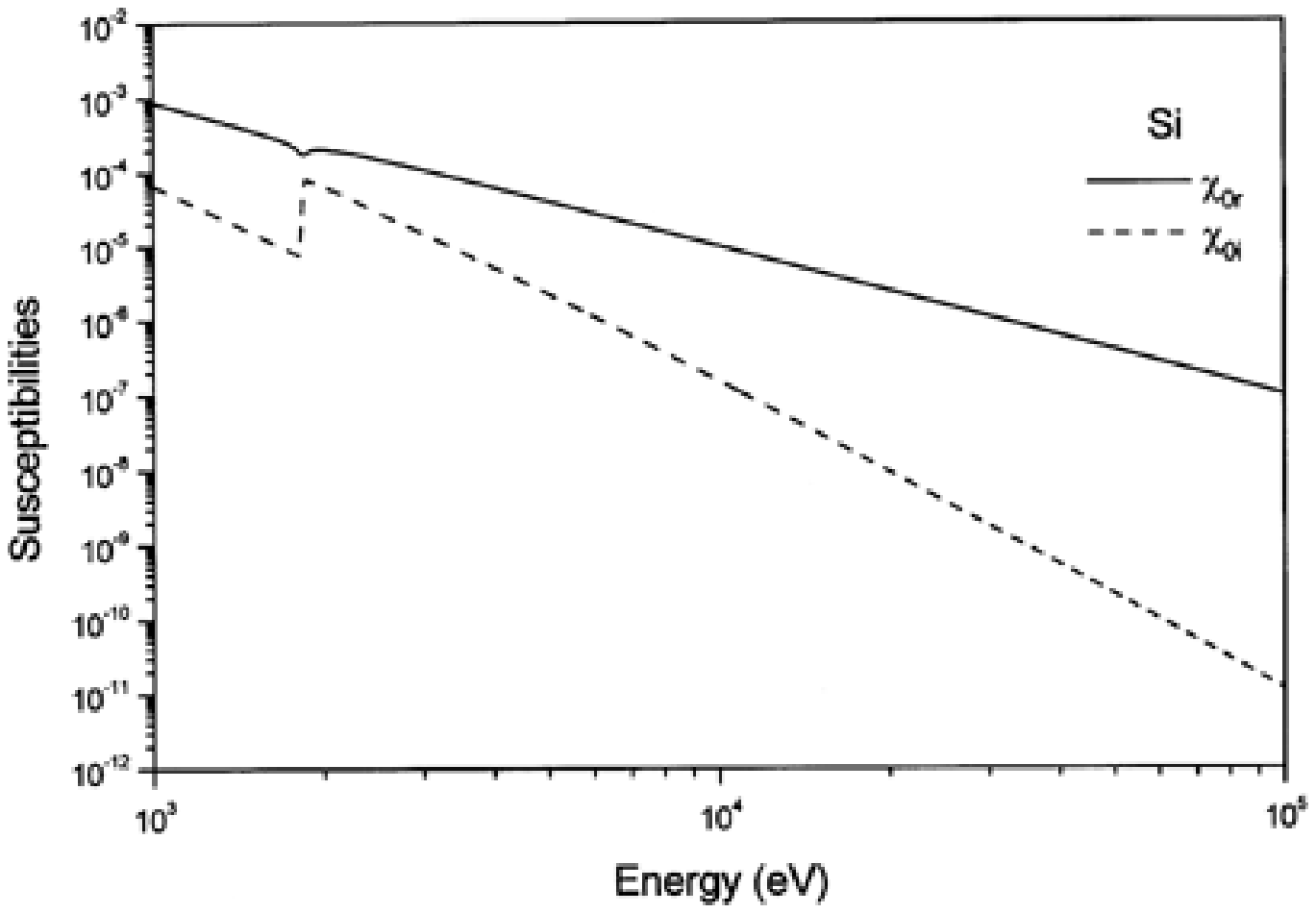}} \\
\bigskip
\resizebox{13cm}{!}{
\includegraphics{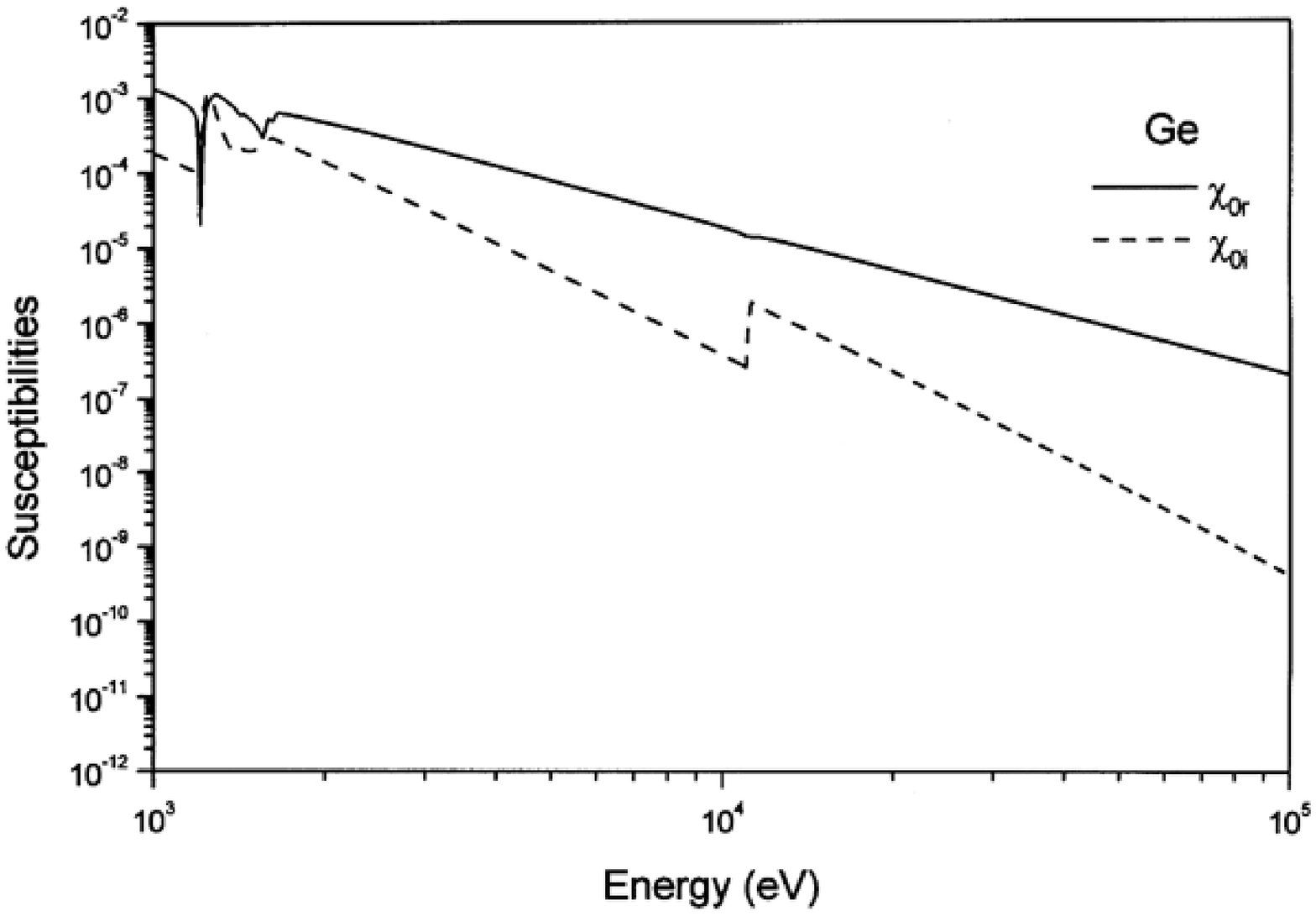}}
\end{center}
\caption{
\textsl{\label{}The values of the zero Fourier components of the
susceptibilities $\chi _{0r}$ and $\chi _{0i}$ calculated for $Si$ and $Ge$
as a functions of energy $E$ \cite{SD98}.}
}
\end{figure}

\clearpage

%
%
%
\begin{figure}[tbp]
\begin{center}
\resizebox{13cm}{!}{
\includegraphics{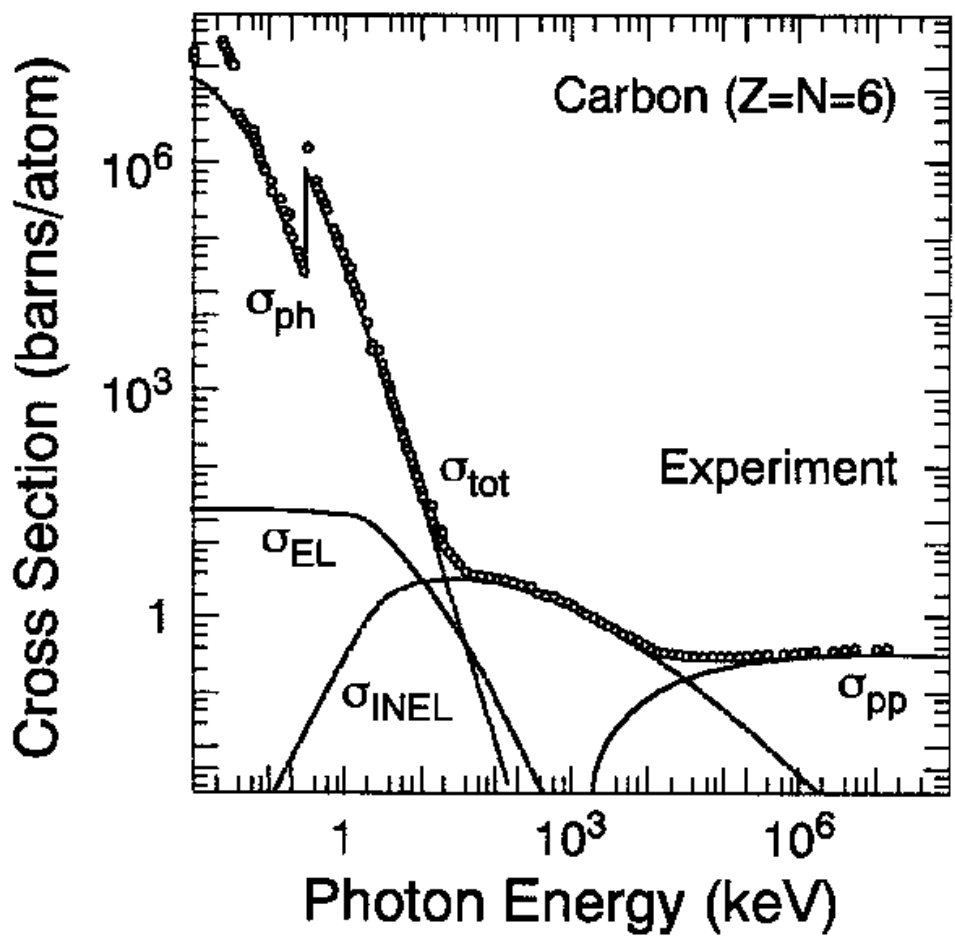}}
\end{center}
\caption{
\textsl{\label{}Calculated cross-sections for photon scattering from carbon
showing the contributions of photoelectric, elastic (Rayleigh), inelastic
(Compton), and pair-production cross-sections. Also shown are the
experimental data (open circles). From Ref. \cite{IT92}. Even for the light
element carbon, the photoelectric cross-section is clearly dominating in the
x-ray energy range up to more than 100 keV.}
}
\end{figure}

\clearpage

%
%
%
\begin{figure}[tbp]
\begin{center}
\includegraphics{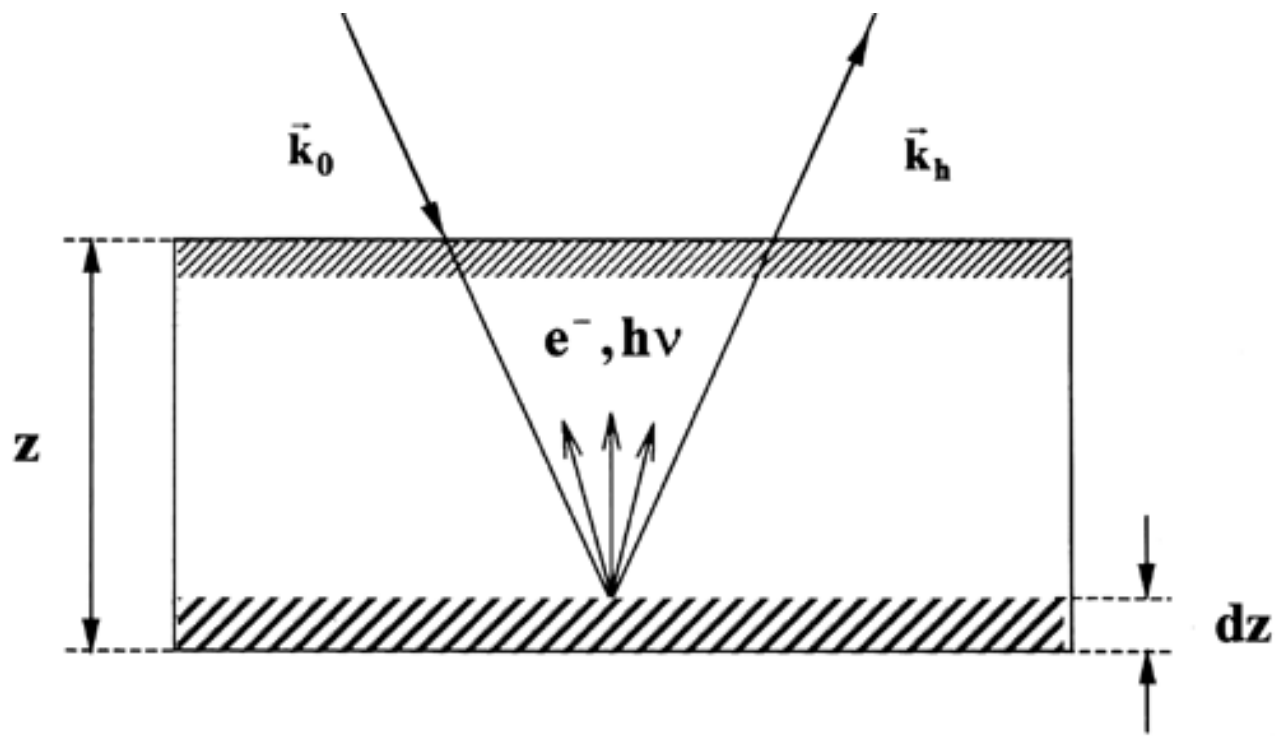}
\end{center}
\caption{
\textsl{\label{}Part of x-ray radiation while scattering on crystal is
absorbed in a layer with the thickness $dz$ at the depth $z$ and gives rise
to the yield of the secondary radiation (photoelectron $e^{-}$, fluorescence 
$h\nu $, \textit{etc}.).}
}
\end{figure}

\clearpage

%
%
%
\begin{figure}[tbp]
\begin{center}
\includegraphics{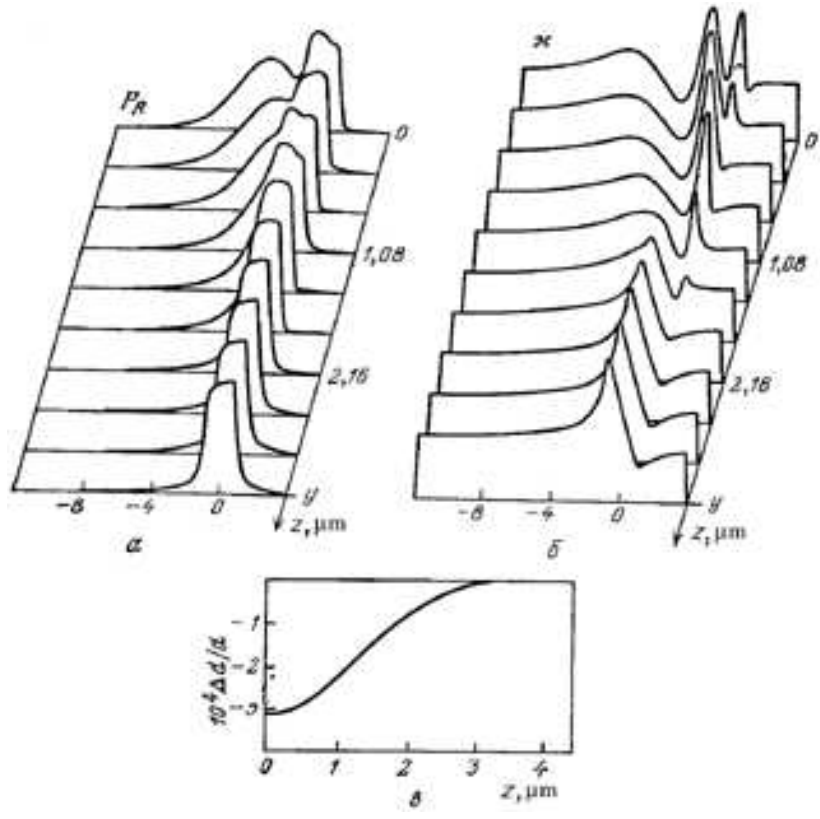}
\end{center}
\caption{
\textsl{\label{}Reflectivity $P_R(y)$ (a) and photoemission $\kappa (y)$
(b) curves calculated for a $Si$ crystal with a deformed surface
layer under conditions of (111) diffraction of a plane monochromatic wave of 
$CuK_\alpha $ radiation. Also shown (c) the profile of deformation $\Delta
d/d(z)$ in the surface layer. From Ref. \cite{KK81}.}
}
\end{figure}

\clearpage

%
%
%
\begin{figure}[tbp]
\begin{center}
\resizebox{9cm}{!}{
\includegraphics{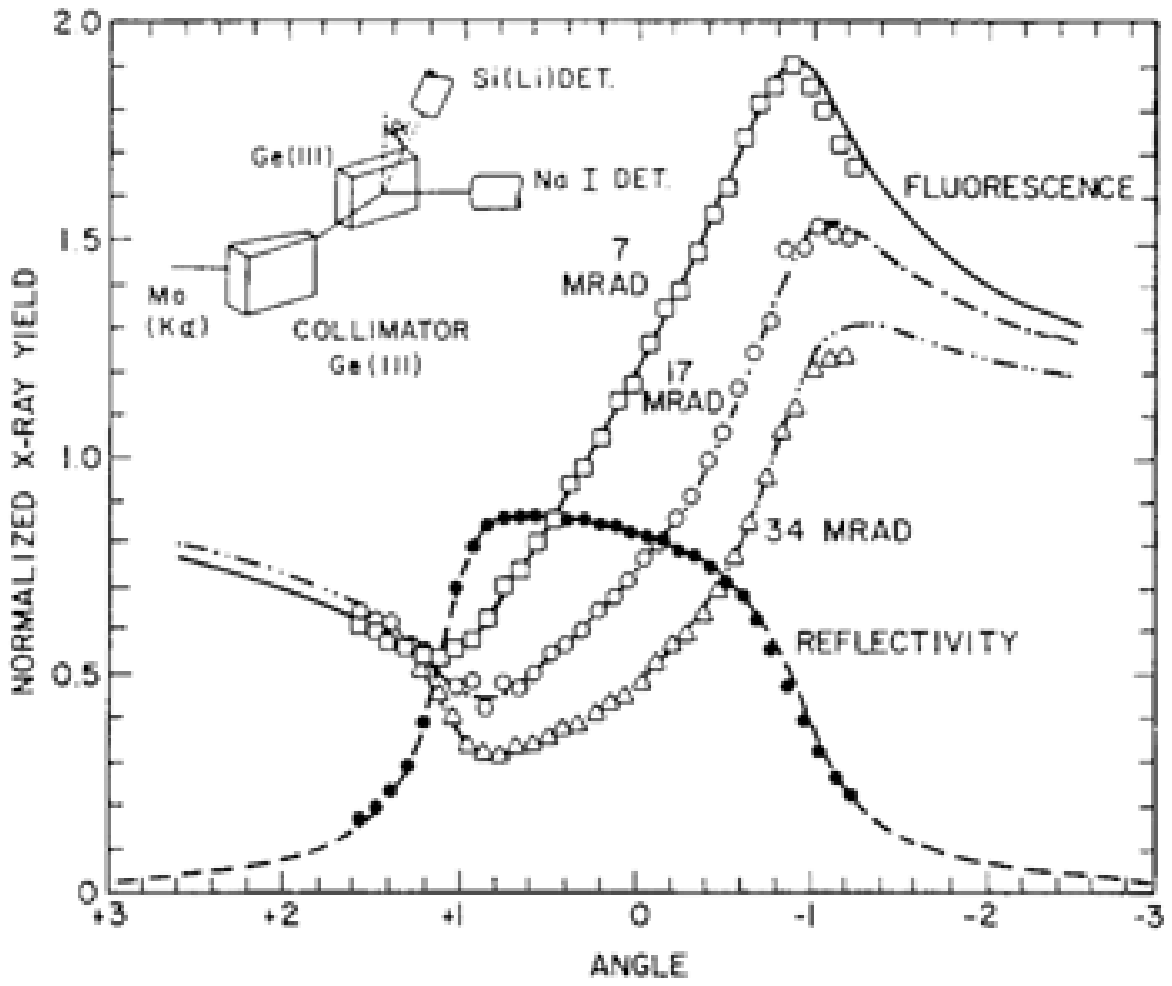}} \  
\resizebox{!}{9cm}{
\includegraphics{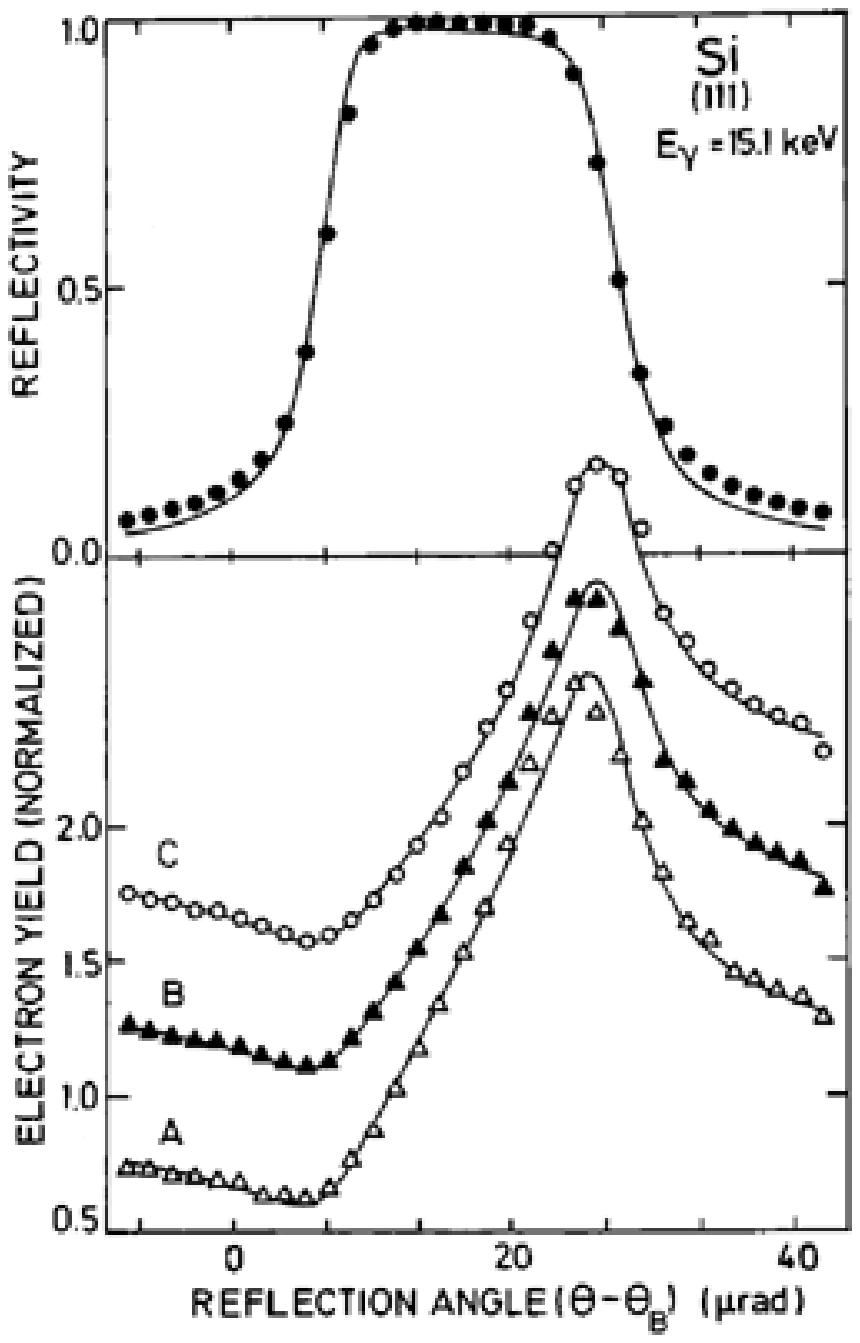}}
\end{center}
\caption{
\textsl{\label{} Left: Angular dependence of the yield of $GeK_\alpha $
fluorescence and of x-ray reflection in (111) diffraction of $MoK_\alpha $
radiation. The fluorescence data are measured with a $Si (Li)$
detector situated at a glancing angle to the surface, the value of which is
indicated in milliradians. From Ref. \cite{PG83};
Right: The angular variation of the perfect $Si$ (111) reflectivity
and photoelectron yields while diffraction of x-ray radiation with energy $%
E=15.1$ $keV$. Photoelectron curves correspond to electrons with different
energy losses $\Delta E$ ($\Delta E=0$ (A)$,2.5$ (B) and $5$ $keV$ (C)
respectively). From Ref. \cite{BMK84a}.}
}
\end{figure}

\clearpage

%
%
%
\begin{figure}[tbp]
\begin{center}
\includegraphics{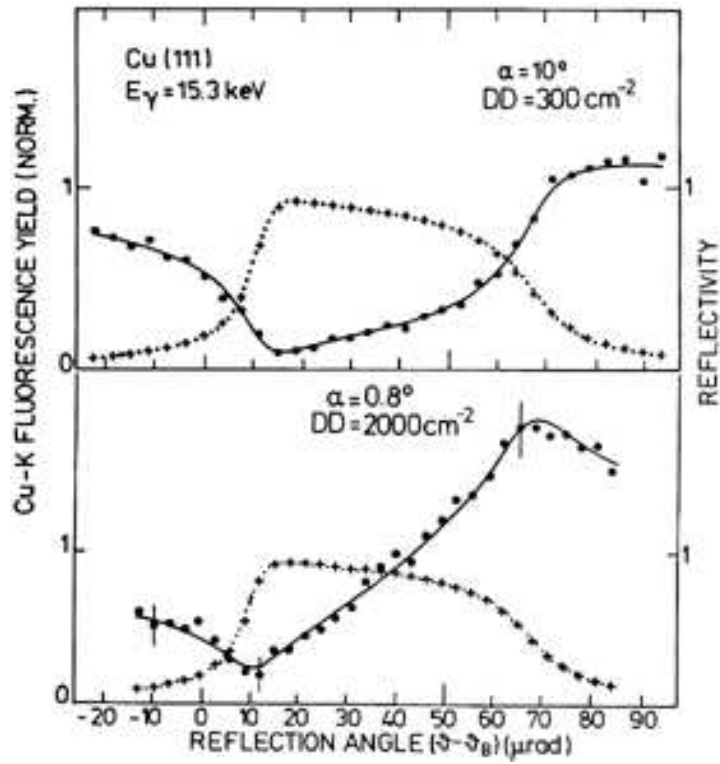}
\end{center}
\caption{
\textsl{\label{}XSW measurement of $CuK$ fluorescence radiation from a $Cu$
crystal. Measured (+++) and calculated ($\cdot \cdot \cdot $) $Cu(111)$
reflection curve; ($\bullet $) experimentally determined fluorescence yield;
(---) fit to the experimental points. Two different dislocation densities DD
and takeoff angles $\alpha $ were used. Fluorescence yield curve for the big
exit angle ($\alpha =10^{\circ }$) coincide with the reverse reflectivity
curve. From Ref. \cite{ZMU90}.}
}
\end{figure}

\clearpage

%
%
%
\begin{figure}[tbp]
\begin{center}
\includegraphics{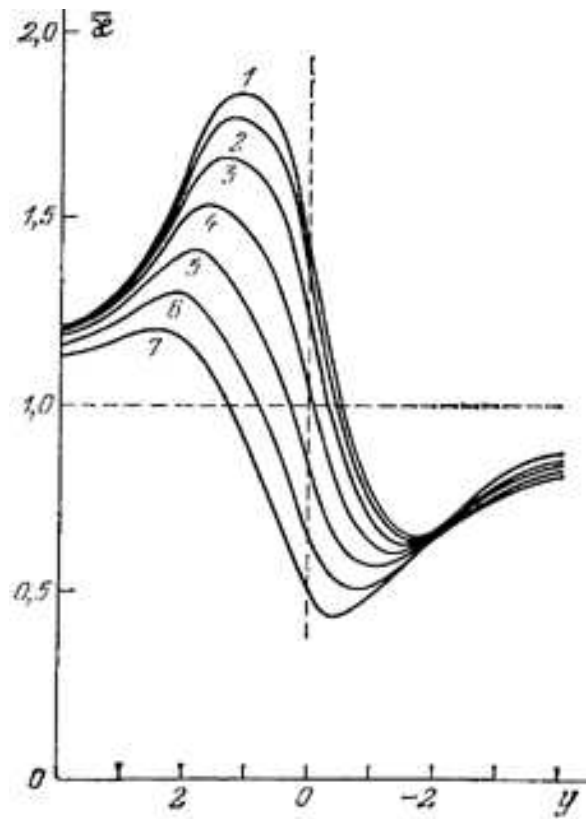}
\end{center}
\caption{
\textsl{\label{} Angular dependences of the secondary radiation yield
calculated for $Si$ (400) reflection and $CuK_\alpha $ radiation
for different values of the ratio $L_{ex}/L_{yi}$ equal to: 12 (1), 7.2 (2),
3.6 (3), 1.8 (4), 0.9 (5), 0.45 (6), and 0.225 (7), in a system with a
symmetric monochromator. From Ref. \cite{KK86}.}
}
\end{figure}

\clearpage

%
%
\begin{figure}[tbp]
\begin{center}
\includegraphics{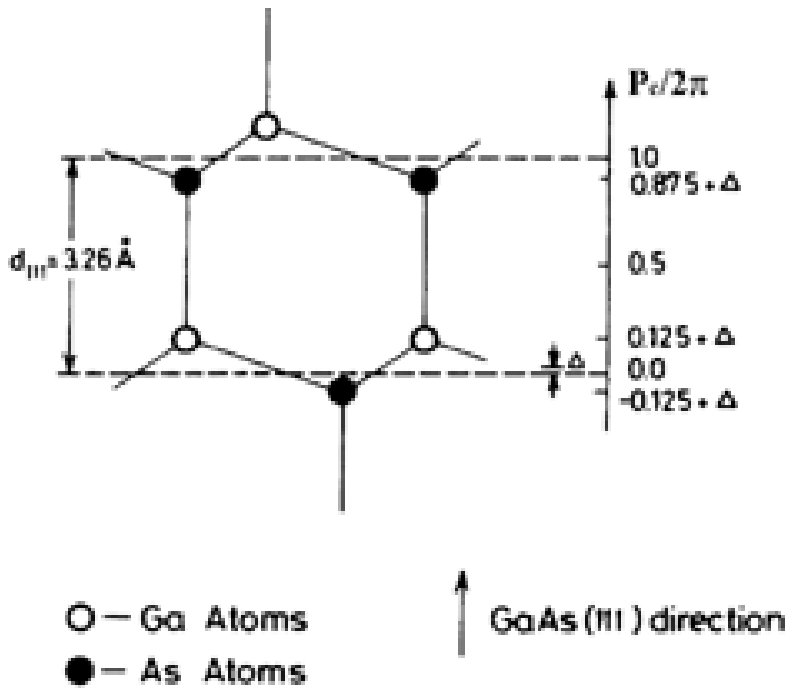} \\
\bigskip
\includegraphics{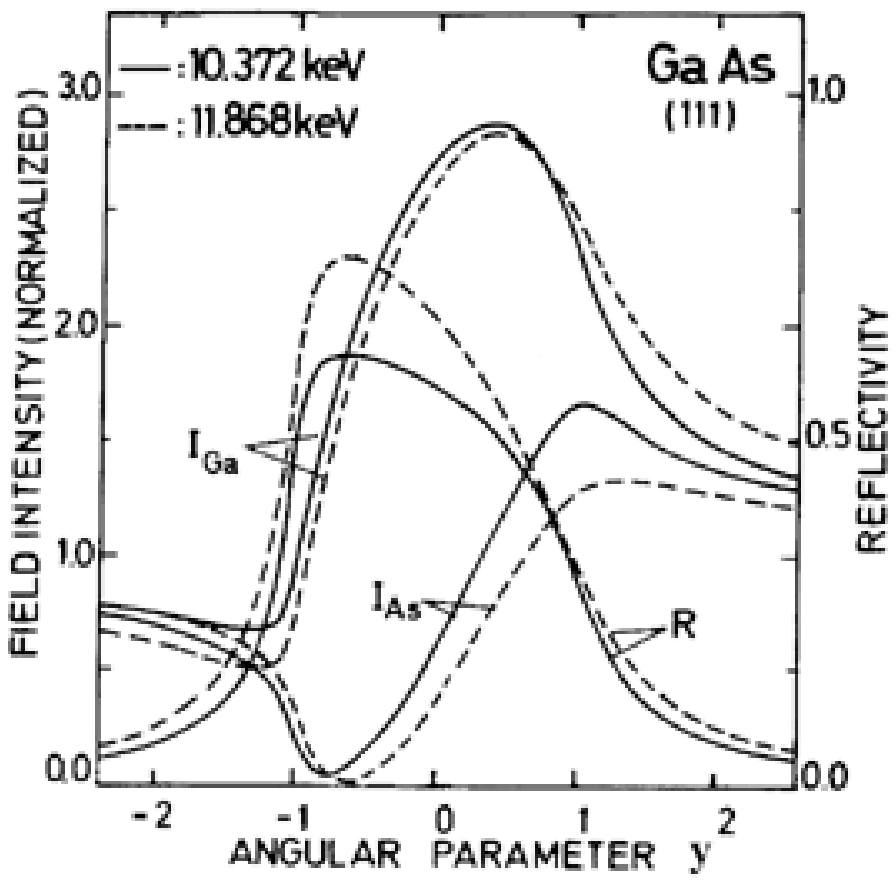}
\end{center}
\caption{
\textsl{\label{} Top: Schematic view showing the position of the
noncentrosymmetric $GaAs(111)$ diffraction planes (dashed lines) relative to
the $Ga$ atoms (open circles) and $As$ atoms (closed circles). Parameter $P_c
$ locates positions in this structure in the (111) direction relative to a
(111) diffraction plane, which is shifted by an amount $\Delta $ relative to
a centrosymmetric (111) diffraction plane;
Bottom: Angular dependence of the wave field intensity in a $GaAs(111)$ crystal
at the $Ga$ ($I_{Ga}$) and $As$ ($I_{As}$) atomic sites for x-ray
diffraction with an energy of $10.372$ $keV$ (solid line) and $11.868$ $keV$
(dotted line) for $\sigma -$polarization. The energies of radiation are
chosen to be $5$ $eV$ above the $GaK$ and $AsK$ edges. From Ref. \cite
{BMK84b}.}
}
\end{figure}

\clearpage

%
%


%
%
\begin{figure}[tbp]
\begin{center}
\includegraphics{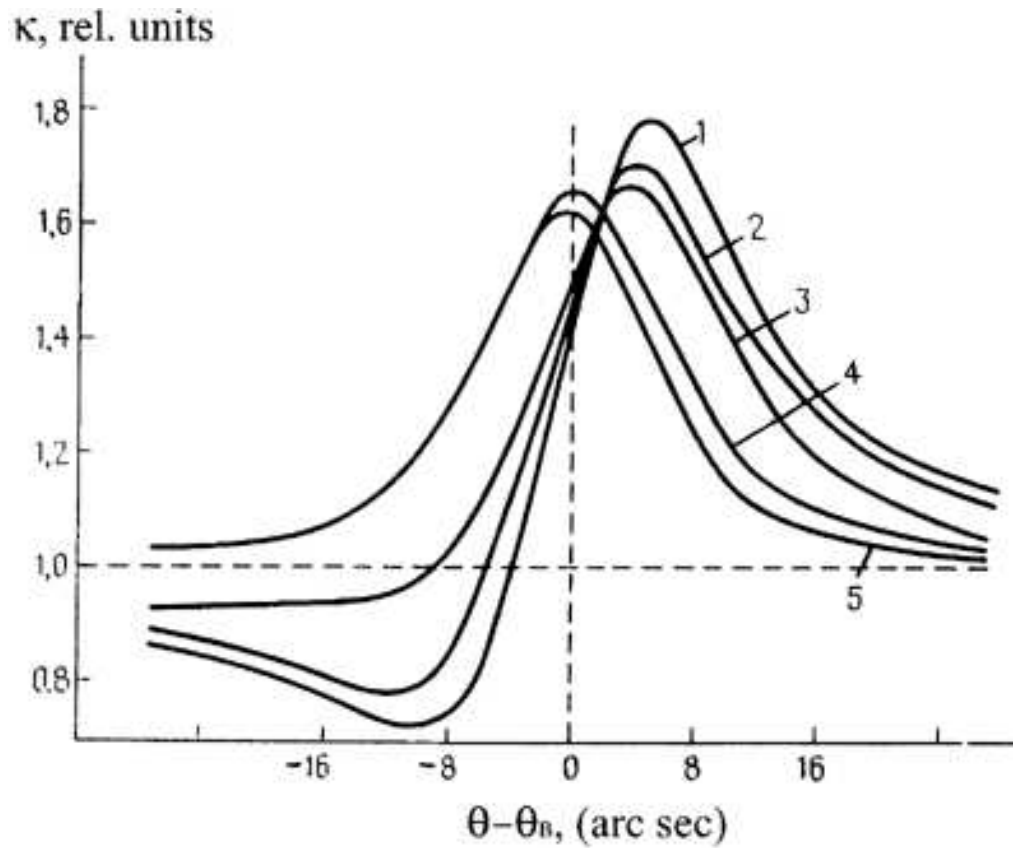}
\end{center}
\caption{
\textsl{\label{} Angular dependences of the photoelectron yield for $Ge(220)
$ diffraction, $CuK_\alpha $ radiation with different thickness of amorphous
layers on the surface: 0 (1), 50 (2), 100 (3) 250 (4) 450 (5) $nm$. From
Ref. \cite{KS77}.}
}
\end{figure}

\clearpage

%
%
\begin{figure}[tbp]
\begin{center}
\includegraphics{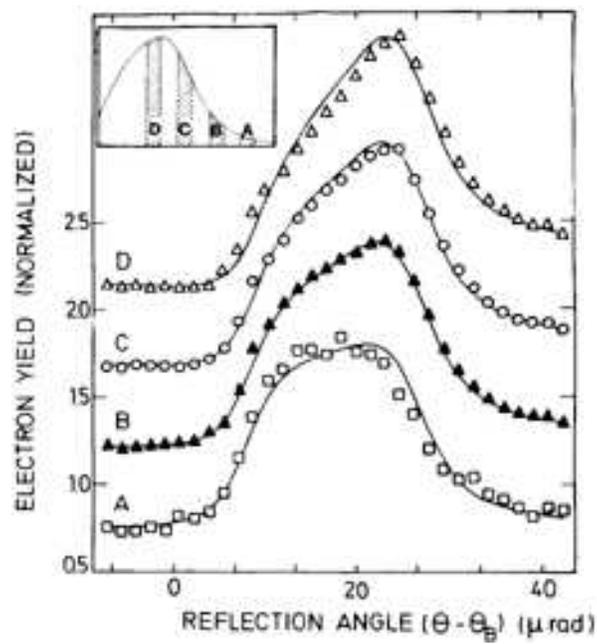}
\end{center}
\caption{
\textsl{\label{} The angular variation of the 0.6-$\mu m$ $Si O_2/%
 Si (111)$ photoelectron yields for energy regions A-D of the
emission spectrum (shown in the inset). The solid lines are theoretical
curves. From Ref. \cite{BMK84a}.}
}
\end{figure}

\clearpage

%
%
\begin{figure}[tbp]
\begin{center}
\resizebox{!}{10cm}{
\includegraphics{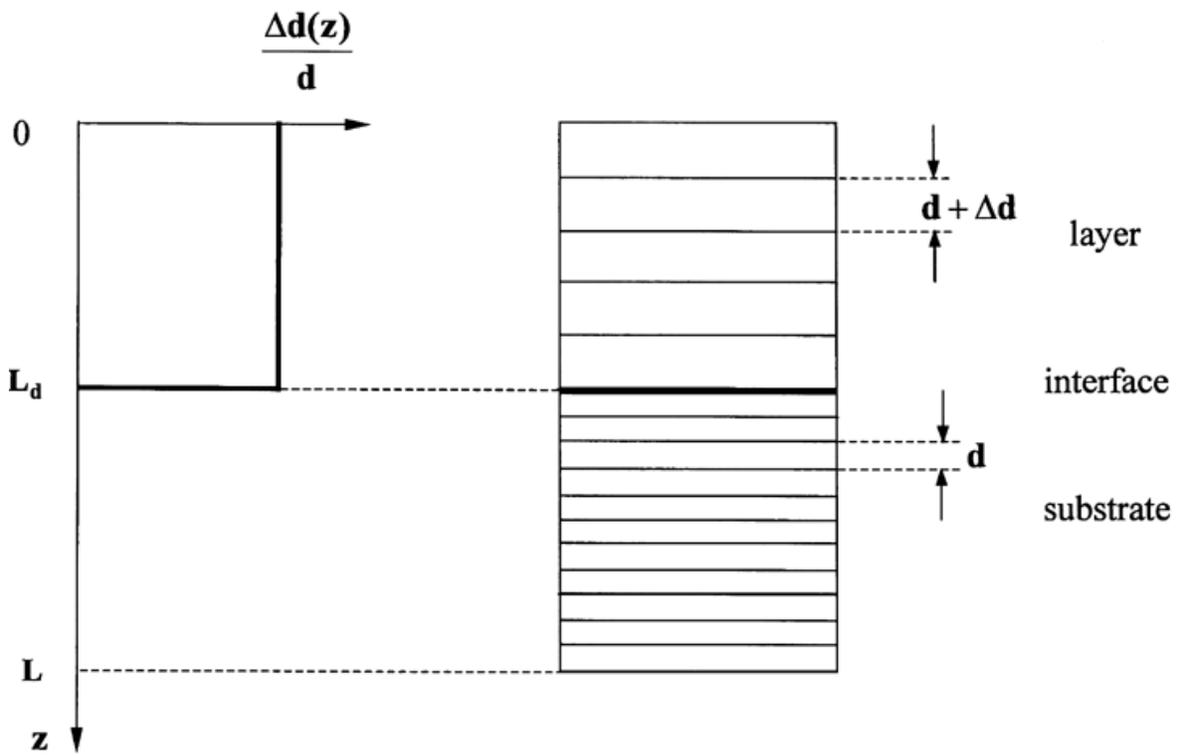}}
\end{center}
\caption{
\textsl{\label{} The profile of deformation $(\Delta d(z)/d)$ in a crystal
for the bicrystal model. Interplaner distance in the layer $d+\Delta d$
differs from that in the layer on a constant value $\Delta d$. In this model
interface is a sharp (step) function located on the depth $L_d$.}
}
\end{figure}

\clearpage

%
%
\begin{figure}[tbp]
\begin{center}
\resizebox{!}{10cm}{
\includegraphics{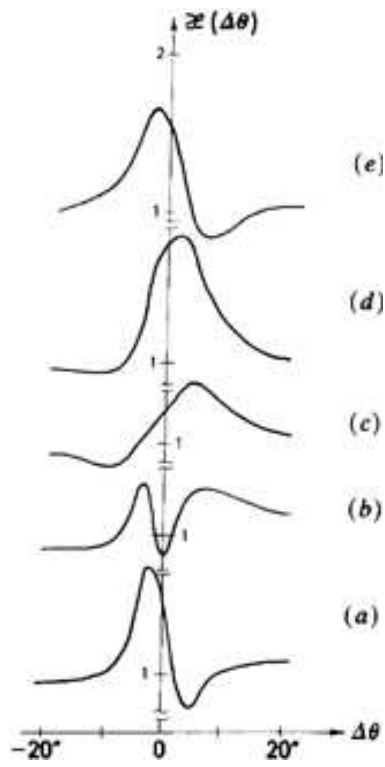}}
\end{center}
\caption{
\textsl{\label{} Photoemission angular dependence for $Si$ single
crystals with autoepitaxial $Si$ film, doped with $B$ and $Ge$
with varying $Ge$ concentrations: $0$ (curve a), $3.7\times 10^{19}$ (b), $%
7\times 10^{19}$ (c), $10^{20}$ (d), $1.5\times 10^{20}$ (e). From Ref. \cite
{KVK87}.}
}
\end{figure}

\clearpage

%
%
\begin{figure}[tbp]
\begin{center}
\includegraphics{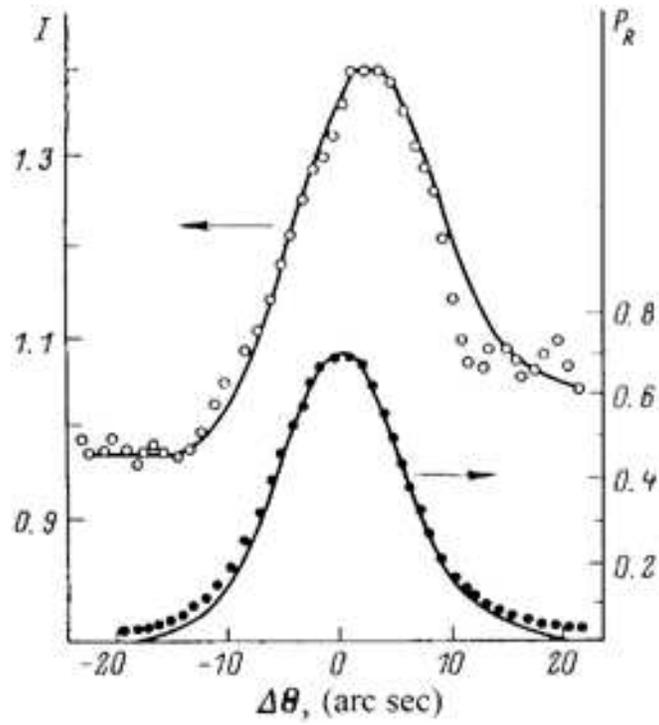}
\end{center}
\caption{
\textsl{\label{} Angular dependence of $InL_\alpha $ fluorescence yield
from the film of $In_{0.5}Ga_{0.5}P$ (thickness of the film $\sim 100$ \AA )
grown on the surface of $GaAs$ (111) single crystal (upper curve) and
reflectivity (bottom curve). Points are experimental data and lines result
of the theoretical fitting. From Ref. \cite{KKK88}.}
}
\end{figure}

\clearpage

%
%
\begin{figure}[tbp]
\begin{center}
\includegraphics{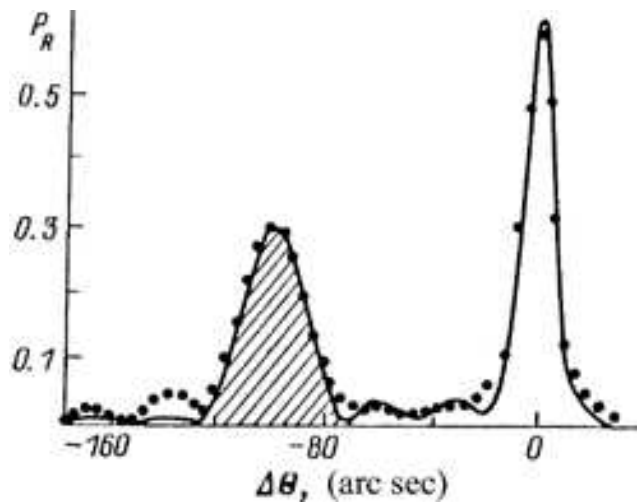} \\
\bigskip
\includegraphics{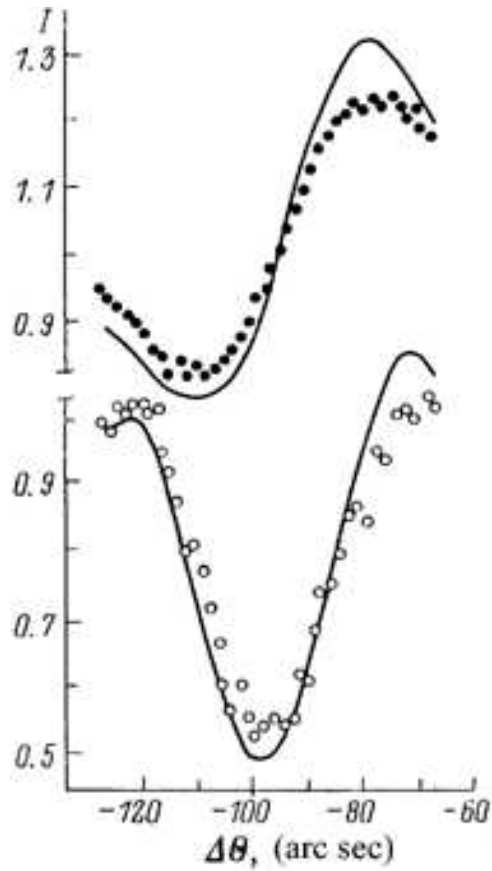}
\end{center}
\caption{
\textsl{\label{} Reflectivity (Top) and angular dependence of $InL_\alpha $
(upper curve) and $PK_\alpha $ (bottom curve) fluorescence yield (Bottom) from the
film of $In_{0.5}Ga_{0.5}P$ (thickness of the film $\sim 0.6$ $\mu m$) grown
on the surface of $GaAs$ (111) single crystal. The shadowed region on the
reflectivity curve is the region, where angular dependence of the
fluorescence yield was measured. Points are experimental data and lines
result of the theoretical fitting. From Ref. \cite{KKK88}.}
}
\end{figure}

\clearpage

%
%
\begin{figure}[tbp]
\begin{center}
\resizebox{!}{7cm}{
\includegraphics{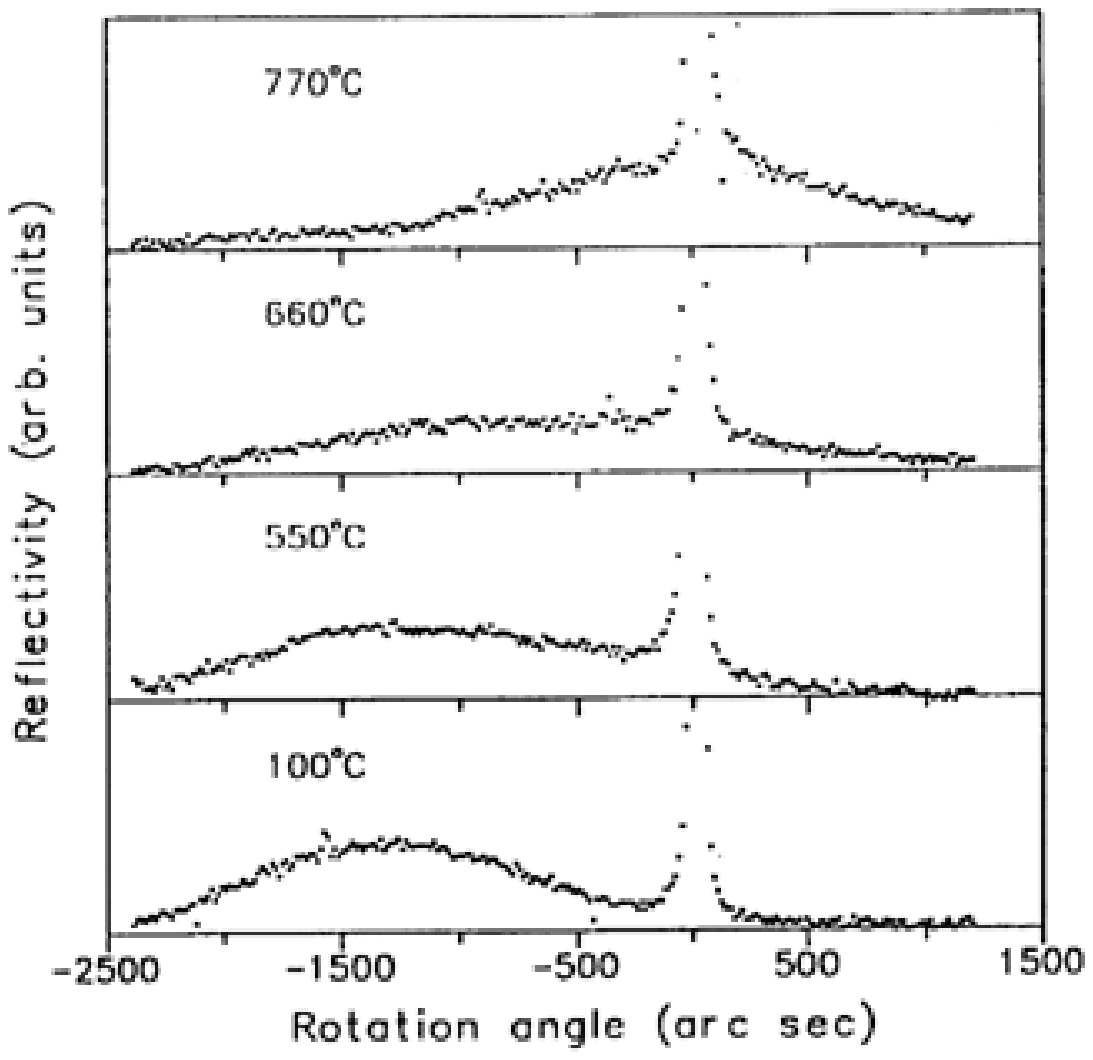}} \   
\resizebox{!}{7cm}{
\includegraphics{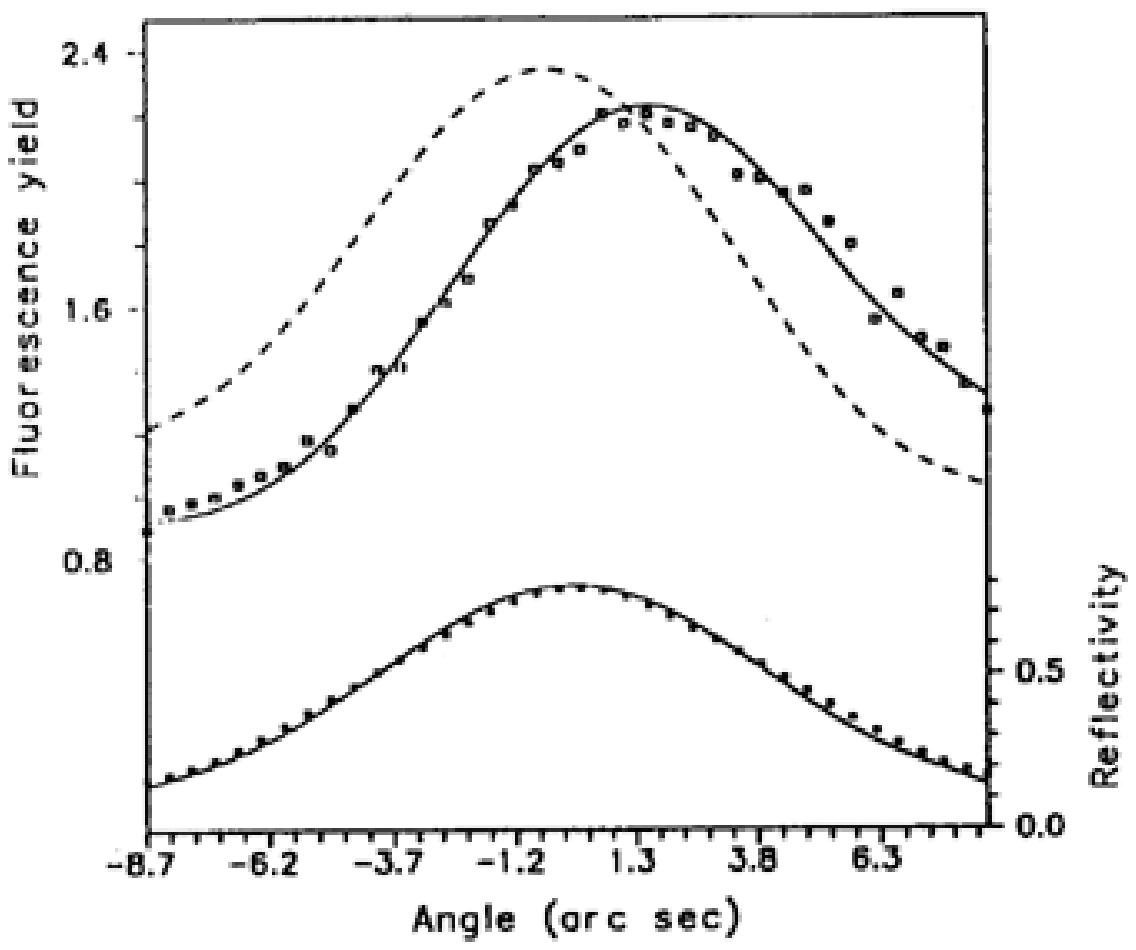}}
\end{center}
\caption{
\textsl{\label{}(Left:) The (222) x-ray reflectivity curves of $CaF_2$
epitaxial $10$ nm layers grown at different temperatures;
(Right:) Angular dependence of the total photoelectron yield $\kappa $ (upper
curve) and reflectivity $P_R$ (bottom curve) for the (888) diffraction of $%
CuK_\alpha $ radiation in a perfect GGG garnet single crystal. Points are
experimental data and lines are theoretical fitting. From Ref. \cite{ZZK88}.}
}
\end{figure}

\clearpage

%
%
\begin{figure}[tbp]
\begin{center}
\includegraphics{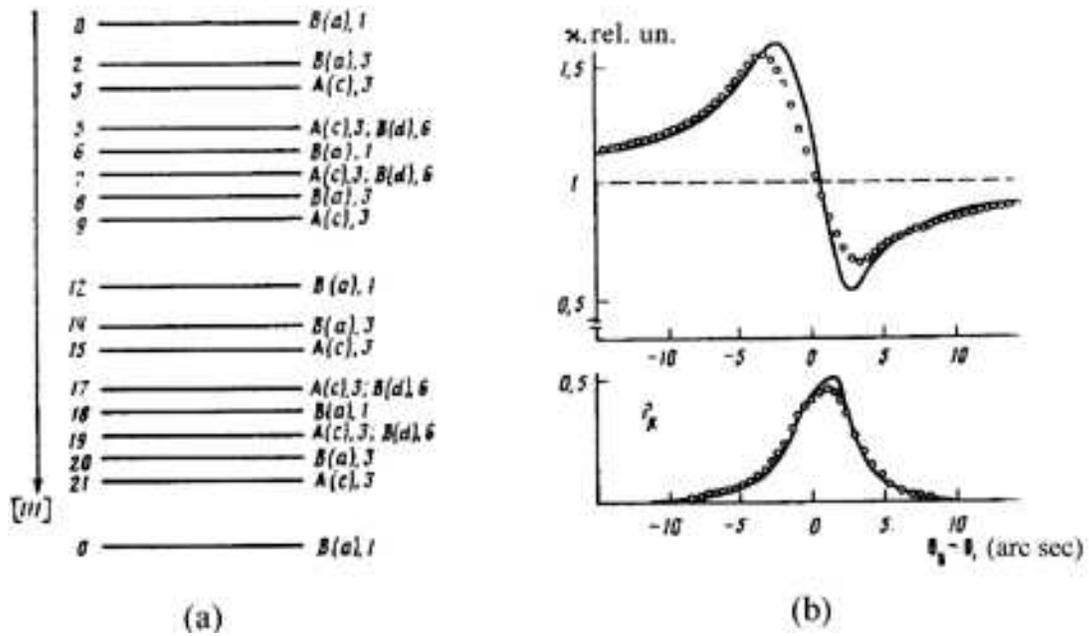}
\end{center}
\caption{
\textsl{\label{}(a) Schematic arrangement of the planes consisting from
the different sort of atoms in positions \textit{a}, \textit{c} and \textit{d%
} in a single garnet crystal in the direction [111] in the unit cell;
(b) Angular dependence of the total photoelectron yield $\kappa $ (upper
curve) and reflectivity $P_R$ (bottom curve) for the (888) diffraction of $%
CuK_\alpha $ radiation in a perfect GGG garnet single crystal. Points are
experimental data and lines are theoretical fitting. From Ref. \cite{ZZK88}.}
}
\end{figure}

\clearpage

%
%
\begin{figure}[tbp]
\begin{center}
\resizebox{!}{6cm}{
\includegraphics{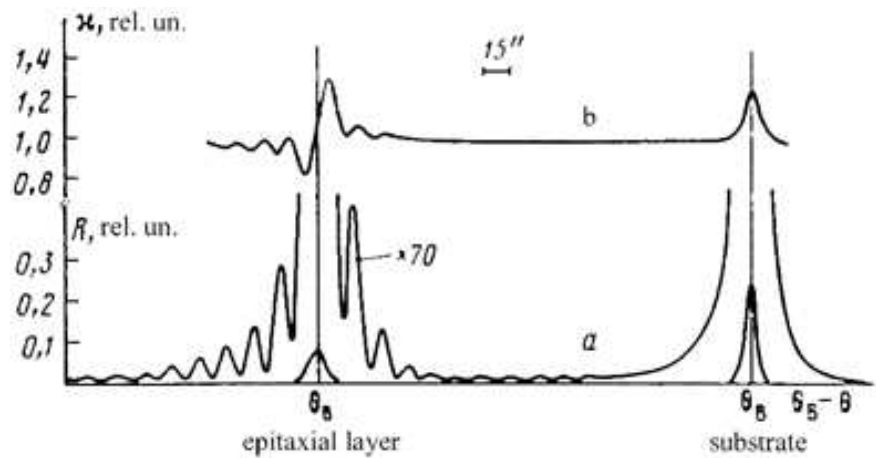}}
\end{center}
\caption{
\textsl{\label{} Experimental results of the photoelectron yield $\kappa $
(upper curve) and reflectivity $P_R$ (bottom curve) for (888) diffraction of 
$CuK_\alpha $ radiation in a perfect GGG garnet crystal with an epitaxial
film of FYG garnet crystal with the thickness 2 $\mu m$. From Ref. \cite
{ZZK88}.}
}
\end{figure}

\clearpage

%
%
\begin{figure}[tbp]
\begin{center}
\resizebox{!}{10cm}{
\includegraphics{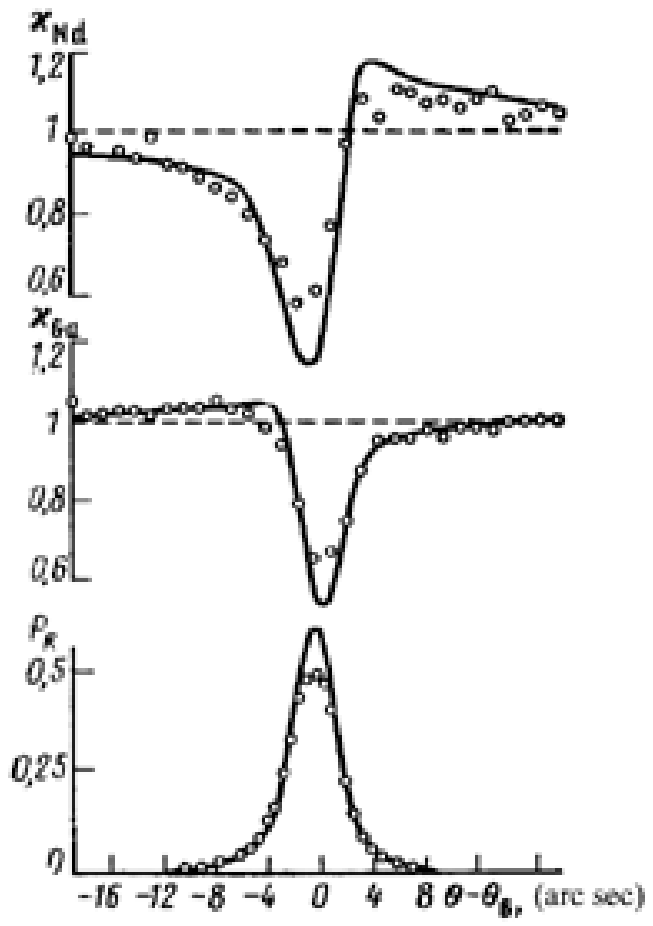}} \  
\resizebox{!}{10cm}{
\includegraphics{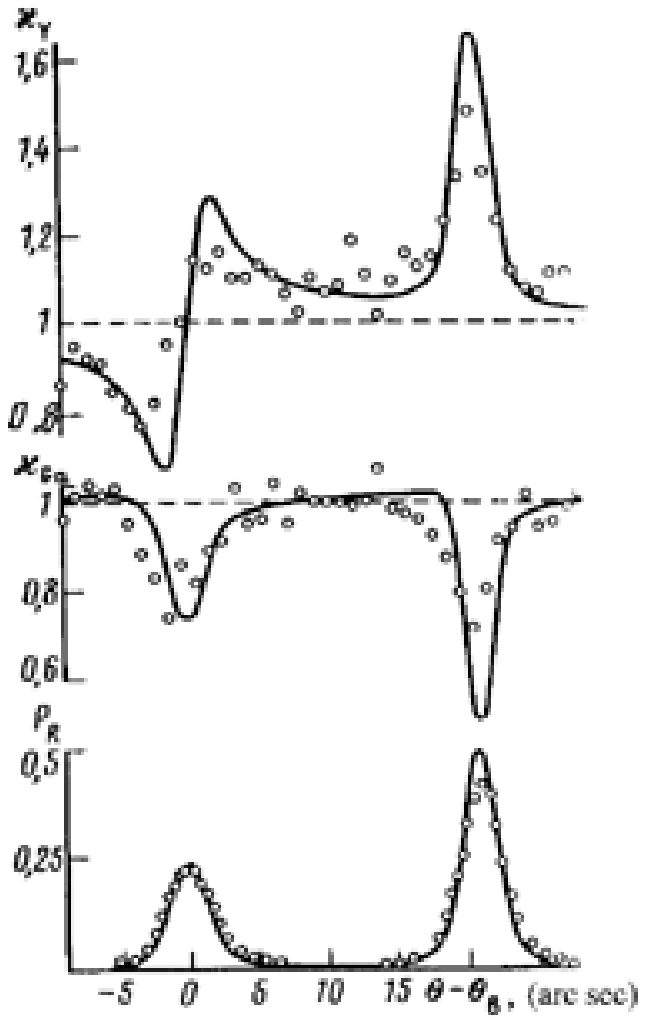}}
\end{center}
\caption{
\textsl{\label{} (Left:) Angular dependence of the fluorescence yield of $%
NdL_\alpha $ (upper curve) and $GaK_\alpha $ (middle curve) for the (444)
diffraction of $MoK_\alpha $ radiation (reflectivity is shown on the bottom
curve) for the perfect GNG garnet crystal;
(Right:) Angular dependence of the $YK_\alpha $ fluorescence yield from the film
(upper curve) and $GaK_\alpha $ fluorescence yield from the substrate
(middle curve) for the (444) diffraction of $AgK_\alpha $ radiation
(reflectivity is shown on the bottom curve) for the perfect GGG garnet
crystal with an epitaxial film of FYG crystal with the thickness 1.6 $\mu m$%
. Points are experimental data and lines are theoretical fitting. From Ref. 
\cite{ZZK88}.}
}
\end{figure}

\clearpage

%
%
\begin{figure}[tbp]
\begin{center}
\includegraphics{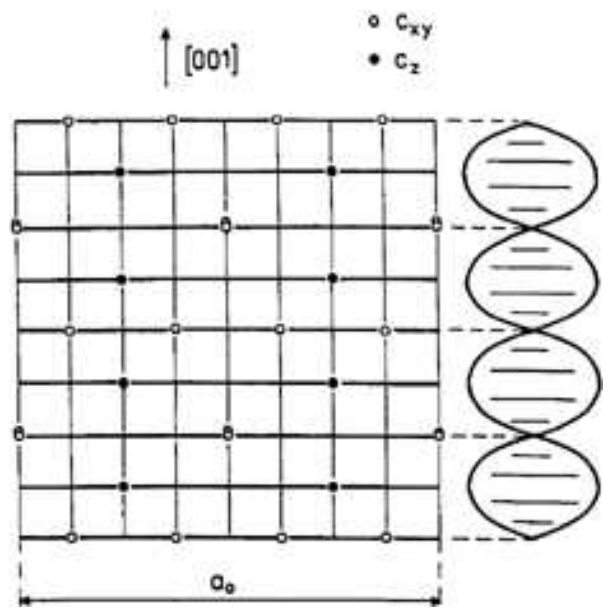}
\end{center}
\caption{
\textsl{\label{} Projection of the dodecahedral positions $c_z$ (full
circles) and $c_{xy}$ (open circles) on the plane (001). The (004) x-ray
standing wave is also shown schematically. From Ref. \cite{KKS92}.}
}
\end{figure}

\clearpage

%
%
\begin{figure}[tbp]
\begin{center}
\includegraphics{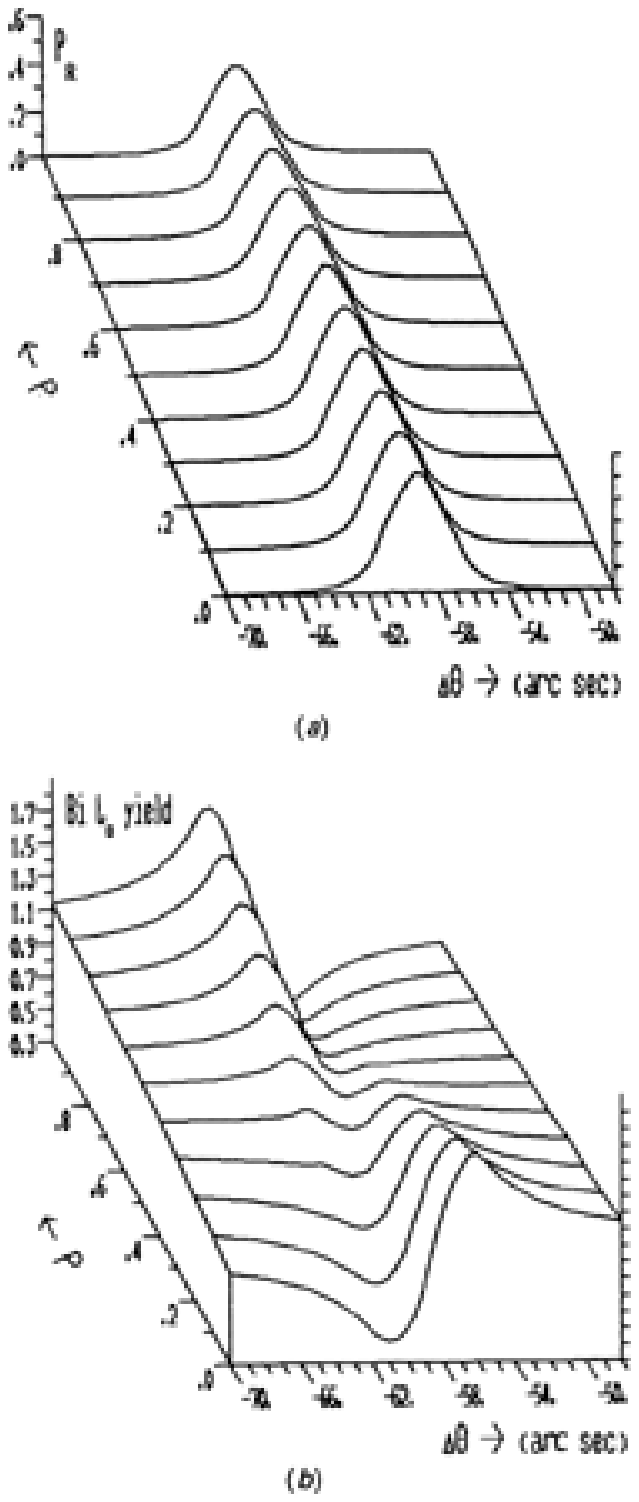}
\end{center}
\caption{
\textsl{\label{} Calculated angular dependencies of the reflectivity $%
P_R(\theta )$ (a) and fluorescence yield of the $Bi$ (b) for the (004)
diffraction of $MoK_\alpha $ radiation on $Y_{3-x}Bi_xFe_5O_{12}$ thin film
for the values of $p$ from 0 to 1 in the angular range of the x-ray
diffraction on the film. From Ref. \cite{KKS92}.}
}
\end{figure}

\clearpage

%
%
\begin{figure}[tbp]
\begin{center}
\resizebox{!}{7cm}{
\includegraphics{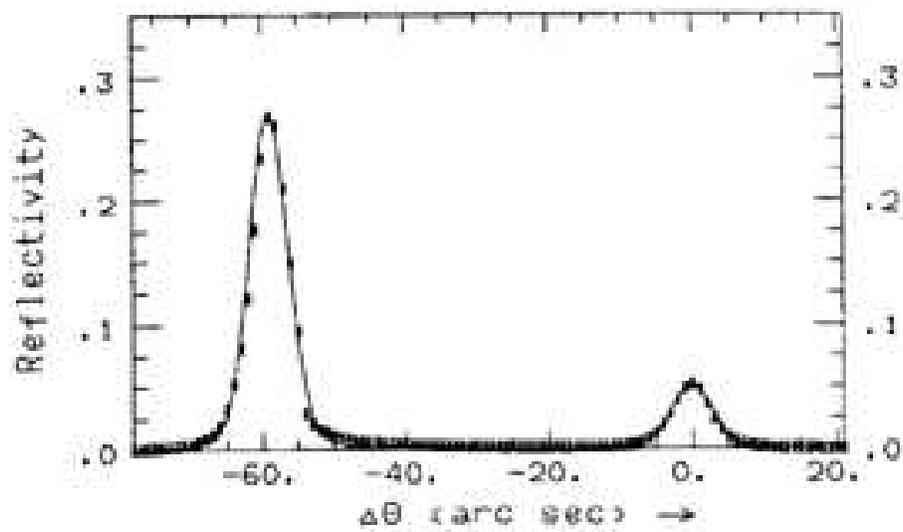}}
\end{center}
\caption{
\textsl{\label{} Experimental (points) and calculated (line) diffraction
curves in the angular range of the x-ray diffraction on the film and on the
substrate. From Ref. \cite{KKS92}.}
}
\end{figure}

\clearpage

%
%
\begin{figure}[tbp]
\begin{center}
\includegraphics{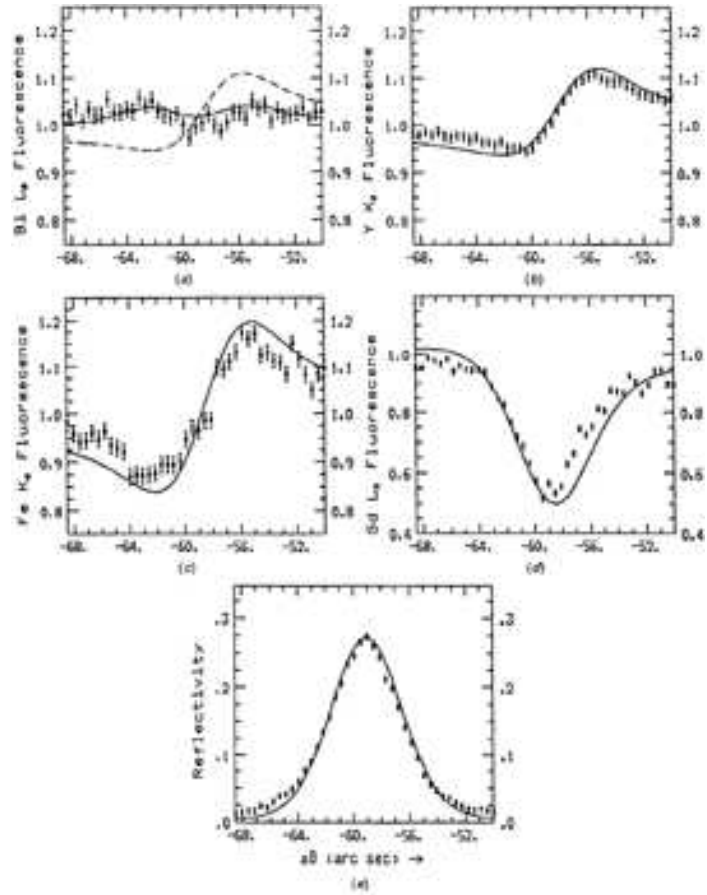}
\end{center}
\caption{
\textsl{\label{} Experimental (points) and calculated (lines) for the value
of $p=0.44$ angular dependencies of the fluorescence yield of $Bi^{3+}$ (a), 
$Y^{3+}$ (b), $Fe^{3+}$ (c) ions from the film and $Gd^{3+}$ ions from the
substrate (d) and the x-ray reflectivity (e). For comparison the
fluorescence yield curve for $p=0.33$ (uniform distribution of $Bi^{3+}$
ions over $c_z$ and $c_{xy}$ sites) is also shown in (a). From Ref. \cite
{KKS92}.}
}
\end{figure}

\clearpage

%
%
\begin{figure}[tbp]
\begin{center}
\resizebox{!}{4cm}{
\includegraphics{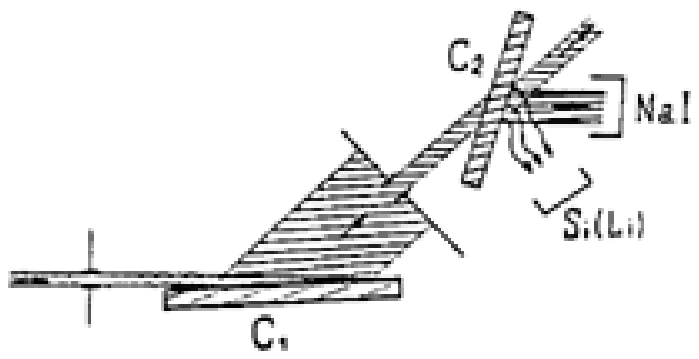}} \\
\bigskip
\resizebox{!}{15cm}{
\includegraphics{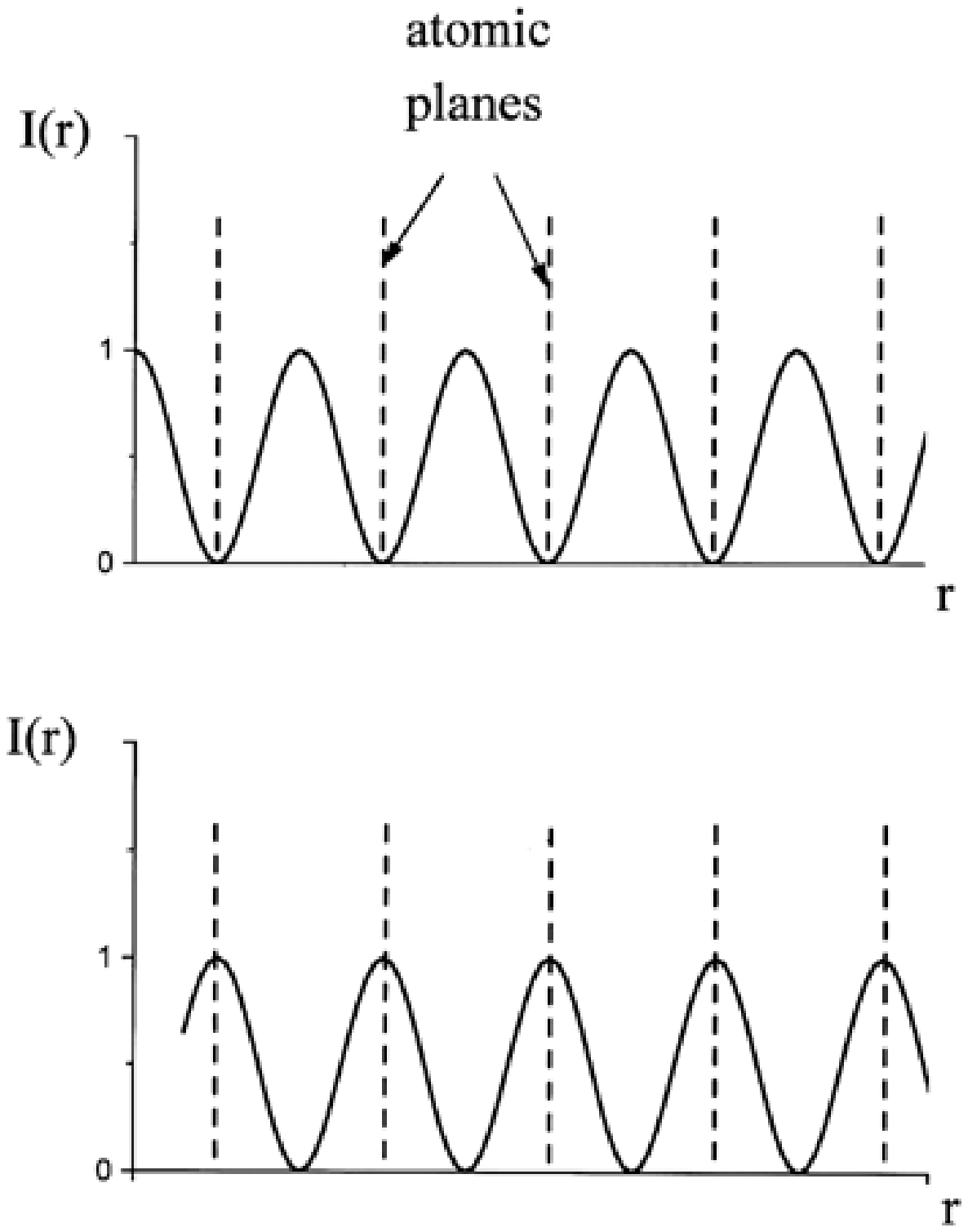}}
\end{center}
\caption{
\textsl{\label{} (Top:) Experimental set up for SR yield measurement in Laue
geometry: $C_1$ crystal monochromator, $C_2$ crystal under investigation, $%
NaI$ detector measuring reflectivity, $Si (Li)$ detector measuring
fluorescence yield.
(Bottom:) Distribution of x-ray standing wave field in a crystal for Laue
geometry. Nodes of the weakly absorbing field (upper curve) coincide with
the atomic planes and maximum of field intensity for highly absorbing
(bottom curve) coincide with atomic planes giving maximum absorption of this
field on the thickness of the crystal $L$.}
}
\end{figure}

\clearpage

%
%
\begin{figure}[tbp]
\begin{center}
\includegraphics{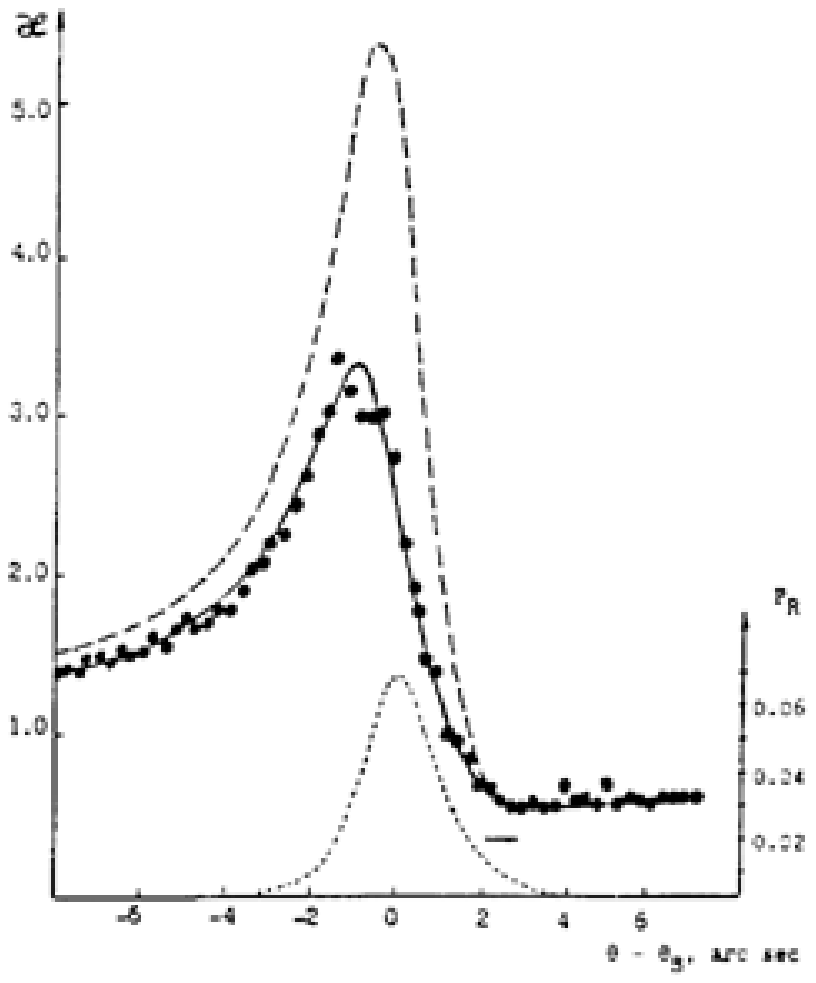} \  
\includegraphics{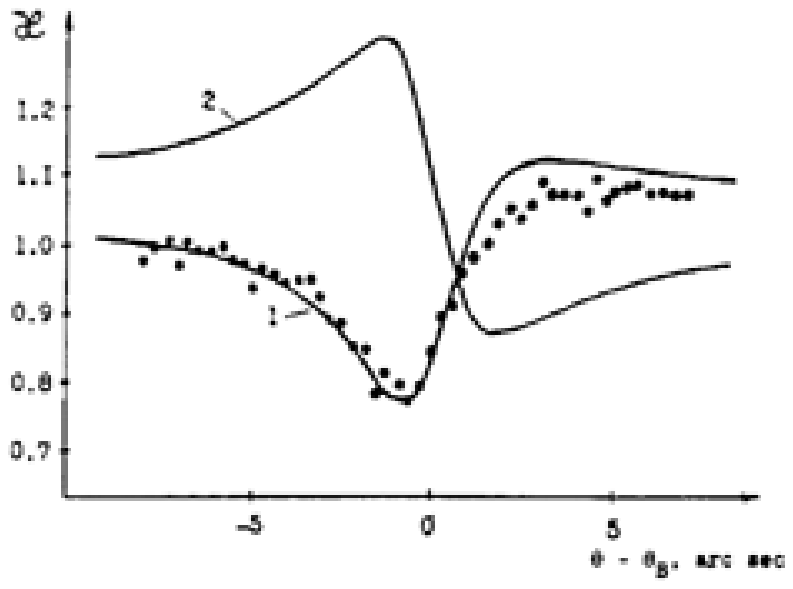}
\end{center}
\caption{
\textsl{\label{}Angular dependence of the $GeK_\alpha $ fluorescence yield
from the $Si$ crystal uniformly doped with $Ge$ with different
thickness of the crystal. Points are experimental data, lines are
calculations.
(Left:) Case of the thick crystal ($L=2.2$ $mm$). Solid line is the calculation
for substitutional impurity atoms (coherent position $P_c^{111}=0$); dashed
curve is the calculation for impurities with coherent position $%
P_c^{111}=0.15$. The x-ray reflectivity curve $P_R$ is also shown.
(Right:) Case of the thin crystal ($L=0.49$ $mm$). Curve 1 is the calculation for
the substitutional impurity atoms and curve 2 is the calculations for the
randomly distributed impurities. From Ref. \cite{KKK90a}.}
}
\end{figure}

\clearpage


%
%
\begin{figure}[tbp]
\begin{center}
\resizebox{!}{10cm}{
\includegraphics{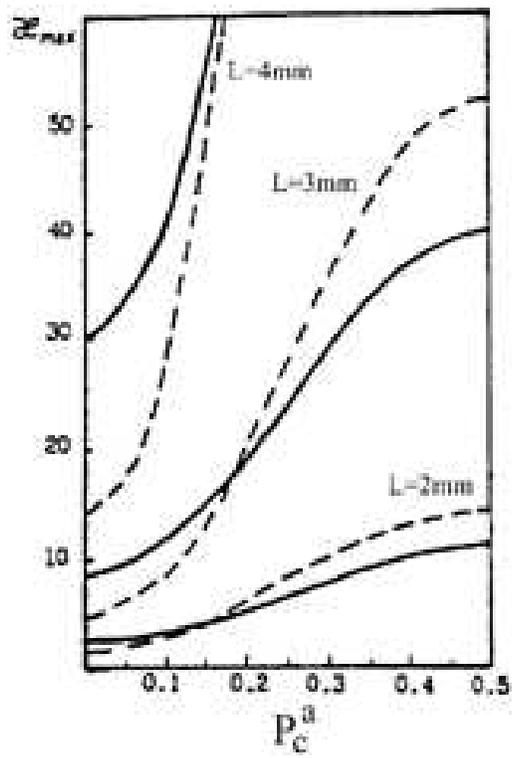}}
\end{center}
\caption{
\textsl{\label{}Theoretical dependencies of the maximum fluorescence yield
for substitutional impurity atoms in silicon crystals with different
thickness. Solid lines are 111 reflection and dashed lines are 220
reflections. From Ref. \cite{KKK90a}.}
}
\end{figure}

\clearpage

%
%
\begin{figure}[tbp]
\begin{center}
\includegraphics{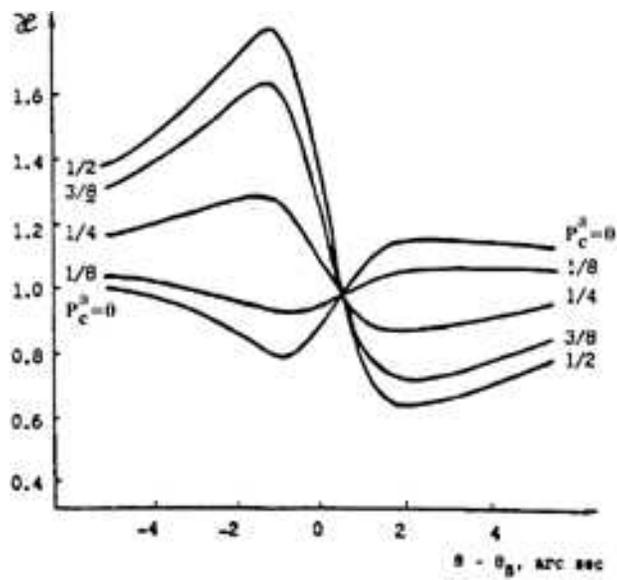}
\end{center}
\caption{
\textsl{\label{} Calculated angular dependencies of the impurity
fluorescence yield from a thin $Si$ crystal for a different
impurity positions. From Ref. \cite{KKK90a}.}
}
\end{figure}

\clearpage

%
%
\begin{figure}[tbp]
\begin{center}
\includegraphics{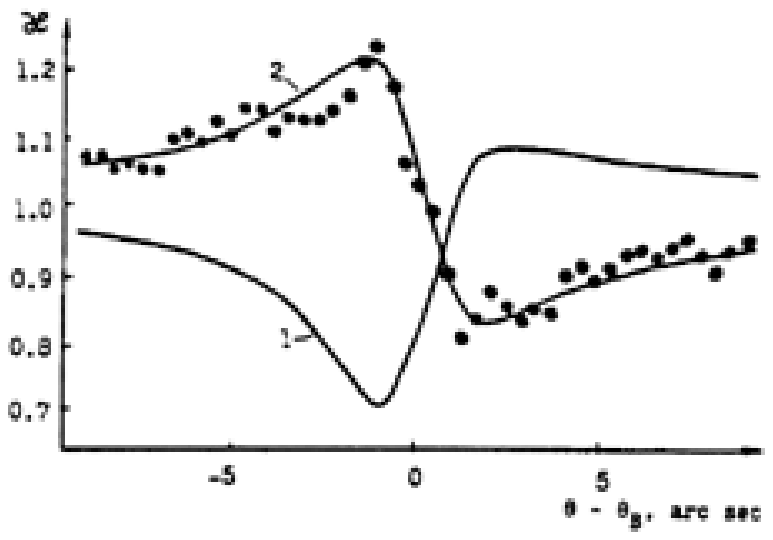} \  
\includegraphics{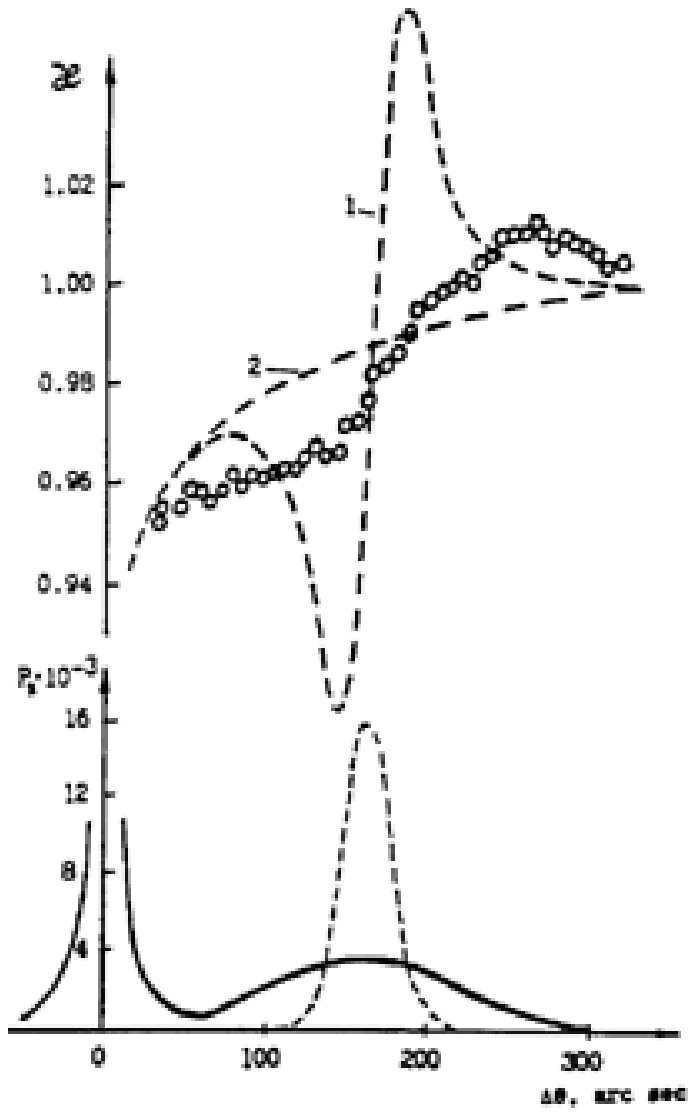}
\end{center}
\caption{
\textsl{\label{}Angular dependence of the $GeK_\alpha $ fluorescence yield
from the $Si$ crystals with an epitaxial layer with thickness $%
0.35 $ $mm$.
(Left) Angular region corresponding to the x-ray diffraction on the substrate.
Curve 1 is the calculation for the substitutional impurity atoms; curve 2 is
the calculation for the randomly distributed impurities; circles are the
experimental data.
(Right) Angular region corresponding to the x-ray diffraction on the layer.
Fluorescence yield is presented on the upper curve and reflectivity on the
bottom. Dashed lines are calculations for substitutional impurity atoms (1)
and for the randomly distributed impurities (2). From Ref. \cite{KKK90a}.}
}
\end{figure}

\clearpage

%
%
\begin{figure}[tbp]
\begin{center}
\resizebox{!}{10cm}{
\includegraphics{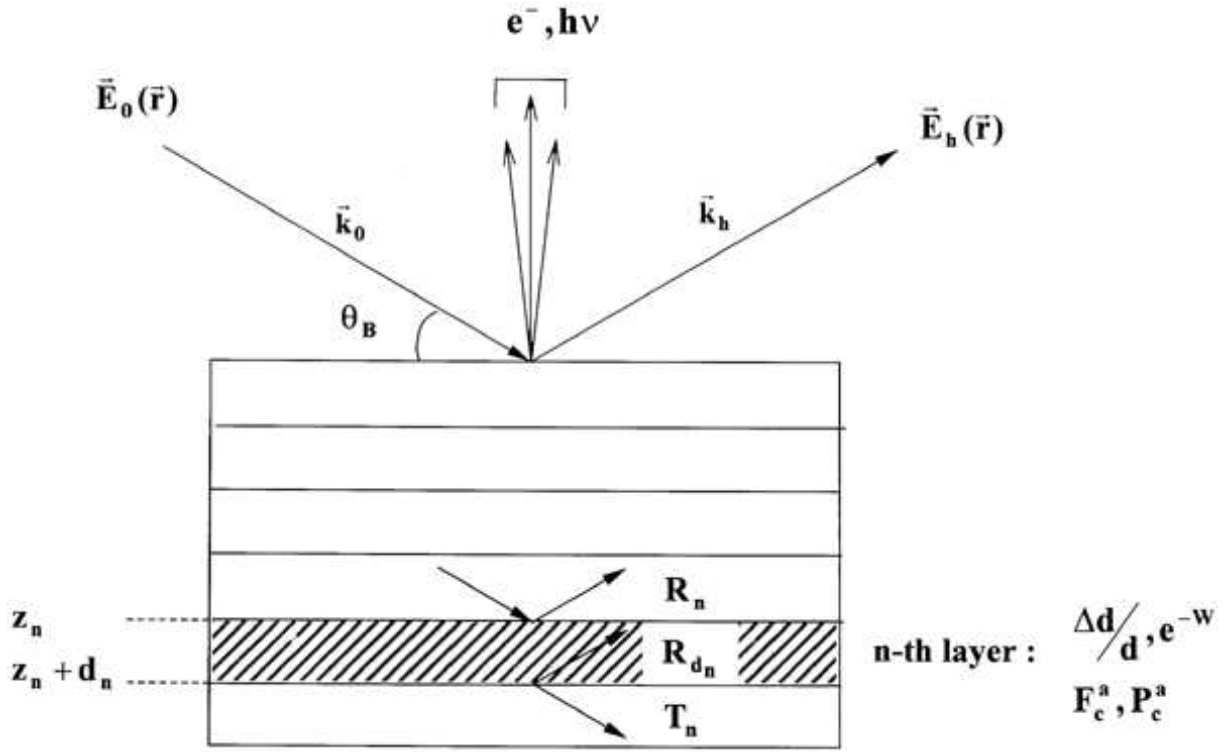}}
\end{center}
\caption{
\textsl{\label{} Schematic view of the SR yield from the multilayer crystal
while the Bragg diffraction of x-rays. The n-th layer of such crystal is
characterized by constant values of deformation $\Delta d/d$ and
amorphization $e^{-W}$. The x-ray scattering from each layer is given by its
reflection $R_n$ and transmission $T_n$ amplitudes. The SR yield from the
n-th layer and the atoms of the sort $a$ is determined by the values of
coherent fraction $F_c^a$ and coherent position $P_c^a$.}
}
\end{figure}

\clearpage

%
%
\begin{figure}[tbp]
\begin{center}
\includegraphics{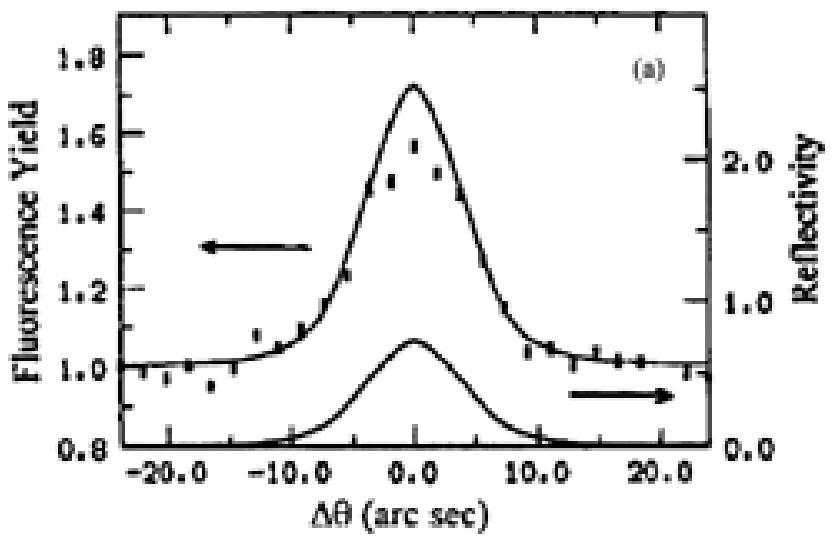} \\
\bigskip
\includegraphics{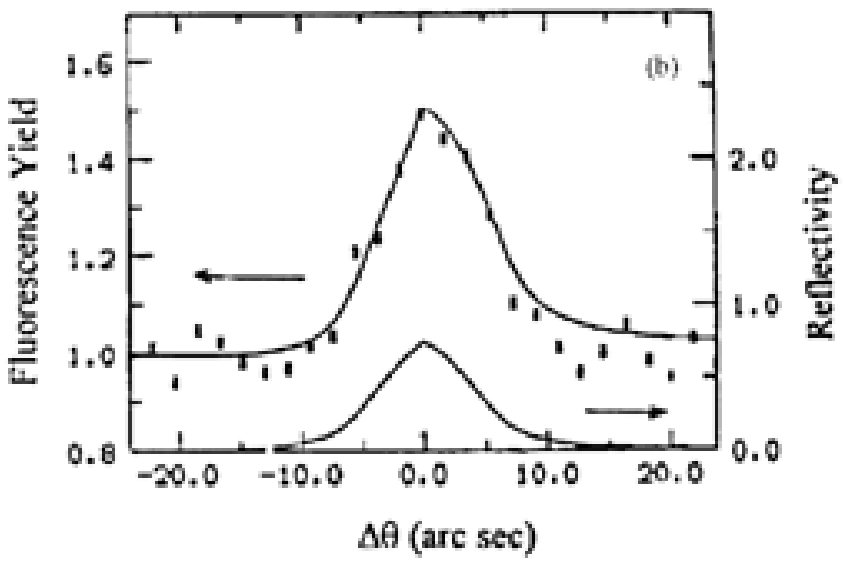} \\
\bigskip
\includegraphics{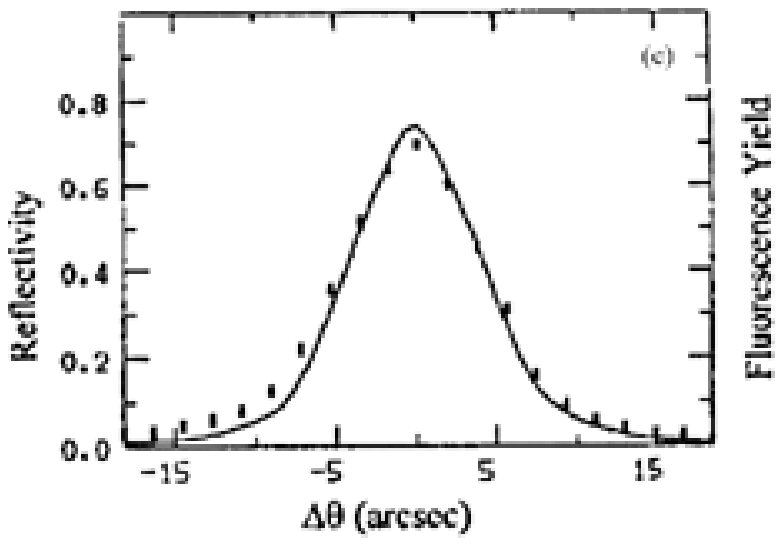}
\end{center}
\caption{
\textsl{\label{} The fluorescence $FeK_\alpha $ yield for silicon crystals
implanted with $Fe$ ions: just after implantation (a) and after annealing at 
$750^{\circ }$ (b). Points are experimental data and lines result of
theoretical fitting (fitting parameters are summarized in Table 3). The
reflectivity curve (c) for the sample ($Si$ (111) reflection, $%
CuK_\alpha $ radiation) is the same before and after annealing. From Ref. 
\cite{VAN96}.}
}
\end{figure}

\clearpage

%
%
\begin{figure}[tbp]
\begin{center}
\includegraphics{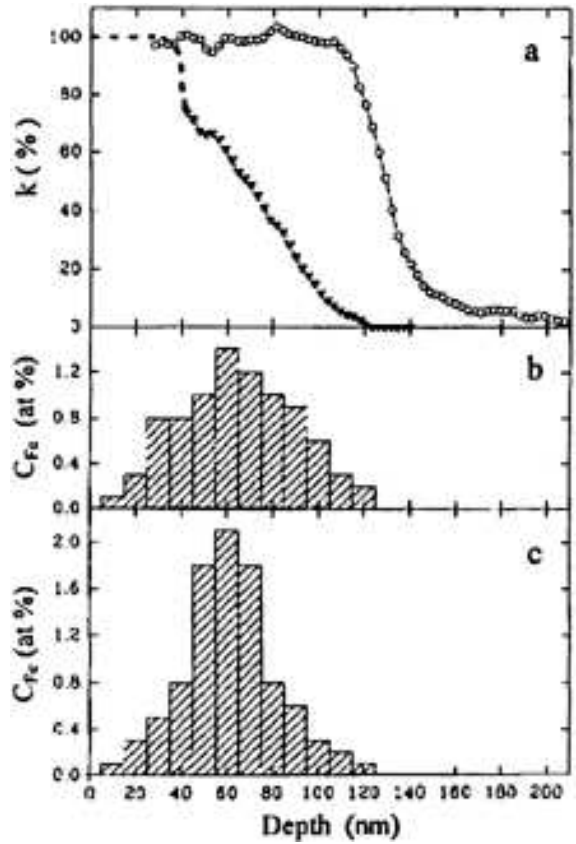}
\end{center}
\caption{
\textsl{\label{} (a) Damage density in silicon due to $80$ $keV$ $Fe$ ion
implantation for the sample just after implantation(circles) and after
annealing (triangles).
(b) Profile of $Fe$ impurity concentration for the sample just after
implantation.
(c) Profile of $Fe$ impurity concentration for the sample after annealing.
From Ref. \cite{VAN96}.}
}
\end{figure}

\clearpage

%
%
\begin{figure}[tbp]
\begin{center}
\includegraphics{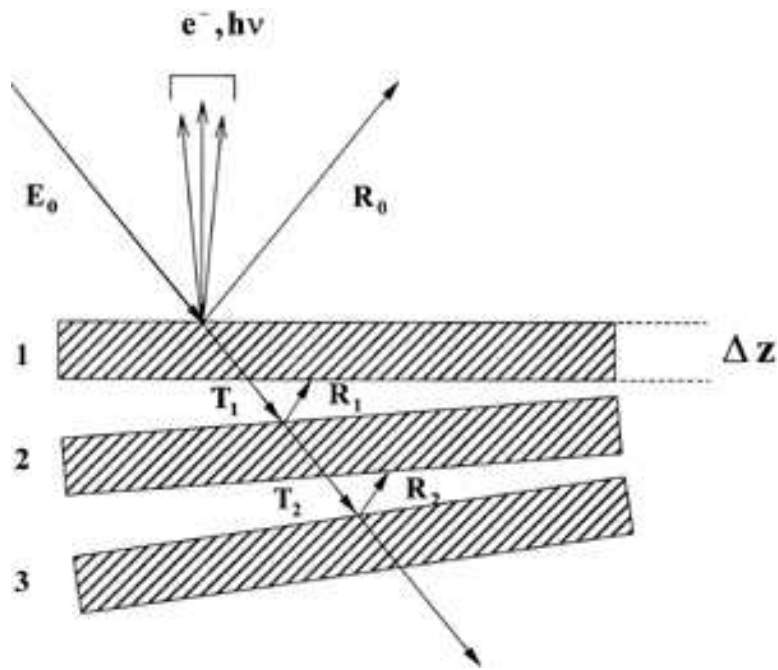}
\end{center}
\caption{
\textsl{\label{} Schematic view of the multilayer model used for the
calculation of the x-ray wavefield in a crystal with USG. Thickness of each
layer is $\Delta z=L_{ex}/(mC)$. The angular deviation parameter $y_i(\theta
)$ is constant in each layer and differs from layer to layer by the
magnitude $\Delta y(\theta )=1/m$. The amplitudes $E_0$, $R_n$, $T_n$ are
the amplitudes of incoming, reflected and transmitted beams from the n-th
layer respectively.}
}
\end{figure}

\clearpage

%
%
\begin{figure}[tbp]
\begin{center}
\resizebox{!}{15cm}{
\includegraphics{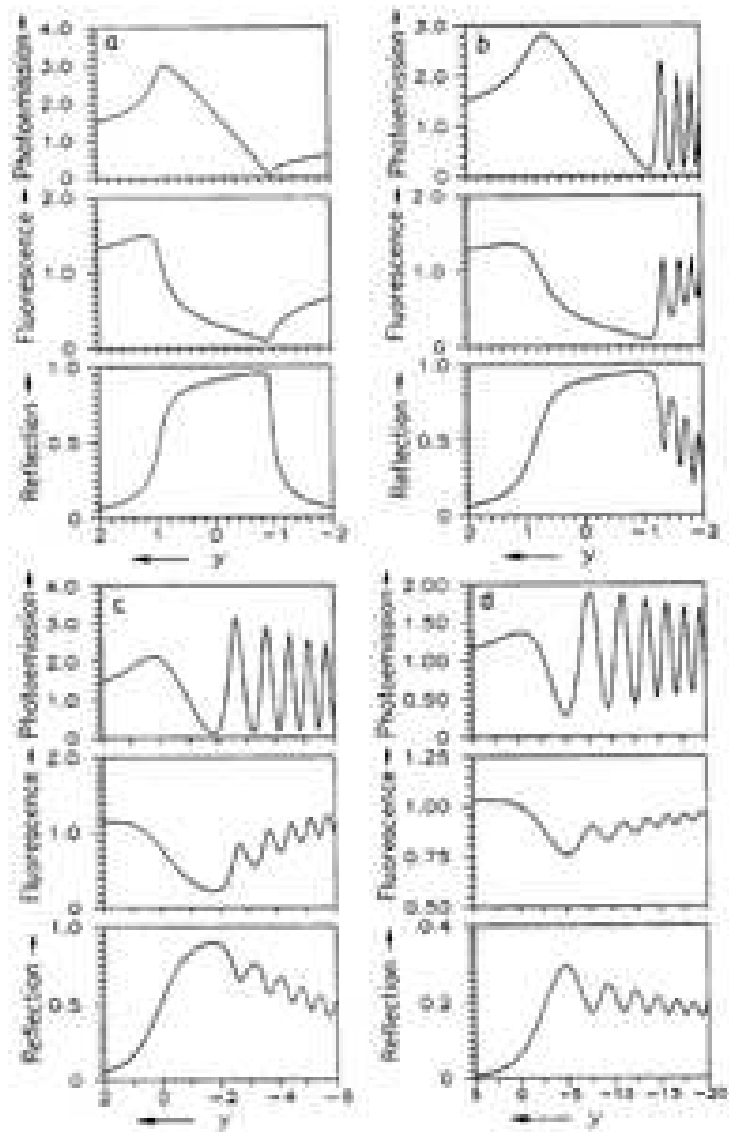}}
\end{center}
\caption{
\textsl{\label{} Angular dependences of the reflectivity, fluorescence and
photoemission yield calculated for a cylindrically bent $Si$ (400)
crystal, $CuK_\alpha $ radiation, at different values of the curvature
parameter $C$: (a) $0$ (perfect crystal), (b) $0.1$, (c) $1.0$, (d) $10$.
From Ref. \cite{VKB93a}.}
}
\end{figure}

\clearpage

%
%
\begin{figure}[tbp]
\begin{center}
\includegraphics{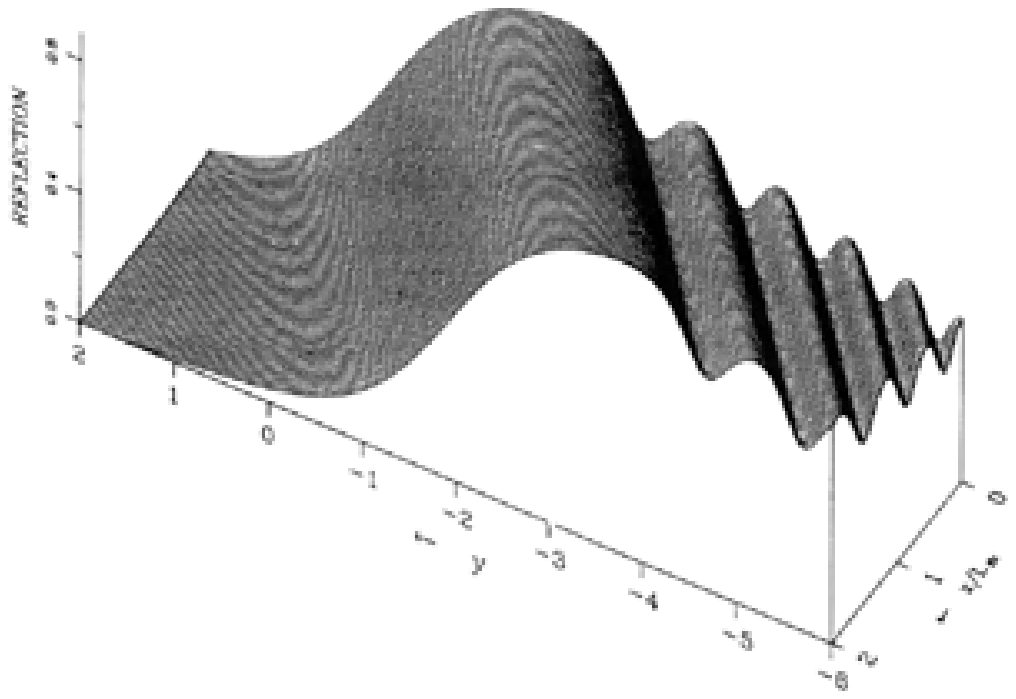} \\
\bigskip
\includegraphics{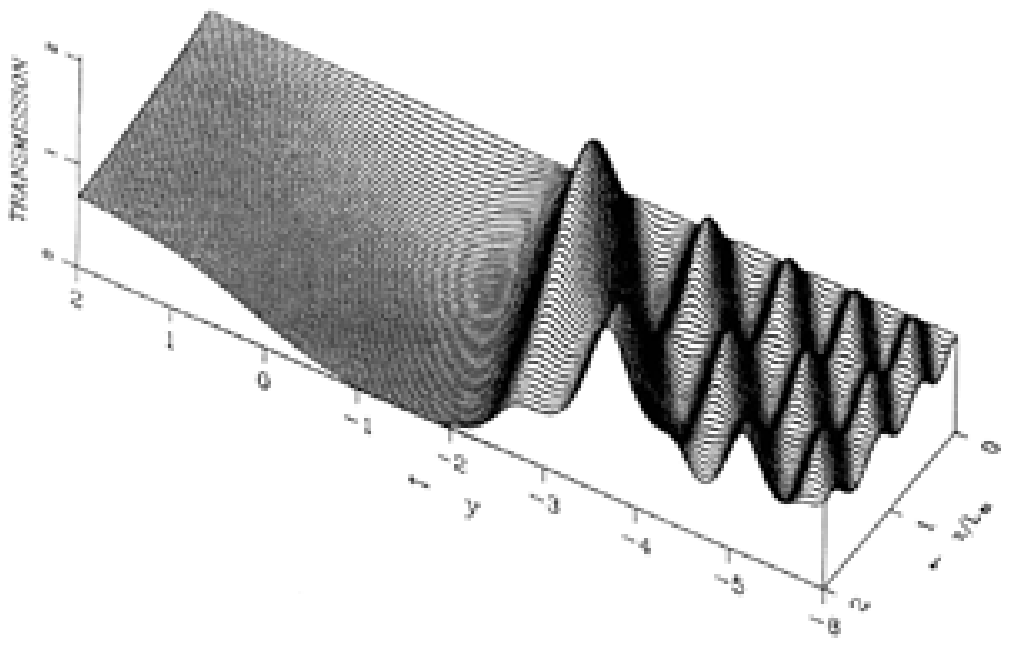}
\end{center}
\caption{
\textsl{\label{} The two-dimensional distribution of the reflectivity $%
P_R(z,y)$ (Top) and transmitted wave intensity $T^2=|E_0(z,y)|^2$ (Bottom)
calculated at the depths $z=0\div 2L_{ex}$ for the curvature parameter $C=1$%
. From Ref. \cite{VKB93b}.}
}
\end{figure}

\clearpage

%
%
\begin{figure}[tbp]
\begin{center}
\includegraphics{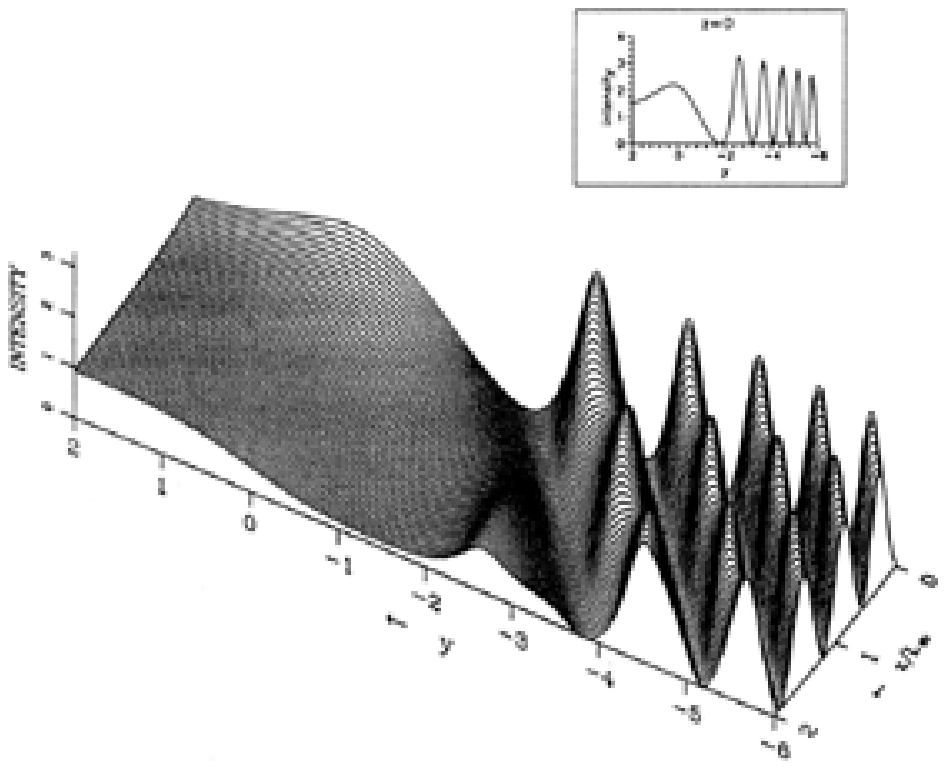} \\
\bigskip
\includegraphics{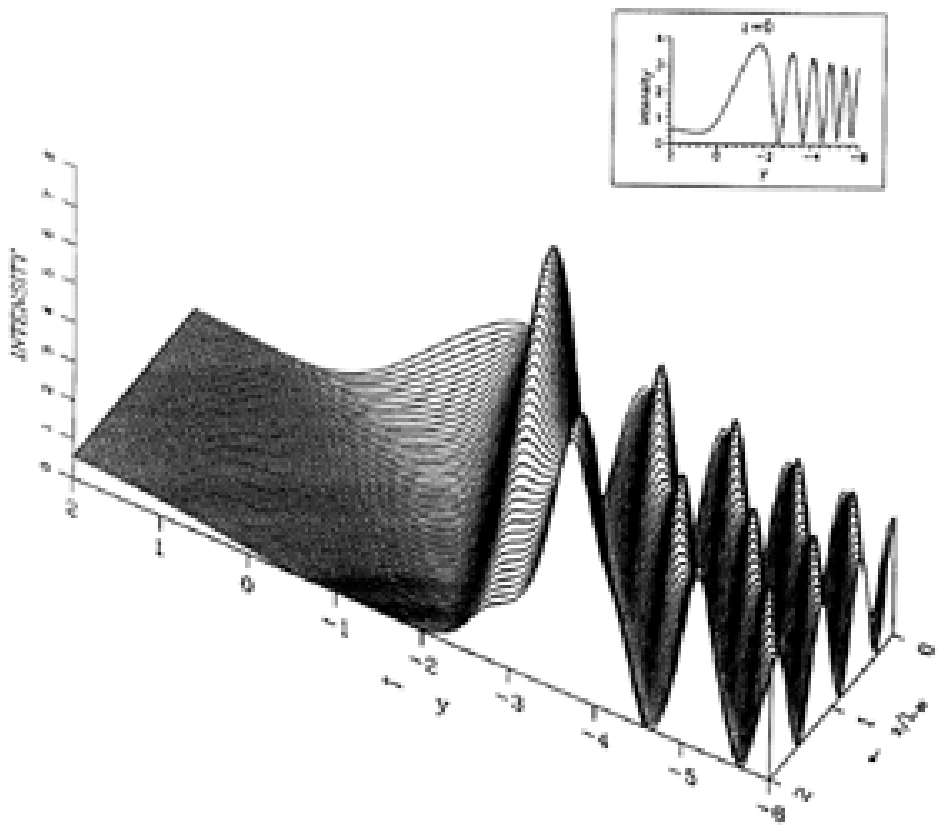}
\end{center}
\caption{
\textsl{\label{} The two-dimensional distribution of the standing wave
intensity $I(z,y)$ calculated at depths $z=0\div 2L_{ex}$ for the curvature
parameter $C=1$. In the inset the angular dependence of the intensity for $%
z=0$ is plotted.
(Top) Intensity values calculated on diffraction planes ($\Delta \varphi _c^a=0
$).
(Bottom) Intensity values calculated between diffraction planes ($\Delta \varphi
_c^a=\pi $). From Ref. \cite{VKB93b}.}
}
\end{figure}

\clearpage

%
%
\begin{figure}[tbp]
\begin{center}
\includegraphics{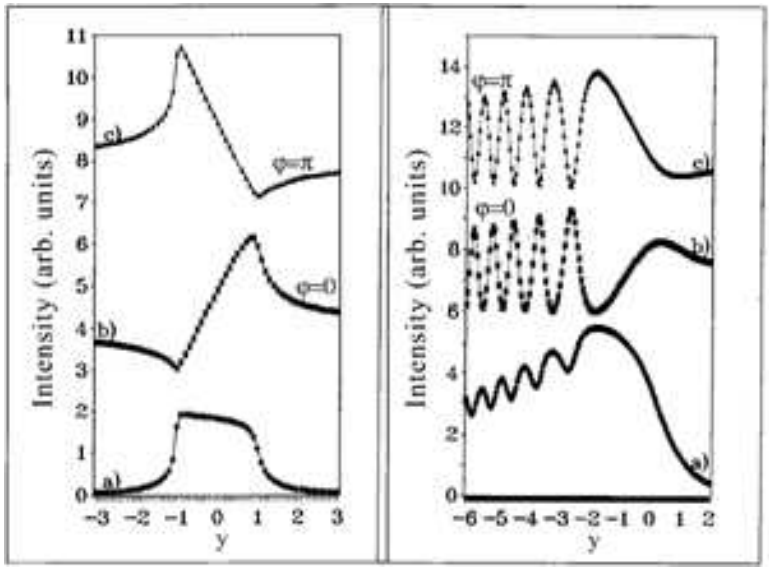}
\end{center}
\caption{
\textsl{\label{} The angular dependence of the x-ray reflectivity (a) and
the x-ray standing waves (b and c) calculated on the surface of the $%
Si$ (400) crystal without deformation (left figure) and with a USG
(right figure). The XSW\ curves correspond to different position (the phase
shift $\Delta \varphi _c^a$) of the impurity atoms on the surface of a
crystal. From Ref. \cite{VKU94}.}
}
\end{figure}

\clearpage

%
%
\begin{figure}[tbp]
\begin{center}
\includegraphics{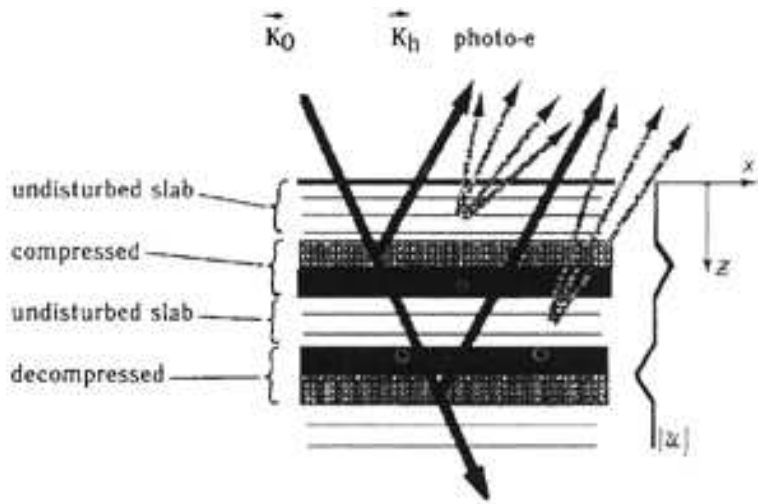}
\end{center}
\caption{
\textsl{\label{} Schematic view of the secondary radiation yield from a
vibrating crystal. This crystal is presented as a set of compressed and
decompressed layers.}
}
\end{figure}

\clearpage

%
%
\begin{figure}[tbp]
\begin{center}
\resizebox{!}{5cm}{
\includegraphics{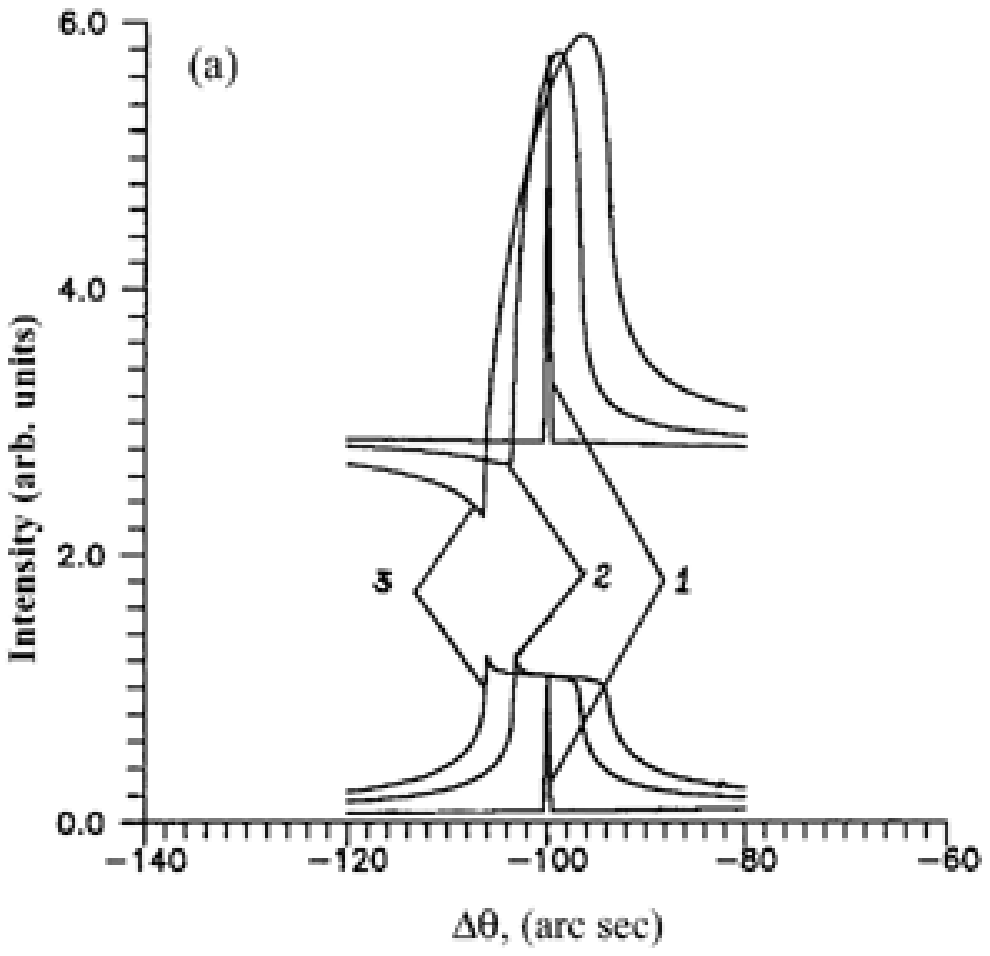}} \\
\bigskip
\resizebox{!}{5cm}{
\includegraphics{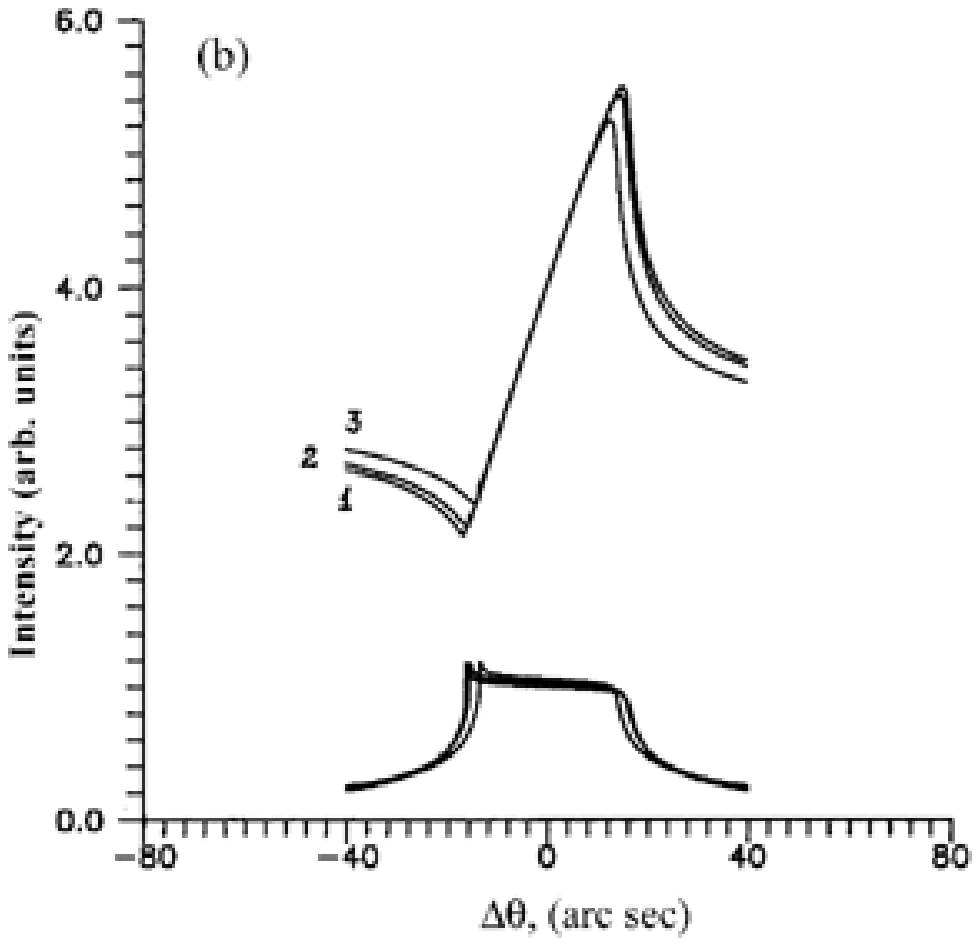}} \\
\bigskip
\resizebox{!}{5cm}{
\includegraphics{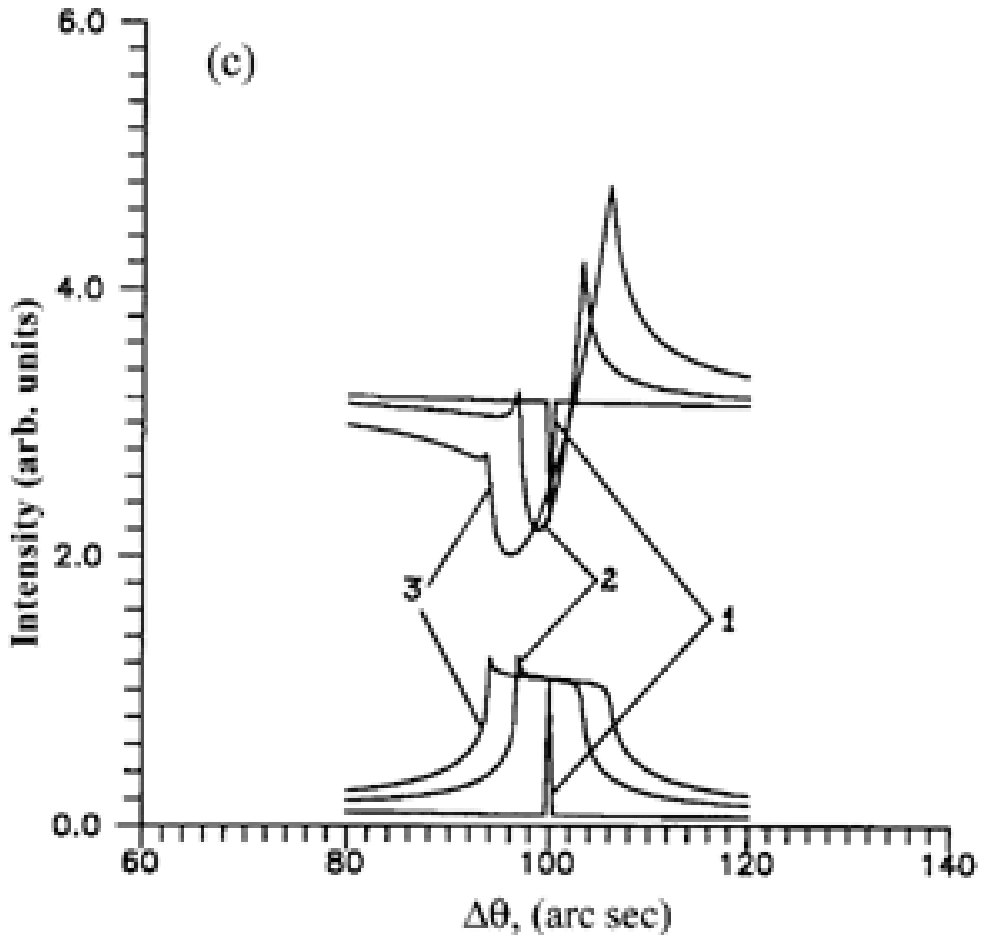}}
\end{center}
\caption{
\textsl{\label{} The angular dependence of the photoelectron yield and
reflectivity calculated for the values of the amplitude $\mathbf{hw}$ equal
to $0.1$ (curve 1), $0.5$ (curve 2), $0.9$ (curve 3). Figures (a) and (c)
correspond to satellites with $N=\mp 1$ and (b) to the main reflex ($N=0$%
).From Ref. \cite{NK98}.}
}
\end{figure}

\clearpage

%
%
\begin{figure}[tbp]
\begin{center}
\resizebox{!}{10cm}{
\includegraphics{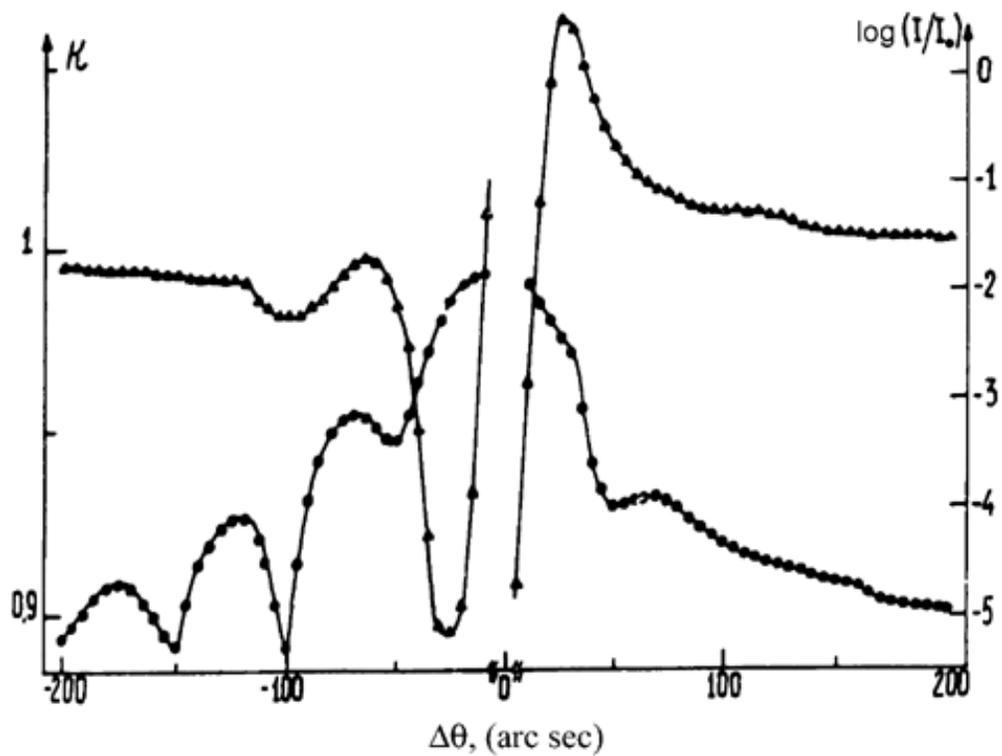}} \\
\bigskip
\resizebox{!}{10cm}{
\includegraphics{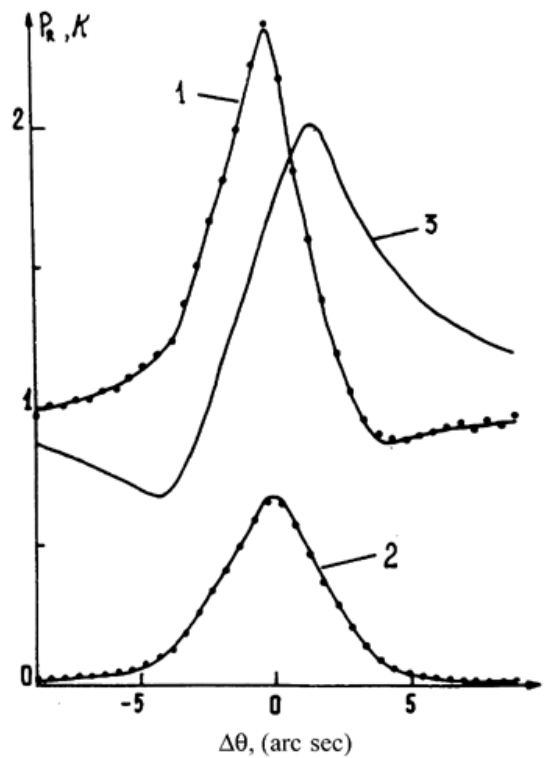}}
\end{center}
\caption{
\textsl{\label{} Photoemission yield (upper curves) and x-ray reflectivity
(bottom curves) measured at big deviations from exact Bragg condition (Top)
and in the region of the strong Bragg diffraction (Bottom). Points are
experimental data and lines are theoretical calculations. Curve 3 on Bottom Figure
 correspond to the photoemission yield from an ideal silicon crystal.
From Ref. \cite{VKK89}.}
}
\end{figure}

\clearpage

%
%
\begin{figure}[tbp]
\begin{center}
\includegraphics{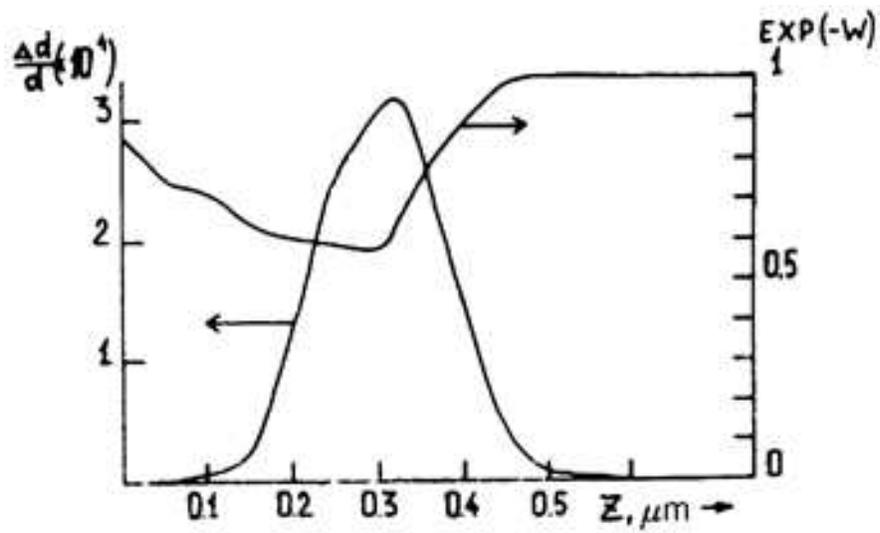}
\end{center}
\caption{
\textsl{\label{} Profiles of the deformation $\Delta d/d(z)$ and of the
static Debye-Waller factor $f(z)=\exp (-W)$ obtained as a result of the
phase retrieval from the angular dependence of the photoemission yield and
reflectivity (Fig. 42a). From Ref. \cite{VKK89}.}
}
\end{figure}

\end{document}